\documentclass[aps,amsmath,prb, a4paper,floatfix,twocolumn, longbibliography]{revtex4}
\usepackage{graphicx}
\usepackage{subfigure}
\usepackage{stackengine}
\usepackage{amsmath}
\usepackage{amsfonts}
\usepackage{amssymb}
\usepackage{mathtools}
	\usepackage[latin1]{inputenc}
	\usepackage[T1]{fontenc} 
	\usepackage{fancyhdr}
	\usepackage{dsfont}
	\usepackage{setspace}
	\usepackage[italian]{varioref}
	\usepackage[colorlinks=true,linkcolor=black,urlcolor=black,citecolor=black]{hyperref}
	\usepackage{makeidx}

\newcommand{\beq}{\begin{equation}}
\newcommand{\eneq}{\end{equation}}
\newcommand{\bea}{\begin{eqnarray}}
\newcommand{\enea}{\end{eqnarray}}
\newcommand{\met}{\frac{1}{2}} 
\newcommand{\freccia}{ \quad \quad \mathbf{\Rightarrow} \quad }

\newcommand{\RR}{\mathbb{R}}
\newcommand{\wt}{\widetilde}
\newcommand{\caj}{{\cal{J}}}

\usepackage{xr}

\begin{document}

\title{Superconducting critical temperature in the extended diffusive SYK model}
\author{ F.Salvati, A.Tagliacozzo$^{1,2,3}$}
\affiliation{$^{1}$ INFN-Sezione di Napoli, Complesso Universitario di Monte S. Angelo Edificio 6, via Cintia, I-80126
Napoli, Italy}
\affiliation{$^{2}$ Dipartimento di Fisica "E. Pancini", Universit\`{a} degli Studi di Napoli Federico II, via Cintia, I-80126 Napoli, Italy}
\affiliation{$^{3}$ CNR-SPIN, Monte S.Angelo via Cintia, I-80126 Napoli, Italy}

\begin{abstract}
Models for strongly interacting fermions in disordered clusters forming an array, with electron hopping between sites,  reproduce the linear dependence on temperature of the resistivity, typical of the strange metal phase of High Temperature Superconducting materials (Extended Sachdev-Ye-Kitaev (SYK) models). We identify the low energy collective excitations as  neutral, energy excitations, diffusing in the lattice of the thermalized, non  Fermi liquid phase. However,  the diffusion is heavily hindered by coupling to the pseudo Goldstone modes of the conformal broken symmetry SYK phase, which are local in space. The imaginary time evolution of the extended model in the strong  interaction and  $1/N$ expansion limit  is presented, in the incoherent non chaotic regime. On the other hand, a  Fermi  electronic liquid at low energy becomes marginal   when perturbed by  the SYK dots. A critical temperature for superconductivity is derived, which is not BCS-like,  in case the collective excitations are assumed to mediate an attractive Cooper-pairing.

\end{abstract}

\maketitle


\vspace{0.5cm}
\section{Introduction}
 Understanding the physics of copper-oxide materials, which undergo the superconducting transition at higher temperature, is still an unsettled topic of Condensed Matter Physics. Recent work suggests the breakdown of the Fermi Liquid (FL) theory at intermediate temperatures in these metals, while FL is the conventional starting point for low critical temperature superconductivity \cite{hill,jain}. New approaches to study high-temperature superconductivity are recently investigated, in particular lattice fermionic models with a strong local interaction \cite{chowdhury20A,lantagne}.

 Recently a $(0+1)-$ dimensional model, the Sachdev-Ye-Kitaev (SYK)\cite{sachdevye,kitaev,KitaevSoft} model, describing random all-to-all  ${\cal{J}}-$interaction between $N$ Majorana fermions, has been extensively studied. In the infrared (IR) limit, when $N$ is large and the temperature  $T$ is low, the model has an emergent approximate conformal symmetry and has become quite popular for its large-$N$ "melons" diagrammatics, which allows for a simple representation of the power-law decay in time of the correlation functions and for the analysis of the thermodynamic and chaotic properties\cite{KitaevSoft,maldacena}, providing a holographic dual for gravity theories\cite{sachdev10,diaz}.

 Generalized SYK models have been proposed with extension to higher space dimensions \cite{guQi,davison,song,berkooz,chew,haldarShenoy,patel18B,chowdhurySenthil} also having in mind applications to High Critical Temperature ($HT_c$) superconducting materials. Indeed, there seems to be widespread consensus that inhomogeneity and strong coupling could be distinguished factors for  the cuprates and their  $2-d$ $CuO$ planes. Moreover, universal features emerge in the high temperature "strange metal" phase,  which  is recognized as a Non Fermi Liquid (NFL) phase\cite{parc,hill,varma,jain,benZion,patel}. The most striking of these is the linear increase with temperature of the electrical conductivity\cite{gurvitch,daou,chaWentzell}. 
 
 The conformal symmetry of the SYK model is spontaneously broken down to the $\wt {SL}(2, \RR)$ group symmetry\cite{kitaevGeom} and Goldstone modes arise which are only approximately gapless, when ultraviolet (UV) corrections are taken into account. The nature and the role of these collective excitations has not been satisfactorily investigated, to our knowledge, up to now, in phenomenological approaches for the description of the low temperature metal phases of extended SYK models\cite{guQi}. 
 \begin{figure}[h]
\begin{center}
\includegraphics[scale=0.4]{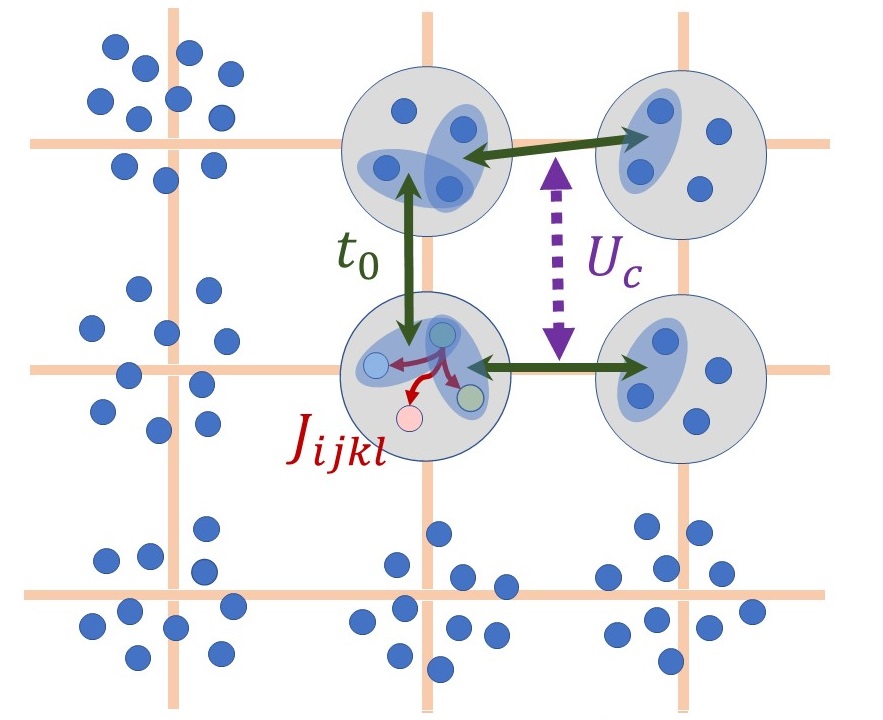}
\caption{A cartoon of the SYK model extended to include hopping between $2-d$ square lattice sites. The neutral fermions are depicted as small blue circles, grouped on the lattice sites (their number is $N$ on each site). In the right upper part of the picture, a magnification of four sites with just four Majorana particles at each site, to represent the all-to-all four fermion interaction $J_{ijkl}$ [{\sl red arrows}] and the residual interaction $U_c$ between quasiparticles (complex superposition of intra-cluster pairs of majoranas) [{\sl dashed arrow}] hopping between sites (with hopping matrix element $t_0$ [{\sl full dark green arrows}]). } 
\label{model}
\end{center}
\end{figure}

  We consider a lattice  of $(0+1)-d$ SYK clusters (or dots), each composed  of  strongly correlated  $N$ neutral fermions, via the SYK interaction. A sketch of the   lattice,   in two space dimensions, is depicted in Fig.\ref{model}. The first part of this work discusses the collective bosonic excitations in the lattice, which arise from the intradot  SYK  fermionic pseudo-Goldstone modes ($pGm$) in the incoherent highly thermalized phase above some threshold  temperature $T_0$.   We propose that these excitations, nicknamed Q-excitations,  could drive the transition to superconductivity, when lowering  $T$ below some temperature $T_{coh}$, at which coherence is established in tunnelling across the  lattice, but not necessarily in the SYK dots. To discuss the superconducting critical temperature $T_c$  of  the coherent phase, we adopt,   in the second part of this work,  an hydrodynamical picture consisting  of a two component system: the  two space dimensional lattice of  (0+1)-d SYK clusters and   a fermionic low energy liquid, weakly interacting with it.  The electronic, one-band  fluid  is  turned into a Marginal Fermi Liquid  (MFL) by the perturbation. The  SYK dots act as charge and momentum sinks. By contrast  the Q-excitations conserve momentum in the lattice, while the quasiparticles of the MFL are badly defined. In driving the superconductive instability, the  Q-excitations could play the same role as the magnons in the $^3He$ superfluidity\cite{bunkov}, though  via an unknown mechanism. As argued in the Conclusions of Section VII,  the validity of this hypothesis  can be experimentally  tested because it could produce anomalous intervortex interaction in presence of magnetic field.  However,   we are unable to describe the crossover between the high temperature and the low temperature phase, which should be further investigated, resorting to the various extended SYK models which have appeared in the literature\cite{haldarShenoy,pengfeiZhang}. 
   
Our approach to the  high temperature phase is one of the possible extensions of the SYK model\cite{song,chowdhurySenthil}, which assumes the SYK properties of the local critical two-point functions on the local scale, but introduces the U(1) symmetry for the ("interdot")  dynamics in the lattice.  It  is not really a complex fermion version of the SYK model\cite{fuSachdev,cha,davison,klebanov}. Indeed, charge is conserved only at low energies, while the  ("intradot") excitations in the SYK clusters are non conserving and  neutral. In this respect, we ignore the possibility of charging of the clusters at the sites of the lattice as if their capacity were infinite. 

Disorder is a distinct feature of the SYK model. Disorder averages make the SYK and its generalizations solvable. We assume random hopping  in the lattice and we assume  that self-averaging restores  space translational invariance.   In our description, the bilocal auxiliary fields $ G_x( \tau_1, \tau_2) $ and $ \Sigma_x( \tau_1, \tau_2) $, in imaginary time\cite{song,guQi}, acquire a slowly varying phase $\varphi_p (\tau)$. Here the subscript $x$ denotes the space coordinate and the wavevector $p = k \tilde{a}$ ($\tilde{a}$ is the lattice parameter) is used as a quantum number in the continuum space limit.  
$ G_x( \tau_1, \tau_2)  \equiv  G_x( \tau_{12}, \tau_+) $ acts as an order parameter which characterizes the SYK-phase in the scaling to strong interaction $ {\cal{J}} \to \infty, N\to \infty$ with  finite  $\beta {\cal{J}}/N $ ratio. Here $\tau_{12}$ is the relative time coordinate,  which takes care of the "intradot" dynamics, while $ \tau_+$ is the center of mass coordinate of the "slow" interdot dynamics.  The first task (Section III) is to study the correlations of the $pGm$'s, when minimal coupling to the compact dynamical  $U(1)$  gauge boson  $\varphi_x (\tau_+)$ is established. 
Fig.\ref{gcofig}  displays a  "dressed"  correlator,  $\langle  \delta G_{x} (\tau_{12}, \tau_+) \delta G_{x'}^*(\tau_{34}, 0)\rangle \approx  \langle \delta G_{x} (0^+, \tau_+) \delta G_{x'}^*(0, 0)\rangle $ compared with a zero order, "naked" one, continued to real time and in the limit $k \tilde{a} <<1$. The naked correlation can be derived with a real time approach in appendix \ref{app:secC}. Both correlators decay with real  time, but the dressed one decays by far faster. This confirms that the extended SYK model at hand describes incoherent dynamics.  However, Fig.\ref{gcofig} proves  that, as long as the gauge boson lacks its own dynamics, correlations cannot be said to be diffusive over the lattice.  Actually,     diffusivity  on a temperature dependent  (and scaling-dependent) space distance $\tilde{a}_\ell (T)$, much larger than the lattice parameter, is expected. In fact, the presence of impurities, low dimensionality, strong interaction and disorder, usually makes the collective excitations diffusive at low frequencies and small momentum\cite{varma}.

  The $pGm$  fermionic  excitations of the SYK dots  generate  fluctuations of the chemical potential in the lattice  $\langle \partial_\tau \varphi _p(\tau_+) \partial _\tau \varphi _p(0) \rangle $, driven by quasiparticle hopping between lattice sites, parametrized by the matrix element $t_0$ and  produce  the bosonic  Q-excitations. 
     
     Our aim is twofold. On the one hand  we want to characterize the quantum diffusion of the Q-excitations   in the lattice\cite{guQi}. On the other hand  we want to study the  response  $D^\beta (p,\Omega_n)$, of these modes to  interdot tunnelling, a $J_Q-J_Q$ response, where $J_Q$ is an energy flux density which is somehow  canonical conjugate to $ \partial_\tau \varphi _p(\tau_+) $. The latter plays the role of a space dependent  chemical potential across the lattice\cite{tagliacozzo}.

 The probability of quantum diffusion, involving retarded and advanced Green's function in real time,  $G^R$ and $G^A$ respectively, is written  in the form:
\bea
\textit{P} \left ( r,r'; \Omega  \right )  \propto  \int  d\omega \:  \overline{ G^R(r,r'; \omega ) G^A (r',r; \omega -\Omega)},
\label{diffuson}
\enea
where overline denotes disorder average. On the other hand  the  retarded density response function $ D^R_{J_QJ_Q}$ involves the retarded  $G^R$ and the Keldysh $G^K$ Green's function. Our approach will be  in imaginary time, but correct time ordering is crucial to guarantee a correct analytical continuation to real times.  A relevant quantity typical of the  diffusion processes is its  Fourier transform in time, denoted as  the heat kernel\cite{akkermans}, which is  defined as the probability $ z(t) $ to return to the origin, integrated over the point of departure.  In Section V we derive a form of it,  $z(t) \sim e^{\tilde{D}_Q\: t \:\nabla ^2 }$, after the Q-excitations have been integrated out  (Eq.(\ref{return})) and  we determine how the  diffusion parameter $\tilde{D}_Q$ depends on the scaling to strong interaction. 
 
 In dealing with the "bad" metal at finite temperature $T$, we concentrate on  two  temperature  scales involved in the extended SYK model,  $T_0$ and $T_{coh}$. 
  At   $ T\lesssim T_{coh} \sim  t_0^2/N {\cal{J}}$, transport in the lattice  is assumed to acquire  coherence.  This crossover  is out of reach in the present work. We expect that  the Q-excitations  merge into the particle hole (p-h) continuum of the   low energy  MFL. A derivation of the Landau damped acoustic plasmon embedded in the p-h continuum, is reported in the appendix \ref{app:secE}.
  At $ T\gtrsim T_0$,  thermalization in the system is very effective and  diffusion is incoherent. $pGm$'s are the intradot excitations which drive the incoherence.  This is a feature of the SYK model and is attained in the present extended version of the model.   The relaxation time is $\sim \hbar \beta$\cite{hartnollNatPhys}. At later times the system evolves toward the scrambled phase  and the chaotic dynamics,  as the analysis of the out-of-time ordered correlator (OTOC) shows.  The fate of the chaotic  single dot  regime in the extended model deserves specific concern\cite{haldarBanerjee,benZion,pengfeiZhang,khveshchenko} beyond the present paper.
    
  Usual hydrodynamical approaches to the response function  $D^R$ do not involve the role of the $pGm$'s  at energies  $\sim k_BT_0$.  This is highly questionable, because  the diffusion constant, $\tilde{D}_Q$, is strongly renormalized by the inverse of four point function of the SYK dots, ${\cal{F}}^{-1}$.  Indeed the  first UV correction plays the role of keeping the $pGm$ propagator  ${\cal{F}}$ finite.  A brief presentation  of this  approximation    to  $D^R$, which does not go beyond the conformal limit \cite{song} and uses the real time Keldysh contour, is reported in the appendix \ref{app:secC}. 
By contrast, our approach is quite simple and even naive, but it  aims to stress the parameter renormalization in the scaling process.   In fact,  the separation in energy of   $T_0$ and $T_{coh}$  allows us to perform a kind of adiabatic factorization, between the  "fast" intradot  $pGm$'s and the  "slow" interdot Q-fluctuations.  We discuss the UV  local space-time correction  and show how they influence   the time correlation  of the Q-excitations.  

 Physically, we  concentrate in distinguishing the two regimes  $ T \lessgtr T_0$.  The $T\gtrsim T_0$ regime, being characterized by  strong thermalization, is governed by the order parameter of the SYK model which, in the UV corrected form,  is described by a complex field $\phi$ in Section IV. The Q-excitations, arising from the minimal coupling with the gauge mode,  are interpreted as energy excitations  induced by the  fluctuations of the chemical potential.  Energy density ${\cal{N}}$ and energy flux density  $\dot{{\cal{N}}} \sim J_Q$ are the  physical dynamical variables\cite{davison}. The corresponding  parameters  which rule the response are  thermal capacitance  $C_{\cal{V}}$ and the  thermal conductivity $\kappa $.  
 
The structure of the paper is as follows. 

In Section II the extended SYK model is presented.  In the conformal symmetry limit of our approach, the SYK clusters acquire an hopping dependent selfenergy of the kind $\sim t_0^2 G_cG_c$, where $G_c( \tau_1, \tau_2) $ is the fermionic propagator of the SYK model\cite{song}. A term of this kind is suggested by a simple derivation of the hopping between two neighbouring SYK sites. The local correlations arising from the kinetic term are obtained by gaussian integration of the $\delta g_m$ fluctuations  in presence of a source term, the chemical potential   $ \partial_\tau \varphi_x ( \tau_+)$. They  are derived in Section III.  In Section IV we clarify that the  proper  dynamics of  the chemical potential fluctuations  should be added to account for the UV corrections which, by giving mass to the $pGm$'s, make the partition functional convergent. This  implies a renormalization  of the correlations  provided by the $pGm$  propagator ${\cal{F}}$, in which the first UV correction is included.  To this  end we  introduce  a  complex local order parameter $\phi (x,\tau_+)$, which is promoted to a bosonic coherent field in Section IV, by means of a more conventional model for the Q-excitations. The inclusion of the dynamics via the local  action   $\tilde{S}_{2}[ \partial_ \tau \varphi (x,\tau_+)]$ of Eq.s(\ref{sdue},\ref{trsm}) implies that the short range, exponentially decaying dependence on real time  $t_+$ of the correlators turns into a diffusive dynamics  for $T \gtrsim T_0$,  the energy window  in which  our approximations are justified (Section V.A). Section V.B discusses qualitatively  how  the transport parameters evolve with scaling  in the incoherent and coherent energy ranges. They can be used to qualify the diffusion parameter $\tilde{D}_Q$ by means of the Einstein relation.  In Section VI we show how a coherent low energy FL, when perturbed by a  higher energy SYK-type  environment,  becomes marginal.  A conventional Eliashberg\cite{marsiglio,chowdhury20B} approach to the gap equation is presented in Section VI, where the Q-excitations constitute a   bosonic virtual pairing mechanism  but with diffusive dynamics. The self-consistent equation for the non BCS  critical temperature $T_c$ is derived.  Additional remarks and a summary are reported in the Conclusions (Section VII).  The Appendices give details of the derivations.

  \section{the extended SYK model} 

    Let the Hamiltonian for the extended model be  $ {\cal{H}}_0 +{\cal{H}}_K$. $ {\cal{H}}_0 $ is the sum of the  neutral fermion Hamiltonians of uncoupled  $0+1$-d SYK dots, ${\cal{H}}_a $,  in a two-dimensional  lattice with  intradot random interaction,  labeled by  the lattice site $a$,  and    $ {\cal{H}}_K$ adds the  kinetic energy of  electrons  with interdot random hopping between neighbouring  dots. $ {\cal{H}}_K$  (given by Eq.(\ref{hamom})) is derived in this Section.  The   Hamiltonian $ {\cal{H}}_0 $ for the uncoupled  $0+1$-d SYK dots  is:
    \bea
    {\cal{H}}_0 =  \sum_a  H_a =   \frac{1}{4!}\:  \sum _a \sum _ {klmn} {\cal{J}}_ {aklmn}    \chi_{a,k}  \chi_{a,l}  \chi_{a,m}  \chi_{a,n},
     \label{je0}
     \enea
     where $ \chi_{a,l} $ are Majorana fermion operators on site $a$ ($ {klmn} \in 1,..,N$).
     
      Electronic quasiparticles  hop from  site $a$ to  a neighbouring  site $b$.  $  c_j^\dagger , c_j  ( j=a,b) $ are the complex fermionic spinless operators for the electrons, which can  
   be   represented   in terms of two  flavours  of  the neutral fermions on the same site:
     \bea 
     c_b = \frac{1}{\sqrt{2}} \left ( \chi_{b1} +i \: \chi_{b2} \right ), \:\:\:  c_b^\dagger  = \frac{1}{\sqrt{2}} \left ( \chi_{b1} -i \: \chi_{b2} \right ).
     \enea
     The kinetic term describing the hopping can be written as  $ h_K =  t_0 \:  c_b^\dagger  c_a  + h.c.$,   where $t_0$ is a constant hopping energy.
     
       The time dependence of the operator $c^\dagger _b$  in the interaction picture is:
      \bea
      - \frac{\partial}{\partial \tau }c^\dagger _b =    e^{ \tau (H_b+ H_a)} \left [ c^\dagger _b,{\cal{H}}_0\right ]\:  e^{- \tau (H_b+ H_a)} .
      \label{inter}
      \enea
   The commutator with the Hamiltonian can be performed  by   applying  the commutation relations for neutral fermions: $    \chi_{a,k}  \chi_{a,l} + \chi_{a,l}  \chi_{a,k} = \delta _{l,k} $ and  $    \chi_{a,k}  \chi_{b,l} + \chi_{b,l}  \chi_{a,k} = 0$ for $ a\neq b$, exploiting the antisymmetry  of  $ {\cal{J}}_ {bklmn} $  in the  permutation of the $ {klmn}$ indices. From Eq.(\ref{inter}) we get:
 
\bea
   \frac{\partial}{\partial \tau } c^\dagger_{b} (\tau) =  i\:  \frac{1}{3!} \sum _{lm}�J_{b12lm}  \chi_{b,l}(\tau)  \chi_{b,m}(\tau)  \:  c^\dagger_{b} (\tau). 
   \enea
       $c^\dagger_a$ commutes  with $H_b$ so that it can be added afterwards.  The hermitian conjugate term $c^\dagger_bc_a$ gives the same result with $b\to a $, $i\to -i$.
       
 This allows to identify the hopping Hamiltonian term  in the interaction representation, from the evolution operator  in a single hopping process, $\delta U(\tau,0)$, to lowest order: 
    \bea
 {\cal{H}}_K (\tau) =   i\: \frac{1}{3!} \sum _{lm,j}�J_{j12lm}  \chi_{j,l}(\tau)  \chi_{j,m}(\tau)  + h.c.  .
 \label{hamom}
\enea
Here $J_{j12lm} $ is random interdot hopping for hopping onto site $j$. Eq.(\ref{hamom}) shows that,  starting from the neutral fermions of the SYK model, a symmetric description of conserving and non-conserving charge processes is provided. This feature sets charge (and spin) dynamics free with respect to energy dynamics, which is the premise for NFL behavior. 

 The disorder average of  the standard SYK model  includes here  the gaussian average of $J_{j12lm}$.  The next step is  the integration over the Majorana fields  $ \chi_{j,l}(\tau)$,  with the help of Hubbard-Stratonovich fields which become complex due to an additional $U(1)$ minimal coupling. The final result is the action in terms of the complex bilocal auxiliary fields $ G_x( \tau_1, \tau_2) $ and $ \Sigma_x( \tau_1, \tau_2) $, with a phase $\varphi _x $  introduced in the next Section \cite{song}:  
\begin{widetext}
  \bea
  \frac{I_{ex}}{N}  =\sum _x\left [ - \ln  Det\left [ \partial _\tau \!- \Sigma_x \right  ] \!+ \!\int \!d\tau\: d\tau ' \!\left \{\! - \frac{ J^2 }{4}\left | G_x(\tau,\tau')\right |^4 \!+\! \Sigma_x(\tau,\tau')G_x^*(\tau,\tau') -\frac{t_0^2}{N} \! \sum _{x'\in  nn} G_x(\tau,\tau')G_{x'}^*(\tau,\tau')\!\right \} \right ].
  \label{aca}
  \enea
 \end{widetext}
 The last  term of the action  is the interdot kinetic term. The  expansion up to quadratic  terms of this action   in $\delta \Sigma_x, \: \delta G_x,\: \partial _{\tau} \varphi _x $ is discussed in Section III and in the appendix \ref{app:secA}.
 The  single dot $0-1$-d SYK  action can be recovered by dropping  $\delta \Sigma_x$,  the last term   and the sum over sites.  The auxiliary fields are now real and the Det has to be substituted with a Pfaffian. In this case the IR limit corresponds to the dropping of $\partial _\tau$ in the Pfaffian.  On the contrary, $\partial _\tau$ plays an important role in the extended model. 
 

\section{Kinetic correlations of the extended SYK model}  
 The single particle  Green's function of the SYK model, in the conformal symmetry limit, is local in space (i.e.  wavevector independent) and, assuming particle-hole (p-h) symmetry and low temperature, it is  given by: 
\bea
 G_c(i\omega_n ) = i \: \frac{sign(\omega_n)}{ \sqrt{ {\cal{J}}} \sqrt{ | \omega _n|}}, 
  \label{galt}
 \enea
 where  $\omega_n$ are fermionic frequencies.
 
\begin{figure}
 \centering
\def\big{\includegraphics[height=5.8cm]{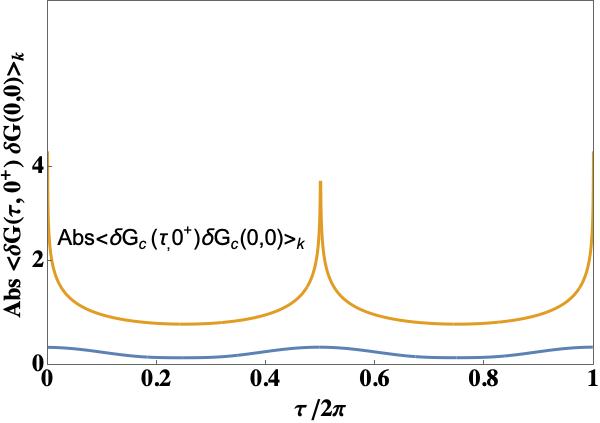}}
\def\little{\includegraphics[height=4.3cm]{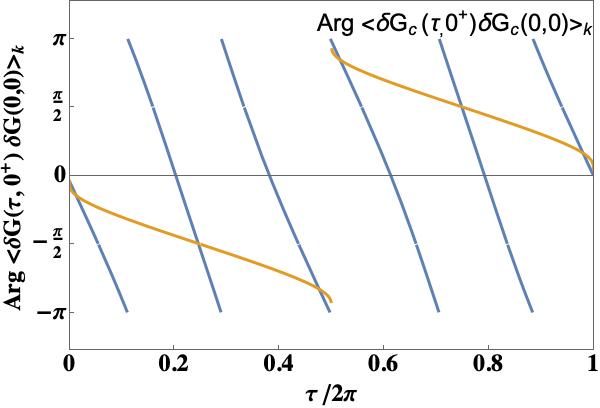}}
\def\stackalignment{l}
\topinset{\little}{\big}{-55pt}{+55pt}
\caption{ The modulus and the phase of  $\langle \delta G( \tau,0^+) \delta G(0,0)\rangle_k$ (blue curves), obtained   in  Eq.(\ref{fifi}), by averaging with  the density matrix of Eq.(\ref{evolmix}),  are plotted vs  the dimensionless intradot  imaginary time $\tau_{12}-\tau_{34} \equiv \tau$, in comparison with  the naked one from Eq.(\ref{gco}) (orange curves).  The prefactor  $k^2{\tilde a}^2\: \beta t_0^2/( 2\pi {\cal{J}})$, which contains the $k$ dependence,    has been dropped in the plots.  } 
\label{imog}
\end{figure}

 Our aim is to include correlations between sites of the lattice, here denoted by the subscript $x$. The Green function and the self-energy become complex fields, 
   $G _x(\vartheta_1,\vartheta_2) ,\: \Sigma_x (\vartheta_1,\vartheta_2)$. They include  space dependent fluctuations of the modulus and of the phase, close to the saddle point  $G_{c }(\vartheta_{12}), \: \Sigma_c (\vartheta_{12}) $:
 \bea
G_{x}(\vartheta_1,\vartheta_2) = \left [G_{c }(\vartheta_{12})+ \delta G(x,\vartheta_{12}, \vartheta_+ )\right] \: e^{i \: \varphi _x(\vartheta_+) },\nonumber\\
 \Sigma_{x}(\vartheta_1,\vartheta_2) = \left [ \Sigma_{c }(\vartheta_{12})+ \delta \Sigma (x,\vartheta_{12}, \vartheta_+)\right ] \: e^{i \varphi _x(\vartheta_+) },
 \label{greo}
\enea
where  $\vartheta_{12}=\vartheta_1-\vartheta_2$ and  $\vartheta _+ = (\vartheta_1+\vartheta_2)/2$ .    We have moved   to the center of mass time coordinate  $\vartheta_+ $   and the relative time  coordinate   $\vartheta_{12} $  of the incoming particles and of the outgoing ones.  Here $ \vartheta= 2\pi \tau / \beta $ is a dimensionless time and the Green's functions and selfenergy are also dimensionless, everywhere, except when explicitly stated.   Nevertheless we will most of the times denote the dimensionless time as $\tau$, unless differently specified.  To spell out the structure of the kinetic term,  we calculate the correlator  of the $\delta G$ fluctuations between neighbouring sites   and Fourier transform  it with respect to space.  Ignoring the relevant role of the  $pGm$'s, we neglect, in the  IR limit, the local correction $ \delta G\left (x,\tau_1-\tau_2, \tau_+\right ) \: e^{i \: \varphi _x(\tau_+) }$ appearing in Eq.(\ref{greo})  and   we consider just    nearest neighbour  $x,x'$  terms  in a lattice of spacing $ {\tilde a}$. We get: 
 \bea
  \delta G_{c,x} (\tau_{12}, \tau_+) \delta G_{c,x'}^*(\tau_{34},\tau_+')\hspace{1cm} \nonumber\\
  =
 \left [  G_{x}(\tau_1,\tau_2)\: G_{x'}^*(\tau_3,\tau_4) -G_{c }(\tau_1-\tau_2)\: G_{c }(\tau_3-\tau_4) \right ] \nonumber\\
   \approx \frac{1}{2}\: G_{c }(\tau_{12}) \:\left ( e^{-i   {\tilde a}\cdot  \nabla_x \left [\varphi _x(\tau_+)-\varphi _x(\tau_+')\right ]} -1 \right ) \: G_{c }(\tau_{34}),  \nonumber\\
 + c.c. \hspace{6cm}  \nonumber
 \enea
where we have  qualified  the  lowest order, originating from the conformal Green's functions, with the label $c$.  Only the quadratic terms  of the exponential are  included in the expansion, to account for the additional complex conjugate contribution, giving 
\bea
\approx  - \frac{1}{2}\: G_{c }(\tau_{12}) \left ( {\tilde a}\cdot  \nabla_x \left [\varphi _x(\tau_+)-\varphi _x(\tau_+')\right ]  \right )\!\!^2\: G_{c }(\tau_{34}).
     \label{pro2}
     \enea
       We now approximate \eqref{reti} $ \left [\varphi _x(\tau_+)-\varphi _x(\tau_+')\right ]  \approx \left  (\tau_+-\tau_+' \right )\partial _\tau \varphi _x(\tau_+)$ and define 
     \bea
      R_c^{-1}\Lambda_c  R_c^{-1}= \nonumber\\
     \frac{1}{2} ( {\tilde a}\cdot   \overset{\leftarrow}{ \nabla}_x  )G_{c }(\tau_{12})(\tau_+-\tau_+')^2 G_{c }(\tau_{34}) ( {\tilde a}\cdot   \overset{\rightarrow}{ \nabla}_{x'} ).
     \label{lalo}
     \enea
      Owing to the
self-averaging established for the SYK model at large N,  translational invariance  allows space Fourier transform:
          \bea
   \frac{1}{N}     FT_p\left [   \delta G_{c,x} (\tau_{12}, \tau_+) \delta G_{c,x'}(\tau_{34},\tau_+') \right ] \nonumber\\
   =\frac{\delta^2}{\delta  \partial_\tau\varphi_p(\tau_+) \: \delta  \partial_\tau\varphi_p(\tau_+')} 	\frac{1}{N} \langle  \partial _{\tau} \varphi _p  |  R_c^{-1}\Lambda_c  R_c^{-1}| \partial _{\tau} \varphi _p  \rangle
                      \label{qrs}
           \enea
           ($FT _p$ denotes  Fourier Transform with respect to the space coordinate of lattice spacing $ {\tilde a}$, with $p= k  {\tilde a}$).
           We now express Eq.(\ref{qrs})  in the frequency space.  The  matrix elements of the kernel are labeled by   $m,m',\ell$ indices. $m,m'$ indices refer to the intradot fluctuations $\delta g$ which are fermionic in the origin, while $\ell$ labels bosonic frequencies $\Omega _\ell$, corresponding to  the spectrum of the Q-fluctuations.
    We get        
           \bea 
  	\frac{1}{N} FT\left [\delta G_{c,x} (\tau_{12}, \tau_+) \delta G_{c,x'}(\tau_{34},0) \right ] | _{ k \neq 0 }  =  \frac{  \beta t_0^2}{N} \:  k^2 \tilde{a}^2 \: \nonumber\\
 \times \sum _{ \ell }  \sum _{ m,m'} e^{i \Omega _\ell \tau_+} \left ( \widehat{{R_c^{-1} \Lambda_cR_c^{-1}}}\right )_{m,m'}^{\ell} e^{i\: \omega _m \tau_{12} }e^{i\: \omega _{m'} \tau_{34} }\!.
 \label{lib}
\enea
 Restricting  ourselves to the IR  limit,  we plot   in Fig.\ref{imog} the time Fourier transform,  keeping just the  dependence on the  relative coordinate $ \tau_{12}- \tau_{34}\:\: \: mod.\:\:2\pi$ ($\omega _{m'}= -\omega _{m}$),
  \bea
 FT_k\left [ \delta G_{c,x} ( \tau, \tau_+)\: \delta G_{c,x'}(0,0)\right ] \nonumber\\
 \approx  k^2 \tilde{a}^2\: \frac{ t_0^2}{N}\: \frac{\beta}{{\cal{J}}}  \sum _{\ell\geq 2}  \frac{ e^{i \Omega _\ell \tau_+} }{\Omega _\ell^2} \: \sum _m \frac{1}{\pi (2 m+1)} \:  e^{i\: \omega _m \tau} ,
\label{gco}
 \enea
 where Eq.(\ref{galt}) has been used.  It  is denoted  as $\langle \delta G_c( \tau,0^+) \delta G_c(0,0)\rangle_k$ in   Fig.\ref{imog}. 
This quantity, together with the dressed correlator of Eq.(\ref{ecco}) (blue curves), is plotted for $\tau_+\to 0^+$.  The prefactors  $k^2{\tilde a}^2\: \beta t_0^2/( 2\pi {\cal{J}})$  have been dropped in the plots.  The real part of the continuation of Eq.(\ref{gco})  to real time $t_+$, when $\tau \to 0^+$, $ \Re e\langle \delta G_c( 0^+,t_+) \delta G_c(0,0)\rangle_k$,  is plotted  in Fig.\ref{gcofig}.   Note the difference in the scale of decay between this correlation  derived from the naked kinetic term and the one of Eq.(\ref{ecco}), including  UV corrections, which  we are going to discuss in detail  in the next Section. 


 Integrating out the  $\delta \Sigma $  fluctuations (see \eqref{gentu}), the functional integral in terms of the fluctuations $\delta g(\tau_{12},\tau_+)$ is
 \begin{widetext}
 \bea
 {\cal{Z}}\left [\partial_\tau\varphi_p(\tau_+) \right ] = \int \left ( \Pi \delta g^*_{\tau_{12},\tau_+} \right ) \left (\Pi  \delta g_{\tau_{12},\tau_+} \right )\: 
  e^{\frac{N}{4} \left [\langle \delta g  | K_c ^{-1}-1| \delta g \rangle\right ]} \: e^{-\frac{N}{2} \Re e \left \{ \left \langle - i\: R_c^{-1} \partial _{\tau} \varphi _p | \delta g \right \rangle \right \} }\nonumber\\
  \times e^{\frac{N}{4} \frac{{ t_0}^2}{N}p^2 \left [\langle - i\: R_c^{-1} \partial _{\tau} \varphi _p  | \Lambda_c | - i\: R_c^{-1} \partial _{\tau} \varphi _p  \rangle\right ]},
 \label{immf}\\
 K_c ( \vartheta _1 ,\vartheta _2, \vartheta _3, \vartheta _4) = R_c (\vartheta _1 ,\vartheta _2) {G}_c( \vartheta _1, \vartheta _3) {G}_c( \vartheta _4, \vartheta _2) R_c(\vartheta _3, \vartheta _4)  \hspace{3cm} \nonumber\\
  = (\beta {\cal{J}} )^2 \: (q-1 ) \left | {G}_c (\vartheta _1 ,\vartheta _2) \right |^{\frac{q-2}{2}} {G}_c( \vartheta _1, \vartheta _3) {G}_c( \vartheta _4, \vartheta _2) \left | {G}_c (\vartheta _3 ,\vartheta _4) \right |^{\frac{q-2}{2}},\nonumber\\
  R_c^{-1}\Lambda_c  R_c^{-1}  =  FT_p\left [ \delta G_{c,x} (\tau_{12}, \tau_+) \: \delta G_{c,x'}^*(\tau_{12}, \tau_+)\right ].
    \label{trasf1}
 \enea

\end{widetext} 
 The forks $\langle ...\rangle$ in Eq.(\ref{immf})  include integration over  $\tau_{12}$ and $\tau_{+}$.  Here $ g(\tau_1,\tau_2) = R_c(\tau_1,\tau_2)\: G(\tau_1,\tau_2)$
 and $ R_c (\tau _1 ,\tau _2)
 = \beta {\cal{J}} \sqrt{(q-1)} \left | {G}_c (\tau _1 ,\tau _2) \right |$ (with $q=4$ in the usual notation). 
 
Integrating out  $\delta g_{\tau_{12},\tau_+},\delta g^*_{\tau_{12},\tau_+}$, the generating functional of the $\delta g$ fluctuations reads: 
   \bea
   {\cal{Z}}\left [\partial_\tau\varphi_p(\tau_+) \right ] 
   =  e^{- 
  \frac{N}{2}  \left \langle - i\: \partial _{\tau} \varphi _p \left | \left ( {\cal{F}} + \: \ \frac{t_0^2}{N} {p^2} R_c^{-1} \Lambda_cR_c^{-1} \right )\right | - i\: \partial _{\tau} \varphi _p \right \rangle } \nonumber \\
  \label{generat}
  \enea
  where
 \bea 
 {\cal{F}}\left ( \tau_{1}, \tau_{2},\tau_{3},\tau_{4}\right ) =  {R}_c^{-1}\: K_c \left [1- K_c \right ]^{-1} {R}_c^{-1} 
\label{effe}
\enea
is the four point function of the  $0+1$$-d$ SYK model. Integration over intermediate times is intended. Here\cite{ordini}  $ {\cal{F}} $ is ${\cal{O}}(1)$, with the meaning of  ${\cal{O}}\left (\left [\frac{\beta {\cal{J}}}{N}\right ]^0\right )$.  As $R_c^{-1} \Lambda_c R_c^{-1} \propto G_c G_c$ is ${\cal{O}}(N/\beta {\cal{J}}) $, it appears from Eq.(\ref{generat})  that we can define a  physical parameter $ \beta t_0^2 / N{\cal{J}}$ of ${\cal{O}}(1)$, to guarantee that  the  hopping across the lattice is not irrelevant in the scaling. It turns out however that both $ {\cal{F}} $ and   $ \frac{\beta t_0^2}{N {\cal{J}}}$ become of ${\cal{O}}\left (\frac{\beta {\cal{J}}}{N} \right )$ when  the UV  correction  is included, which   is crucially important  to give sense to the functional integration of Eq.(\ref{immf}), as we explain here below.  

   Actually, the functional integral of Eq.(\ref{immf}) includes a  divergent  contribution  due to the  Goldstone modes  $\delta g_c$  corresponding to eigenvalues of  $K_c \to 1$, which has to be regularized resorting to the first UV correction  $\langle \delta g  | K_c ^{-1}-1| \delta g \rangle \sim \beta {\cal{J}}$. The Faddeev Popov regularization provides an integration performed in the orthogonal space with respect to the $pGm$, while the smallest eigenvalue of the kernel $1- K_c $ is approximated with its UV correction, given by $1- k_c(h=2,n) \approx \frac{\alpha _K }{ \beta{\cal{J}}} |n|+ ... $ ($\alpha _K\approx 3 $ is a constant)\cite{maldacena}. 
It follows that the large, but finite contribution to  $ {\cal{F}}$ in Eq.(\ref{effe}) with this  UV correction  (i.e. for $ \tau \sim 1/ {\cal{J}} $) is not ${\cal{O}}(1)$ as stated here above, but ${\cal{O}} \left ( \frac{\beta {\cal{J}}}{ N}\right ) $ and the same has to occur for $ \frac{\beta t_0^2}{N {\cal{J}}}$. We will discuss this point in the next Section. The temperature threshold for coherence defined here,   $ T_{coh}= \frac{ t_0^2}{N {\cal{J}}}$,  is recurrent  in the next. 


If we ignore this matter for the time being, the generating functional of Eq.(\ref{generat}) provides the correlator $\frac{1}{N}  FT\left [\delta g_{x} (\tau_{12}, \tau_+) \delta g_{x'}(\tau_{34},0) \right ]| _{ k \neq 0 }$ in imaginary time, inclusive of the hopping in the lattice:
\bea 
\frac{\delta^2}{\delta  \partial_\tau\varphi_p(\tau_+) \: \delta  \partial_\tau\varphi_p(\tau_+')} \frac{1}{N} \sum _{p'} \ln{ {\cal{Z}}\left [\partial_\tau\varphi_{p'}(\tau_+) \right ] }\nonumber\\
\propto  \met \left (1 + \: \frac{t_0^2}{2\:N} {p^2} R_c^{-1} \Lambda_cR_c^{-1} {\cal{F}}^{-1} \right ) {\cal{F}} .
\label{gens}
\enea
This result adds  to the  naked  correlator ${\cal{F}}$ the contribution coming from Eq.(\ref{lib}), so that  the two dynamics are just added together in this approximation. However, one can envisage the present one  as the lowest order of a ladder resummation which will appear more clearly in the next Section. 
The operator 
 $ \widehat{G_cG_c } \hat{ {\cal{ F}}}^{-1}$ appearing in the Kernel  of  Eq.(\ref{gens})  is the inverse matrix of \cite{maldacena}
  \bea
  {\cal{F}}\{G_cG_c\}^{-1}   = \frac{6 \alpha _0 \beta {\cal{J}}}{\pi^2 \alpha _K} \! \! \! \nonumber\\
\times  \sum _{| n | \geq 2, even} \! \! \! \! \!  \frac{e^{ i\: n (y_{12}-y_{34})}}{n^2 (n^2-1)}\:  f_n(\tau_{12}) f_n(\tau_{34} ),
   \label{zozo1}
   \enea
     where     $y_{ij} \equiv  (\tau_i+\tau_j)/2$ in unit of $ \hbar \beta / 2 \pi $.  The basis  functions  $f_n(\tau_{12})$ are defined in appendix \ref{app:secB},
     together with  the spectral representation of the kernel $ K_c^{-1}\: [1-K_c ] $, as well as with  their Fourier transform.  
      \section{dressed correlator of ${pGm}$ modes}
 The derivation of the previous Section  has  assumed that  $\varphi _x(\tau_+)$  is given as an external source. However, continuation to real time requires that   $\varphi _x(\tau_+)$    acquires a dynamics.   Meanwhile,  the symmetry breaking induced by  the UV perturbation source $\sim \partial_\tau$,  couples to  $ G_{IR} $.   $ G_{IR} $ is  derived from a  time  reparametrization under the diffeomorphism   $ e^{i\vartheta }\to e^{i \:  \varphi _x(\vartheta )}$ of the conformal Green's function ($\Delta = \frac{1}{4}$):
\bea
G_{IR}( \vartheta_1,\vartheta_2) = G_c\left ( \! \varphi _x(\vartheta_1) , \varphi _x(\vartheta_2) \right ) \!  \varphi _x'(\vartheta_1) ^\Delta \!  \varphi _x'(\vartheta_2) ^\Delta,
\enea
  where  $ \vartheta = 2\pi \tau /\beta $ and $\varphi '\equiv \partial_\vartheta \varphi$. 
  
The leading correction to the conformal action  arising from this reparametrization (apart for a  shift of the ground state energy) is 
the  Schwarzian\cite{kitaev}:
   \bea
    \frac{ I_{local}}{ N}[\partial_\tau \varphi]= -2\pi \: \alpha _S \varepsilon \: \int \frac{d\vartheta}{2\pi} \left [ \met - \frac{ ( \varphi '')^2 - ( \varphi ' ) ^2}{2} \right]. 
    \label{schwarz}
       \enea
     
 Here $\alpha _S $ is a constant\cite{maldacena} and $\varepsilon =1/ \beta {\cal{J}}$.
  Hence, the full action in place of the one appearing in Eq.(\ref{generat}) reads: 
    \begin{widetext}
   \bea
 I_{\partial\varphi } = \sum _p\left \{ I_{local}[ \partial_\tau\varphi_p]
 -  \frac{N}{2} \int d\tau_{12}   \left \langle - i\: \partial _{\tau} \varphi _p \left | \left ( {\cal{F}} + \: \ \frac{t_0^2}{N} {p^2} R_c^{-1} \Lambda_cR_c^{-1} \right )\right | - i\: \partial _{\tau} \varphi _p \right \rangle  \right \}.
\label{gen2a}
\enea

\end{widetext}

 Now the field  $ \partial_\tau\varphi_x$  has its own dynamics and could be  integrated out, possibly after  adding a source term to get a generating functional   of $ \langle   \partial\varphi\partial\varphi\rangle $ correlators. However, the action of Eq.(\ref{gen2a}) is essentially a "phase only" model for hopping of  $ \partial_\tau\varphi_x$ across the lattice. This is so, because we have neglected $\delta G\!\left (x,\tau_{12}, \tau_+\right ) $ appearing in Eq.(\ref{greo}).  So as it stands, 
 $ I_{local}[ \partial_\tau\varphi_p] \sim {\cal{O}}(N/\beta {\cal{J}})$, while the other term is ${\cal{O}}(1)$. Hence,  the local action $ I_{local}[ \partial_\tau\varphi_p]$ is irrelevant in the $\beta {\cal{J}}\to \infty$ limit and the phase lacks its own dynamics.  Writing down a partition function for the order parameter given by  Eq.(\ref{greo}),with inclusion of its  modulus, gives the chance of extracting correlation functions  with include the UV correction and  can be extended to real  time. Let us denote the complex order parameter  in two space dimension  $\phi(x, \tau _{12},\tau_+) = \sqrt{ \rho_0 + \delta \rho} \: e^{i \theta }$,  for each degree of freedom.  The functional integral, with $\tau$ of  the dimension $time$  in the following, reads: 
   \bea
\int D \phi(x, \tau _{12},\tau_+)^*  D \phi(x, \tau _{12},\tau_+) \: e^{- \tilde{S}\left [\phi ^*, \phi \right ]} .
 \enea
  The action  leading to the one of  Eq.(\ref{gen2a})  can involve also space derivatives:
 \begin{widetext}
 \bea 
\tilde{S}= -\int ^\beta \!\! \!\!d\tau_{12}\:\!\! \int\! d^2x \!\! \int ^\beta \!\! \!\!d\tau\: \frac{ 1}{2}\left [- v^2   \partial _x \partial _\tau \phi ^*   \partial _x \partial _\tau \phi  \!+\! \left (  \phi ^* \partial _\tau\phi -  \phi  \partial _\tau \phi ^*  \right )\! -\!  \frac{1}{2\pi \alpha _{S} \epsilon}�|\phi |^4\right ]
\label{stid}
\enea
\end{widetext}
 (an expression for  the velocity $v$  appearing here,   derived from a Hamiltonian approach, is presented  in  Eq.(\ref{coi3}) of the next Section  and  in the appendix \ref{app:secD}). 
 In fact,  expanding to quadratic order in $\theta$ and $\delta \rho$, we get (we  imply  $\int ^\beta d\tau_{12}$ in the notation in what  follows)  

\bea
\tilde{S}_2=-\int d^2x \: \int ^\beta  d\tau \frac{ 1}{2}\left [-\rho_0  v^2  ( \partial _x  \partial _\tau\theta )^2 \right . \nonumber\\
\left .  + 2 i \: \delta \rho  \:  \partial _\tau\theta - \frac{1}{2 \pi \alpha _S \epsilon}�\delta \rho ^2\right ].
\enea
Integrating out the fast field $\delta \rho$  in the functional integral,
\bea
\int D \delta \rho \: e^{  \int d^2x \:  \int ^\beta d\tau  \left [i \: \delta \rho  \:  \partial _\tau\theta -  \frac{1}{2\pi \alpha _S \epsilon}�\delta \rho ^2\right ]}  \nonumber\\
= e^{-  \pi \alpha _S \epsilon \int d^2x \: \int ^\beta  d\tau \: \met (\partial_\tau  \theta )^2},
\enea
 we obtain
 \bea
 \tilde{S}_2= - \int \!\!d^2x\!\! \int ^\beta \!\!\! \!\!d\tau\!\! \left [- \rho_0\frac{  v^2}{2}  ( \partial _x\partial _\tau \theta )^2 + \pi \alpha _S \epsilon  \met (\partial_\tau  \theta )^2\right ], \nonumber \\
 \label{sdue}
   \enea
   which can be identified with   $\frac{ I_{local}}{N}�[\partial_\tau \varphi]$  of Eq.(\ref{schwarz})  provided  we also introduce space non locality there, by trading   $ \frac{ G_{c }(\tau_{12}) }{\tilde{a}}   v \: \partial_x$, which appears in Eq.(\ref{sdue}),  for  $\partial_\tau $. Identification requires  that   
  \bea
    \rho_0 = |\phi_0|^2  = \frac{ \pi \alpha _S \varepsilon}{\tilde{a}^2}  \left |G_{c }(\tau_{12}) \right |^2, \:\:\:\:\: \partial_\tau\theta \equiv  \partial_\tau \varphi 
    \label{trsm}
    \enea
    (an extra factor $N$ pops up from the number of flavours  in the first equality). Introducing $\tilde{R}_c = \left |G_{c }(\tau_{12}) \right |$ and substituting  $- i\: \partial _{\tau} \varphi _p \to   \: e^{-i\theta}\partial_\tau \phi_p \tilde{R}_c^{-1} \left [  \frac{ \pi \alpha _S \varepsilon}{\tilde{a}^2} \right ]^{-1/2}$, the functional integral  becomes 
 \begin{widetext}
 \bea 
\int D  \phi(x, \tau _{12}, \tau_+)^*  D \phi(x, \tau _{12},\tau_+) \:  e^{- \tilde{S}\left [\phi ^*, \phi \right ]}\: e^{ - 
  \frac{1}{ 2\pi \: \alpha _S  \varepsilon } \sum _p  \left \langle \phi_p \tilde{R}_c^{-1} \left |   \overset{\rightarrow}{\partial_\tau}\left ( {\cal{F}} + \: \ \frac{t_0^2}{N} {p^2}R_c^{-1} \Lambda_cR_c^{-1} \right )   \overset{\leftarrow}{\partial_\tau}\right | \phi _p  \tilde{R}_c^{-1} \right \rangle },
  \label{parto}
 \enea 
 \end{widetext}
  where $\tilde{S}$ is given by Eq.(\ref{stid}).  It is useful to ridefine the $\tilde{\phi} = \phi \:  \tilde{R}_c^{-1}$ in the functional integral. In the change of the integration field,  the action   $\tilde{S}\left [\phi ^*, \phi \right ] \to \tilde{S}\left [\tilde{\phi} ^*, \tilde{\phi} \right ] $ acquires a factor $\left (  G_{c }(\tau_{12}) \right )^2 $, except for the $|\phi|^4$  term which acquires the fourth power.   Note, however,  that in the UV  domain is $\tau_{12} \sim {\cal{J}}^{-1}$, so that, with\cite{maldacena} $b^{-2} = \sqrt{4\pi} {\cal{J}}$, 
 \bea 
\left (  G_{c }(\tau_{12}) \right )^2 =  \frac{ b^2}{ |\tau_{12}|} \sim  {\cal{O}}(1) .
\enea
Hence, the last term of the full action from Eq.(\ref{parto})  is  ${\cal{O}}(\beta{\cal{J}})$ in the large $\beta {\cal{J}}$ limit, while the first contribution to the full action,  given by   $\tilde{S}$, is ${\cal{O}}(1)$.  Actually the $|\phi|^4$  term in $\tilde{S}$ is also ${\cal{O}}(\beta {\cal{J}})$, but we stick to zero order in the anharmonic functional integration.  The evolution of  Eq.(\ref{parto}) is characterized by  an interplay between the dynamics of the  intradot fluctuations and the dynamics of  the  interdot $\partial _\tau \varphi _p $  fluctuations, which is mostly represented by  the action  $\tilde{S}$.  
If   $\tilde{S}$  is dropped alltogether, because it becomes  irrelevant in the scaling, the  gaussian integration of   Eq.(\ref{parto}) can be easily performed,  giving rise to  a  density matrix $ {{\rho}}_{\delta g}$ of the intradot fluctuations at each given time $\tau_+$. When  Fourier transformed with respect to the  intradot times $\tau_{12},\tau_{34}$,  stripping off the unperturbed evolution $\hat{ {\cal{ F}}}$, the result of the functional integration of Eq.(\ref{parto}), in the absence of $\tilde{S}$ is (again, $y_{ij}$ are center-of-mass  times in unit of $ \hbar \beta / 2 \pi $ and  here $ \tau \equiv \tau_+ =   y_{12}-y_{34} $ in unit of $ \hbar \beta / 2 \pi $):
\bea
 {{\rho}}_{\delta g}\left (y_{12}- y_{34} ; p\right )_{m,m'}\:\:\: \:\:\: \:\:\: \:\:\: \nonumber\\
  \sim   \frac{1}{N} \left . \left \{1 + \: \ \frac{t_0^2}{N} \frac{{p^2}}{2} \overset{\rightarrow}{\partial_\tau}\! \left ( \left [ \widehat{ R_c^{-1} \Lambda_cR_c^{-1}}\right ]\hat{ {\cal{ F}}}^{-1}\right ) \!\overset{\leftarrow}{\partial_\tau} \right \}^{-1}\right | _{\tau_+,m,m'}\!\!\!\!\!\!\!\!\!\!\!\!\!\!\!\!\!\!.  \label{evolmix}
  \enea

     We define  
   \bea
   \hat{{\cal{B}}}_{m,m'}(\tau_+) \equiv  \left [\overset{\rightarrow}{\partial_\tau}\! \left ( \left [ \widehat{ R_c^{-1} \Lambda_cR_c^{-1}}\right ]\hat{ {\cal{ F}}}^{-1}\right ) \!\overset{\leftarrow}\,{\partial_\tau}\right ]_{m,m'}, 
   \label{bibbo}
   \enea
   which will be used in the following. 
   
    The correlation function  $\left .\langle  FT\left [\delta G_{x} (\tau_{12}, \tau_+) \delta G_{x'}(\tau_{34},0) \right ] \rangle \right | _{ k \neq 0 }$, corresponding to Eq.(\ref{lib}),  but including the ladder resummation, is obtained  by tracing on the density matrix of Eq.(\ref{evolmix}), after the  $p=0$ term has been subtracted. To  lowest order, we get:
     \bea 
     \left .\langle  FT\left [\delta G_{x} (0^+, \tau_+) \delta G_{x'}(0,0) \right ] \rangle \right | _{ k \neq 0 }\nonumber\\
    \approx \sum _{mm'} G_c(\omega _m) \:  {{\rho}}_{\delta g}\!\left (\tau_+; k \right )_{m,m'}\: G_c(\omega _{m'}).
    \label{necor}
     \enea
    where $G_c(\omega _m) $ is given by Eq.(\ref{galt}).
When  $k \tilde{a} <<1$,  the contribution of the  ladder  can be dropped  and the unnormalized correlator   $ \langle \delta G( \tau_{12},\tau_+) \delta G(\tau_{34},0)\rangle_{k\neq 0} $ reads: 
 \bea
 \langle \delta G( \tau_{12},\tau_+) \delta G(\tau_{34},0)\rangle_{k\neq 0}  \nonumber\\
  \equiv FT_{k\neq 0} \langle \left [\delta G_{x} (\tau_{12}, \tau_+) \delta G_{x'}(\tau_{34},0) \right ] \rangle \nonumber  \\
  = \frac{1}{N} \sum _m \frac{ t_0^2}{2\: N} \:  k^2 \tilde{a}^2   \left [\widehat{ G_c {\cal{B}} G_c }\right ]_{ y ,m,-m} \!\!\!e^{i \: \omega_m (\tau_{12}-\tau_{34} )}.\label{fifi}
\label{ecco}
\enea 
Only  the dependence on the relative coordinate $ \tau_{12}- \tau_{34}\:\:mod.\: 2\pi$  has been retained. 


  In  Fig.\ref{imog}, the   correlator  $  \langle \delta G( \tau,0^+) \delta G(0,0)\rangle_k $  from Eq.(\ref{ecco})  is plotted  and compared with the  naked $ \langle \delta G_c \delta G_c\rangle_k $ correlator given by  Eq.(\ref{gco}). The main panel of   Fig.\ref{imog} displays the modulus while   the phase appears in  the inset of Fig.\ref{imog}. The prefactor  $k^2{\tilde a}^2\: \beta t_0^2/( 2\pi {\cal{J}})$  has been dropped.     
 
    The correlator $\langle \delta G( \tau,0^+) \delta G(0,0)\rangle_k$  has been  calculated as reported  in appendix \ref{app:secB}, using the Fourier Transform of Eq.\ref{zozo1}, with the  inclusion of   ${\cal{F}} ^{-1}$ in the evolution.  We had to truncate  the sum over the (even)  indices  $n$ up to $n=12$, and consequently the sum over  internal  (odd) indices just  includes  up to $m,m' = 5$. Its  modulus and phase, compared to those of  the corresponding  naked  $ \langle \delta G_c \delta G_c\rangle_k $ correlator,  are plotted  in Fig.\ref{imog}.  The modulus of the naked correlator is exponentially decaying  at  the intradot time $\tau \sim 0, \: mod[2\pi]$, while the dressed one is powerlaw, highlighting the criticality of the phase, when the UV correction is included.  The  Fourier transform of the sawtooth phase oscillations  of $ \langle \delta G \delta G\rangle_k $ (blue curves) appearing in Fig.\ref{imog} is not simply  $ \propto 1/i\:  \omega _m$, revealing   the "fast"  intradot time scale induced by  the UV correction, with respect to the phase of the naked correlator. They could have acquired further structure, if larger $n,m$ values had been retained. 
    
\begin{figure}
 \centering
\def\big{\includegraphics[height=8.0cm]{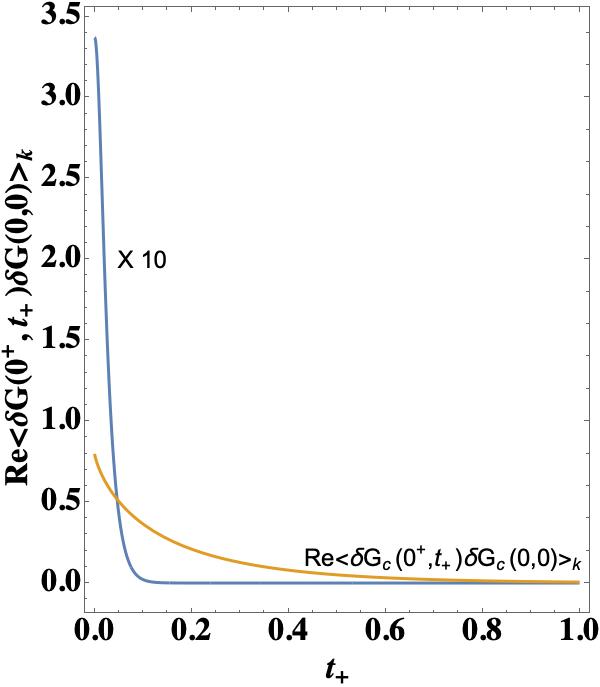}}
\def\little{\includegraphics[height=6.5cm]{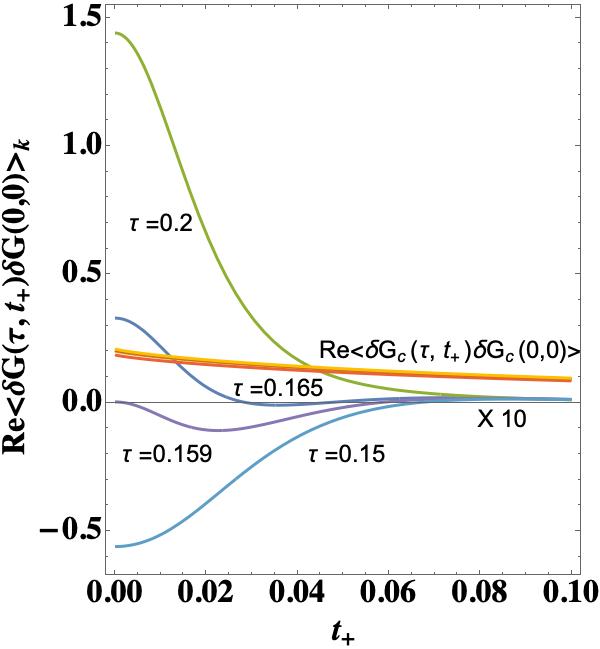}}
\def\stackalignment{l}
\topinset{\little}{\big}{-22pt}{+60pt}
\caption{ The real part of the analytic continuation to real time of the correlator,   $ \Re e \langle \delta G( 0^+,t_+) \delta G(0,0)\rangle_k$,  given by Eq.(\ref{necor}) (blue curves),  compared with the the naked  correlator $\Re e\langle \delta G_c( \tau, t_+) \delta G_c(0,0)\rangle_k$ from Eq.(\ref{gco})(orange curves), are plotted vs  the dimensionless interdot  time $t_+$. While  the intradot time $\tau_{12}- \tau_{34}\equiv \tau  =0^+ $ appears  in the main panel, the  curves for $ \tau  =0.15, 0.159,0.165, 0.2$ are plotted in the inset in an expanded scale. The oscillations of $ \Re e \langle \delta G( \tau, t_+) \delta G(0,0)\rangle_k$  in  $\tau$ follow those  appearing in Fig.\ref{imog}, while the naked correlator $\langle \delta G_c( \tau, t_+) \delta G_c(0,0)\rangle_k$ has negligible dependence on $\tau$. The prefactor  $k^2{\tilde a}^2\: \beta t_0^2/( 2\pi {\cal{J}})$,  which contains the $k$ dependence,  has been dropped in the plots.   } 
\label{gcofig}
\end{figure}

       The real part of  the analytic  continuation to real times of the center of mass  coordinate $t_+$ in  $  \langle \delta G( 0^+,\tau_+) \delta G(0,0)\rangle_k $  is  plotted in the  main panel  of Fig.\ref{gcofig} and compared to the  corresponding naked correlation of Eq.(\ref{gco}).  The same correlators, but keeping  the   dependence on the  relative imaginary time coordinate $ \tau_{12}- \tau_{34},\:\:mod [2\pi]$ as in Eq.(\ref{ecco}), are plotted for various values of  $\tau = \tau_{12}- \tau_{34}\:$  in the inset panel.  The dependence on  $\tau $  is oscillating and we have chosen values for $\tau$ within a single oscillation.  The prefactor  $k^2{\tilde a}^2\: \beta t_0^2/( 2\pi {\cal{J}})$  has been dropped again. The   $  \langle \delta G( 0^+,\tau_+) \delta G(0,0)\rangle_k $'s   appearing in  Fig.\ref{gcofig}  are scaled  by  $\times  10 $  with respect to the correlators arising from the naked kinetic term of Eq.(\ref{gco}).   The  $t_+$ dependence in the presence of UV corrections appears   very localized and highly variable with  the intradot time, as compared with the naked one.   The UV corrections squeeze  the  interdot correlations in time, increasing their "local" nature. This drastic drop in time of the correlations  cannot guarantee quantum diffusion on  an extended spae scale, much larger than the lattice spacing, and we have to resort to a better approximation which retains the dynamics entailed by  the action $\tilde{S}$, which was lost in this result.  
       
Besides,  the strong dependence of the dressed correlations on the intradot imaginary time,  with a  relatively stable  interdot  real time dependence,  confirms that the UV correction introduces a sizeable   time scale separation  between the  intradot and interdot correlations.   This is the basis of the factorization of the two dynamics, which we use  to approximate the quantum diffusion discussed  in the next Section.    

   \section{quantum diffusion}

 In this Section,  we attempt a better approximation for evaluating  the partition function of Eq.(\ref{parto}), to  investigate the quantum diffusion of the Q-excitations across the lattice, induced by  the intradot $pGm$'s.  We want to  extract a diffusion coefficient  $\tilde{D}_Q$  out of the scaling flow, to be related to the  thermal  conductance  $\kappa$  of the "electronic"  carriers and to the thermal "electronic" capacitance $C$ in the lattice. In turn, they are connected to a relaxation time ${\cal{T}}_Q$ and to  the  inverse lifetime of the Q-excitations, $\Gamma$.
 
 \subsection{Partition function of the Q-excitations}
 
   In Section IV we have shown that,  to improve the  $t_+$-dependence  of the correlator $ \Re e \langle \delta G( \tau, t_+) \delta G(0,0)\rangle_k$  of Eq.(\ref{necor}), the UV  local time corrections should be included more carefully. In fact, the result of the previous Section is unsatisfactory, because,  in the  flowing to  the fixed point  of the partition function of Eq.(\ref{parto}),  we had  to drop the order parameter dynamics entailed by the action ${\tilde{S}}$  of Eq.(\ref{stid}).  A semiclassical approach to the diffusion process can be   still envisaged, however, in the results of the previous Section.   
   When   the trace over the intradot frequencies is performed,  the density matrix of Eq.(\ref{evolmix}),  appropriately continued to real time, $ {{\rho}}_{\delta g}\left (y_{12}- y_{34} ; k\right )_{m,m'} \to  {\cal{P}} \!\left ( r,r', t_+\right )$, takes the form of a heat kernel  $ z(t) $, typical of a diffusion process\cite{akkermans}, defined as the probability to return to the origin, integrated over the point of departure. From Eq.(\ref{evolmix}),  in  the  $k \tilde{a} <<1$  limit,  we have :
\bea
z(t_+) =  \frac{1}{N} \int _{\cal{A}} \left [ {\cal{F}}\cdot  
 {\cal{P}}\right ] \!\left ( r,r, t_+\right )\:  d^2r \nonumber\\
 \propto \sum_m \sum _p \:  e^{-  \left .\frac{  \beta t_0^2}{N} \frac{{p^2}}{\hbar} \hat{{\cal{B}}}(t_+) \right |_{m,m}},
 \label{return}
\enea
where we have restored the free intradot evolution. 

Now that we know what  the drawback is, we reconsider the UV correction to the action given by  Eq.(\ref{schwarz}). Its variation with respect to $\varphi '$ gives a simple  equation of motion $  \partial _\tau^2 \varphi ' = - \varphi '$. When derived from the action of  Eq.(\ref{sdue}), this motion equation is  rewritten in the form of lattice space oscillations. In the following we quantize these space extended  excitations by means of a  phenomenological  2-d Lagrangian with  canonical conjugate variables, introduced in appendix \ref{app:secB}:
 \bea
 \dot \theta = \left ( \frac{\kappa C}{\hbar T}\right )^{1/2} \!\! \frac{ J_ Q {\cal{T}}_Q}{k_B},\:\:
 \nabla \theta = \left ( \frac{\hbar}{\kappa C\: T}\right )^{1/2} \!\! \frac{\kappa}{T} \nabla T 
 \label{can}.
 \enea
 Here $J_Q$ is the thermal energy current density. 
 The corresponding Lagrangian is 
 \bea
 {\cal{L}}= \met \int d^2 x\: \left [\frac{k_B}{T}\: \left (\frac{ J_ Q  {\cal{T}}_Q}{k_B}\right )^2+ \frac{\hbar }{\kappa C} \: \left (\frac{ \kappa}{T} \nabla T \right ) ^2\right ]
 \nonumber\\ \equiv \met \int d^2 x\: \left [  \frac{\hbar k_B}{\kappa C}\: \dot \theta ^2+ T \: ( \nabla \theta )^2 \right ].\label{lagx}
 \enea
The
terms in the square brackets  have dimension $ {\cal{E}}/\ell ^2 $ ($ {\cal{E}} \equiv \: energy,\: \ell \equiv\:  length$).

This Lagrangian  is of course conserving,  but we have introduced the relaxation time $ {\cal{T}}_Q$, so that we can reproduce a diffusive motion equation of the form $ J_ Q= -\kappa\: \nabla T$, if we  approximate the time derivative of the energy current fluctuations $\dot {J}_ Q \approx  {J}_ Q/ {\cal{T}}_Q $.
Here $ \kappa = \frac{C \: v\: \ell}{\tilde{a}_\ell^2} $ is the thermal conductivity in $2-d$, where   $\ell $ and $v$ are typical mean free path and velocity, respectively, while $\tilde{a}_\ell^2 \sim \tilde{D}_Q {\cal{T}}_Q $ is the area over which the thermal capacitance $C$ is defined and will be introduced here below. 

 We quantize  the corresponding Hamiltonian, in terms of the creation and destruction bosonic operators $ { a}^\dagger _{k}, { a}_{k}$:
  \bea
 \pi _k = -i\: T^{1/2} \frac{1}{ (2 \Omega_k)^{1/2}} |k| \: \left (  a_{-k} - { a}^\dagger _k \right ), \nonumber\\
 \theta _k = T^{-1/2} \frac{ (2 \Omega_k)^{1/2}}{ |k|} \: \left (  a_{k} + { a}^\dagger _{-k} \right )\nonumber\\
  \Omega_k = \tilde{a} \left [\frac{ \kappa }{k_B} \frac{ C T}{\hbar }\right ]^{1/2}|k| \equiv v\: |k| , \label{coi1}\\
{H}_0^{\tilde{D}}= \sum _k  \Omega_k \: { a}^\dagger _{k} a_{k} + cnst .\label{coi3}
 \enea
$  \Omega_k $ is the linear dispersion law of these modes with velocity $v$ defined in Eq.(\ref{coi1}). From the damped fluctuations of these modes, the response function $D^\beta( \omega )$ is derived  in eq \eqref{dbet}, within this Lagrangian approach.  On the contrary, here in the following,  we aim to deriving the quantum diffusion probability, stressing the interplay between intradot  $\delta g_m$ modes and the kinetics  of the Q-fluctuations in the lattice. 

 From Eq.(\ref{parto}), we recognize  the coupling Hamiltonian $\hat{ {\cal{H}}}_{\tilde{D}}$, which, in the interaction representation of $ \hat{H}_0^{\tilde{D}}$, takes the form: 
  \bea
\hat{ {\cal{H}}}_{\tilde{D}}(\tau) = - \frac{\tilde{a}^2}{\pi \alpha _S N \varepsilon} \sum _{p \neq 0} {p^2} \: \frac{\beta t_0^2}{2 N} \times \:\:\:\: \label{amo} \\
 \sum _k \! \hat{{\cal{B}}} (\tau)\: { a}^\dagger _{k+p}(\tau) { a}_{k}(\tau).\nonumber
 \enea 
  $\hat{ {\cal{H}}}_{\tilde{D}}(\tau _+)$ of Eq.(\ref{amo})
  represents an "effective  interaction Hamiltonian" for  energies in the incoherent phase. We remind that  $  \hat{{\cal{B}}} (\tau) $, defined in Eq.(\ref{bibbo}),  is $ \sim { \cal{O}}\left ( \frac{N}{\beta {\cal{J}}}\right )$  and that the $hat$ denotes the $m\times m'$ matrix structure.  As $\frac{\beta t_0^2}{2 N} \sim {\cal{O}}\left (\frac{\beta {\cal{J}}}{N}\right )$, the additional factor $( \pi \alpha _s N \epsilon)^{-1}$ in Eq.(\ref{amo}) makes \cite{ordini} $\hat{{\cal{H}}}_{\tilde{D}} $ of   ${\cal{O}} \left (\frac{\beta {\cal{J}}}{N}\right )$ and allows us to define a scaled length $ \tilde{a}_\ell \sim \tilde{a} \:( \pi \alpha _s N \epsilon)^{-1/2} >> \tilde{a}$, which is the length   scale for diffusion in the lattice.


 The partition function of  Eq.(\ref{parto}),  represented in the bosonic coherent field $\tilde{\phi} = \phi R_c^{-1}$, can be expressed as
  \bea
  {\bf Tr} \: e^{-\beta H^0_{SYK}}\!\!\left \{\! {\large \bf tr _{\tilde \phi}} \!\! \left ( e^{ -\beta H_0^{\tilde{D}} } \!{\bf T_{\tau_+}}\!\! \left [ e^{-\int _0^{\beta} \hat{{\cal{H}}}_{\tilde{D}} (\tilde{\phi}^*,\tilde{\phi}, \tau _+) \: d \tau _+ }\right ]\!\right ) \!\right \}\nonumber \\
   \label{secd}
    \enea
  and  the full quantum dynamics, is included (we drop the tilde on $\phi$  henceforth). 
Here   $ { \bf tr _{\phi}}$ denotes  the trace of a  time ordered functional integral (${\bf T_{\tau_+}}$ is time ordering in $\tau_+$), while we keep the symbol $ {\bf Tr} $ for the trace of the $m\times m$ matrices. $\hat{{\cal{H}}}_{\tilde{D}} (\phi^*,\phi, \tau _+)$ is the matrix element  derived from Eq.(\ref{amo}),  in the coherent basis representation. 
  In performing the trace, we assume  $ \hat{{\cal{H}}}_{\tilde{D}} \!( \phi_p^*,\phi_p, \tau)$ to be diagonal in the $p$ label.  
  
  As we are dropping the $|\phi|^4$ term appearing in the original action $\tilde S$ of Eq.(\ref{parto}),  our toy model involves non interacting bosonic fields only.  The partition function can be written down straightforwardly by slicing the trace ${\bf  tr _{\tilde \phi}}$ into $\frac{\beta}{M}$   time  slices ($M$ integer)\cite{negele}:
  \bea 
  {\cal{Z}} =   \lim _{M\to \infty} {\bf Tr} \left \{e^{-\beta H^0_{SYK}}  \right . \nonumber\\
\times \left .   \prod_k   \left [ 1-  \left \{ e^{  \frac{ \beta}{M} k^2\frac{ \tilde{a}^2 }{\pi \alpha _SN\epsilon} \frac{\beta t_0^2}{N}\: \hat{{\cal{B}}}\left (\frac{\beta}{M} \right )}\right \}^M   \right ] ^{-1} \right \}.
\label{zeta0}
  \enea
  In Eq.(\ref{zeta0}) the dynamics of the intradot fluctuations $\delta g_m $ and their  interdot extension to the lattice are fully entangled.  In view of some simplification, we limit ourselves to the regime in which the inverse  timescale of the Q-fluctuations  in the lattice, $\overline{\tau}^{-1} \equiv - i\:  {\cal{T}}_Q^{-1}$, is much smaller than  the typical frequency scale of the intradot evolution (which includes  the dominant term of the UV corrections). In  this regime 
   we factorize the $\left (\frac{\beta}{M} \right )$ slices of the intradot  propagator generated by $H^0_{SYK}$ which is   $ \hat{{\cal{ F}}}$,  while $ \hat{{\cal{B}}}$ includes  the kernel   $ \widehat{G_cG_c {\cal{ F}}^{-1} }$ of the Q-fluctuations.  The factorization amounts to  a kind of "non interacting blip approximation"\cite{chakravarty,roman} and  can be justified as long as  the thermalization is very effective.   
   
 With this approximation, the functional integration of the partition function of Eq.(\ref{zeta0}) can be cast in the form: 

    \bea
  \lim _{M\to \infty} \prod_k \!{\bf Tr} \!\left \{\hat{ {\cal{ F}}}(\beta)\left [ 1\!-\! \left ( 1- \frac{\beta}{M} k_BT_0\hat{ f}_{\overline{\tau}}(k)\right )^M \right ] ^{-1}\! \right \},
 \:\: \:\:\label{zeta}
  \enea
 \begin{figure}
  \includegraphics[height=48mm]{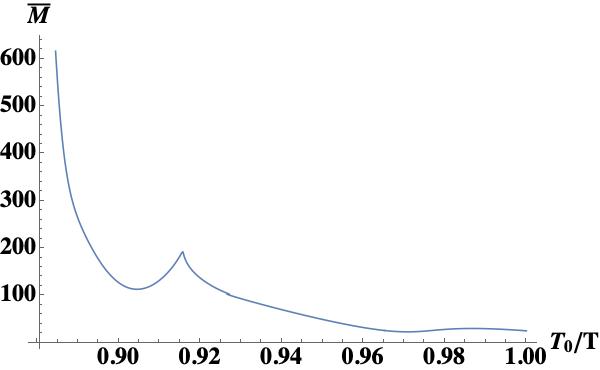}
\caption{ Plot of the approximate lowest $\overline{M}$ value which fulfills the unitarity condition of the partition function in Eq.(\ref{tracca}), up to $10^{-5}$, vs. $T_0/T$ values, for $ k \tilde{a} =1$. } 
 \label{overM} 
\end{figure}
 where $\frac{\beta}{M} k_BT_0\: \hat{ f}_{\overline{\tau}}\left (k\right ) $ is a linearized  $m\times m$ matrix, for a small increment $\frac{\beta}{M}$ of $\tau_+$, arising from the correspondence 
 \bea 
- k^2\frac{ \tilde{a}^2 }{\pi \alpha _SN\epsilon} \frac{\beta t_0^2}{N}\: \hat{{\cal{B}}}\left (\frac{\beta}{M}\right )
 \to \frac{\beta}{M} k_BT_0\:{\hat{ f}}_{\overline{\tau}}\! \left (k\right )\! .
 \label{foi}
 \enea
  The subscript $\overline{\tau}$ is to remind that  the factorization of the traces is only justified  in a limited temperature range in which the separation of the  time scales holds. 
 We have  extracted a temperature scale $T_0$ from the left hand side,   of  ${ \cal{O}}\!\left ( \frac{\beta {\cal{J}}}{N}\right )$, and introduced  the function   $ \hat{ f} _{\overline{\tau}}$ of ${ \cal{O}}\left ( 1\right )$.  
  
 As we are on a closed time contour\cite{kamenev}, the partition function should be unity.  The intradot propagation should be periodic in $\tau _+$, as well:  $ {\bf Tr} \left \{\hat{{\cal{ F}}}(\beta )\right \} =1$. As  both  $\hat{{\cal{ F}}}(\beta )$ and ${\hat{ f}}_{\overline{\tau}}$  in Eq.(\ref{zeta})  are  $m \times m$ matrix of  rank $\tilde{r}_m$,  the limit of the trace is costly from the numerical point of view. It can be done straightforwardly if we trade $ 1/\tilde{r}_m$  for the stripping of  $ \hat{{\cal{ F}}}(\beta )$ off the trace.  Once done this, we have checked what is  the minimal $M$ value, $\overline{M} $, which  fulfills unitarity, at a given approximation order:
 \bea
 {\bf z}_{\overline{M}} \left (k\right ) = \frac{1}{\tilde{r}_m}  {\bf Tr} \left \{ \frac{ 1}{\hat{1}- \frac{1}{\tilde{r}_m} \: \left ( e^{- \frac{1}{\overline{M}} \frac{T_0}{ T}{{\hat{f}}}_{\overline{\tau}}}\right )^{\overline{M}}} \right \} \approx 1.	 
 \label{tracca}
 \enea
  In Fig.\ref{overM}, we plot an interpolated smoothed curve of the (approximate) lowest $\overline{M}$ value, which  satisfies Eq.(\ref{tracca}),  vs. $T_0/T $, for  $ k^2 \tilde{a}_\ell^2 =1$. Precision is up to $ > 10^{-5}$.   $\overline{M}$  is practically constant, when $ \frac{T_0}{T} \gtrapprox 1 $, but  it increases strongly when $T$ takes values $T>T_0$. The trend is only meaningful for $T_0/T \sim 1 $, because $T$  values  larger than $T_0$ require $n >12 $ in the spectral representation of Eq.(\ref{zozo1}) and matrices $ m\times m$ of rank $ \tilde{r}>3$, i.e. higher than the ones used here. 
 Fig.\ref{overM} is  the numerical proof that $T_0$ represents the temperature above which the thermalization is more efficient and our factorization between evolutions breaks down. The threshold temperature scale  $T_{0} $ introduced  in Eq.(\ref{foi})  and the space scale   $\tilde{a}_\ell $ defined after  Eq.(\ref{amo})  are  discussed in Subsection C.  

 \subsection{$\langle\overline{ \delta G^R\delta G^A }\rangle $ diffusion probability}
 The generating functional  to obtain  the correlator  of the field $ \phi _p(\tau)$ at different times $y^{r}-y^s $ ($ y^r \equiv 2\pi r /\overline{M}; \: r,s\: integers$) can be derived from Eq.(\ref{secd}) by  adding  a source term. Its  ${r,s}$ matrix element  can be denoted as $ {\bf Tr} \left (\left [\partial _\tau + \hat{{\cal{H}}}_{\tilde{D}}\right ]^{-1} \right )_{r,s} $:
 \begin{widetext}
\bea 
  \lim_{M\to \infty} {\bf Tr} \left (\left [\partial _\tau + \hat{{\cal{H}}}_{\tilde{D}}\right ]^{-1} \right )_{r,s}   \approx \lim_{M\to \infty} {\bf Tr} \left \{\left [ \hat{{\cal{ F}}}\left (\frac{\beta}{M} \right )\right ]^{r-s}
\frac{\left ( 1-\frac{1}{M} \ \frac{T_0}{ T} \hat{f}_{\overline{\tau}} \right )^{(r-s)}}{1-  \left  [\hat{{\cal{ F}}}\!\left (\frac{\beta}{M}\right ) \left ( 1-\frac{1}{M} \frac{T_0}{ T} \hat{f}_{\overline{\tau}} \right )^M\right ]} \right \} \nonumber\\
\approx \lim_{M\to \infty} {\bf Tr} \left \{\left [ \hat{{\cal{ F}}}\!\left (\frac{\beta}{M} \right )\right ]^{r-s} \right \}\:{\bf Tr} \left \{\frac{ \left [ e^{- \frac{T_0}{ T}\hat{f}_{\overline{\tau}}}\right ]^{ ( y^r-y^s)}}{\hat{1}-\frac{ 1 }{\tilde{r}_m} \left [ e^{- \frac{T_0}{ T}\hat{f}_{\overline{\tau}}}\right ]}\right \} \!\! \approx\lim_{M\to \infty} {\bf Tr} \left \{\left [ \hat{{\cal{ F}}}\!\left (\frac{\beta}{M} \right )\right ]^{r-s} \right \}\:  {\bf Tr} \left \{\left [ e^{- \frac{T_0}{ T}\hat{f}_{\overline{\tau}}}\right ]^{ ( y^r-y^s)}\right \},
 \label{ric} 
\enea
\end{widetext}
  to be compared  with the  correlators of Eq.(\ref{necor}) and Eq.(\ref{fifi}) ( here the term $k=0$ has not been subtracted, yet). 
  
\begin{figure}
   \includegraphics[height=58mm]{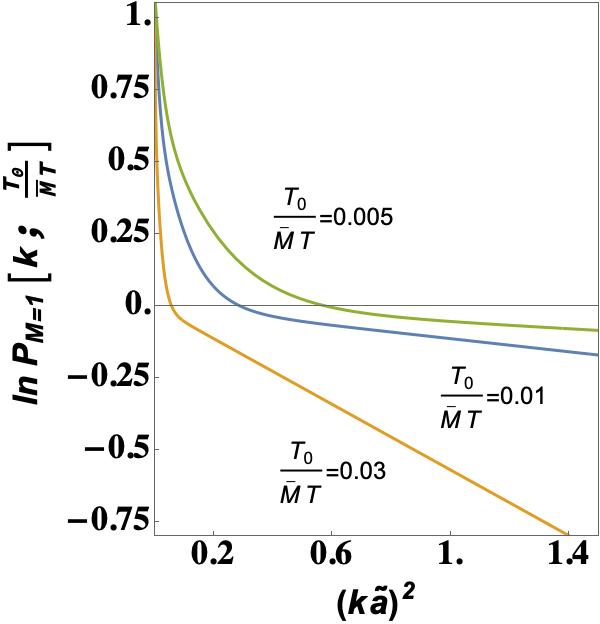}
   \caption{ Plot of the logarithm of the propagator appearing in Eq.(\ref{logo}) for $y^r-y^s=1$ (i.e.. $M=1$) , $ \ln \left ({\bf Tr}  \left [ e^{- \frac{T_{0}}{\overline{M}T}\hat{f}_{\overline{\tau}}}\right ]^1 \right )$, vs $(k \tilde{a}_\ell)^2$, for $\frac{T_0}{\overline{M}T} = 0.005,\: 0.01,\: 0.03$, displaying an  approximately linear behavior as a function of $(k \tilde{a})^2$, for $(k \tilde{a})^2 > 0.5$. } 
  \label{zozo} 
\end{figure}

 According to Eq.(\ref{return}), our aim is to  define  a scalar diffusion coefficient $\tilde{D}_Q$ such that,  when moving from euclidean to real time,  $ {\bf Tr}  \left [ e^{- \frac{T_{0}}{T}\hat{f}_{\overline{\tau}}}\right ] \to e^{ -i\: \tilde{D}_Q \: k^2 {\cal{T}}_Q }$. To accomplish this,  we have to check that $ {\bf Tr}  \left [ e^{- \frac{T_{0}}{\overline{M}T} \hat{f}_{\overline{\tau}}}\right ]$  provides an exponential with a $k^2$ factor in the exponent, when the trace has been performed.  In Fig.\ref{zozo}, we plot ,  vs $(k \tilde{a}_\ell )^2$, the logarithm  of the last term of the propagator on the right hand side of Eq.(\ref{ric}) for $ y^r-y^s=1$, 
 \bea
  \ln \left ({\bf Tr}  \left [ e^{- \frac{T_{0}}{\overline{M}T} \hat{f}_{\overline{\tau}}}\right ]^1 \right ),
  \label{logo}
  \enea 
   for $\frac{T_0}{\overline{M}T} = 0.005,\: 0.01,\: 0.03$ and we see that it is a linear function of $(k \tilde{a}_\ell)^2$, for $(k \tilde{a}_\ell)^2 > 0.5$. The linear dependence confirms that, not only the single matrix element contributions of Eq.(\ref{foi}), but also the logarithm of the trace appearing in Eq.(\ref{logo}) has the linear dependence on $(k \tilde{a}_\ell)^2$. 
This linear dependence on $(k \tilde{a}_\ell)^2$ is the signature of the diffusivity of the Q-excitation modes which sets in at larger values of $(k \tilde{a}_\ell)^2$. 
 The scale $k \tilde{a}_\ell >1 $ characterizes the virtual Q-fluctuations, which we are investigating, by including  the UV corrections.

  To sum up, the steps of  the logical inference starting   from Eq.(\ref{foi})  are
 \bea
 \left . - k^2\frac{ \tilde{a}^2 }{\pi \alpha _SN\epsilon} \frac{ \beta t_0^2}{N}\: \hat{{\cal{B}}}\:  \right |_{\tau_+ = \beta /M}
 \to \frac{\beta}{M} k_BT_0\:{\hat{ f}}_{\overline{\tau}}\! \left (k; \frac{\beta}{M}\right )
 \nonumber\\
\to   \ln \left ({\bf Tr}  \left [ e^{- \frac{T_{0}}{\overline{M}T}\hat{f}_{\overline{\tau}}}\right ]^{\frac{\beta}{M}=1} \right ) \to \frac{T_{0}}{\overline{M}T} \: {\overline{f}}_{\overline{\tau}}  
  \to\tilde{D}_Q \: k^2 \frac{{\cal{T}}_Q}{\overline{M}} .
  \label{dif0}
  \enea
   As the left hand side is ${\cal{O}} \left (\frac{ \beta {\cal{J}}}{N}\right )$, the product $\tilde {D}_Q {\cal{T}}_Q$  is    ${\cal{O}} \left (\frac{ \beta {\cal{J}}}{N}\right )$, as well.  In fact we will put  $\tilde {D}_Q {\cal{T}}_Q = \tilde{a}_\ell^2=\tilde{a}^2 T_0/T$.   The diffusion coefficient $\tilde{D}_Q$  and the relaxation  time scale  of the diffusion process  $i \: {\cal{T}}_Q\equiv \overline{\tau} $ are discussed in Subsection C.
   
   Our simplified  approach to the diffusive constant ${\tilde D}_Q$  provides an analytical 
 approximate expression for $\langle \delta G(  0^+,\tau_+) \delta G(0,0)\rangle_k$. From the left hand side of  the second line of Eq.(\ref{ric}) we write:
\bea
\left . \langle \delta G(  0^+,y) \delta G(0,0)\rangle\right |_{k\neq 0}\propto  \frac{e^{-{\overline{f}_k }'\: y}}{1-e ^{-2 \pi \:  {\overline{f}_k }' } }, 
\enea
with ${\overline{f}_k}' =\beta \tilde{D}_Q k^2/2\: \pi $. 
By Fourier transforming to Matsubara Bose frequencies ($n$ stands here for $\beta\Omega _n /2\: \pi $) we obtain: 
\bea
\int_0^{2\pi} dy  \frac{e^{-{\overline{f}_k }' y}}{1-e ^{-2 \pi \: {\overline{f}_k }' }} \: e^{i \: n \: y} =\frac{ 1- e^{- 2\pi \: {\overline{f}_k }' + 2\pi\: i \: n }}{ {\overline{f}_k }'-i \: n} \frac{1}{1-e ^{-2 \pi \: {\overline{f}_k }'} } \nonumber
\enea
so that:
\bea
\left .\left \{ \langle \delta G( 0^+,\tau_+) \delta G(0,0)\rangle\right \}  \right |_{k\neq 0, n } 
 \propto \frac{2\pi}{\beta} \frac{1}{ \tilde{D}_Q k^2 -i \:\Omega _ n}.
  \label{acori}
\enea
This result highlights the diffusive pole in the Fourier transform of  the Q-fluctuation correlator\cite{guQi}.  
 
 Identifications of the  threshold  temperature,  $T_0$, and of the  space  parameters,  $\tilde{a}_\ell$,   $\tilde{D}_Q {\cal{T}}_Q $,  introduced as  scales in the previous derivation, require a modelization of the damped dynamics of the Q-excitations,  which we  derive in the next Subsection C.  While these parameters,  in the course of the derivation, have been  recognized  as marginal in a renormalization group sense, as they are   ${\cal{O}} \left (\frac{ \beta {\cal{J}}}{N}\right )$ (in the limit ${\cal{J}}, N\to \infty$, ${\cal{J}}/ N\to cnst$), they should rest on phenomenological  fundamental quantities, like the  thermal capacitance $C$ per unit mass and  the thermal conductivity $\kappa$.  These quantities will be related  to  two parameters,  i.e. the damping  of the  neutral  Q-excitations, $\Gamma$, and their  propagation velocity $v$, given by Eq.(\ref{coi1}) in our model. The velocity $v$, appears in  the linear spectrum of the Q- excitations given by  Eq.(\ref{coi3}), while  $\Gamma$ is introduced as a broadening of their spectral peak.  Subsection C is devoted to the presentation and  discussion 
 of these relations.  
  \subsection{Thermalized  and coherent energy processes}
   In the previous Sections we have shown how  
   the  $pGm$  within each SYK dot, $ \delta g_m $,  generate  energy modes  diffusing in the lattice of  the extended SYK model.   The validity of our approach, involving  the  partial factorization that we have adopted in our traces, rests on different  temporal dependence  scales of the center of mass times $y^r-y^s$  on one side and  of the intradot fluctuations  on the other. The "interdot"   dynamical  time scale is discussed  phenomenologically in this Subsection.

 The "interdot"  time scale is  the thermalization  time ${\cal{T}}_Q $, introduced in Eq.(\ref{dif0}). ${\cal{T}}_Q $,  the  threshold temperature scale for thermalization  $T_0$,  and the space scale $\tilde{a}_\ell$, are connected with the velocity  $v$, given by Eq.(\ref{coi1}), and with the phenomenological damping  $\Gamma$,  which is the inverse lifetime of the  Q-excitations. In turn, these quantities depend on   the  thermal capacitance $C$ per unit mass and  the thermal conductivity $\kappa$, which are the phenomenological, experimentally measurable quantities. 
 
  When adopting  our rough approximations, we cannot ignore that these parameters depend on  temperature.   In particular,  the two sets of scales should be considered, $(T_{coh}, {\cal{T}}_{Q})$ related to particle transport in the coherent phase and $\left (T_{0} \cdot \left ( \frac{ N }{ \beta {\cal{J}}}  \right ), \hbar \beta \right )$ for a thermalized system  in the incoherent phase, when $  T \gtrsim T_{0} >T_{coh}$.  The parameters $T_{coh}, {\cal{T}}_{Q}, T_0 $  and $\tilde{a}_\ell^2$  are  $ {\cal{O}} \left ( \frac{\beta {\cal{J}}}{  N }\right )$. We will discuss the two regimes in this Subsection. At finite $T$ the small gap of the $Q-$excitations can be disregarded.  We assume that both regimes have gapless and chargeless bosonic excitation modes of energy $ \Omega _k =\hbar v |k| $, given by  Eq.(\ref{coi1}). Indeed, we exclude charging effects in transport.  In the incoherent regime the gaussian action of Eq.(\ref{sdue}) involves energy  density $ {\cal{N}} \propto  \partial_\tau \theta $  fluctuations and energy flux  density $  \dot{{\cal{N}}} \propto J_Q$ fluctuations\cite{davison}, the Q-excitations. In the coherent phase, bosonic excitations are particle-hole excitations with fluctuations of the particle number $N_e$ and first sound excitations. In the appendix \ref{app:secE} we show that the sound mode survives when the interaction with the SYK dots is turned on perturbatively, embedded in the p-h continuum. We attribute an inverse  lifetime  $\Gamma \propto T $ to these excitations. 
  
  We proceed with the incoherent regime first, at $T \approx T_0$.
 The thermal conductivity, in presence of a damping  $\Gamma$, is  derived in Eq. \eqref{kdis} from the $J_Q-J_Q$ response: 
 \bea
  \kappa = k_B \: \Gamma \: \left ( \frac{ h \: \Gamma } { C T }� \right )^{1/2}. 
    \label{caco}
  \enea
  From one of the Einstein relations, the diffusivity $\tilde D_Q$ is related to the thermal capacitance $C$ and to the density $\rho_0$, according to (p-h symmetry is assumed)
    \bea
 \tilde{D }_Q =\frac{ \kappa}{C \rho_0}.
 \label{ded}
\enea
 As the chemical potential  $\mu$ is assumed to vanish,  the    2-d particle density  $\rho_0$  involved in these excitations, given by  Eq.(\ref{trsm}), is not well phenomenologically defined.   We will estimate it as   $\rho _0= \Gamma ^2/v^2$, a choice that will turn out to be consistent with our results of this Subsection. 
We proceed now by deriving an estimate of $T_{0}$. In this case, energy diffusion is mainly due to heat transport in a highly thermalizable  environment  and  we use the first temperature dependent correction to the energy of the SYK model\cite{maldacena}: $\delta E= c/ (2 \beta^2)  = CT $ in Eq.(\ref{caco}),  where $ c = 4 \pi ^2 \alpha _S N / {\cal{J}}$. From Eq.(\ref{caco}), we get:
  \bea
 \kappa  =  k_B \: \Gamma ^{3/2} \left ( \hbar \beta \right )^{1/2} \: \left ( \frac{ \beta {\cal{J}}} { \pi \alpha _S N }� \right )^{1/2} ,
  \label{kapo}
 \enea
 which, inserted in Eq.(\ref{ded}), with $\rho _0= \Gamma ^2/v^2$  gives:
   \bea
 \tilde{D}_Q 
  = \Gamma  \left ( \frac{ \beta {\cal{J}}} { \pi \alpha _S N }� \right ) \: \tilde{a}^2 ,
  \label{dif1}
  \enea
 where Eq.(\ref{coi1}) has been used.   On the other hand, the last inference in Eq.(\ref{dif0}), together with  Eq.(\ref{foi}),  suggests  that   ${\overline{f}}_{\overline{\tau}} \propto k^2 \tilde{a}^2 $, with   $ \tilde{a}_\ell^2 \sim  \tilde{a}^2 {T_{0}}/{T} $. We conclude  from Eq.(\ref{dif1}) that $ \tilde{D}_Q  \propto \Gamma \:   \tilde{a}_\ell^2$   and, as  $ \tilde{D}_Q{\cal{T}}_Q = \tilde{a}^2_\ell $, the relaxation time\cite{hartnollNatPhys} ${\cal{T}}_Q \sim  \Gamma ^{-1}\sim \hbar \beta $. Thermalization is better handled in euclidean time. Putting $  {\cal{T}}_Q \to  \hbar \beta $ in   $ \tilde{a}^2 \frac{T_{0}}{T}  \sim   \tilde{D}_Q{\cal{T}}_Q$  and using  Eq.(\ref{dif1}), we conclude  that  
 \bea
k_B T_{0} \sim 2\pi\: \hbar \: \Gamma \left ( \frac{ \beta {\cal{J}}} { N \pi \alpha _S }� \right ) .
\label{tdisst}
  \enea
   This equation qualifies  $ k_BT_{0} $  as a threshold energy for efficient   thermalization  and confirms that $T_{0}$ is  ${\cal{O}} \!\left ( \frac{ \beta {\cal{J}}} { N \pi \alpha _S }� \right )$, if just the zero order for $\Gamma$ is retained.  As we have assumed  that $\Gamma \propto T$,  both   $\tilde{D}_Q $  of  Eq.(\ref{dif1}) and  $T_{0}$ of   Eq.(\ref{tdisst}) are  temperature independent.

  Our approximations, which involve some kind of adiabatic factorization, do not allow us to discuss the  coherent carrier transport regime, $ T  \lesssim \Omega _n  \ \lesssim T_{coh} < T_0$, except for a  very qualitative bird's eye. Indeed,  the convergence of the   'normalization' of  Eq.(\ref{tracca})  in Fig.\ref{overM}  is misleading, as one should keep in mind that just the  dominant UV contribution  of   ${\cal{F}}$ has been retained and all the regular contributions (belonging to the fluctuation domain   orthogonal to the $pGm$'s) have been neglected. These include low energy contributions  and their evolution cannot be factorized.
 Anyhow,  back to Eq.(\ref{ded})  for  this case, an approximated expression for  the specific heat arising  from the gapless modes  of the model given by  Eq.s(\ref{coi1},\ref{coi3}) is given by Eq \eqref{heat}: 
  \bea
  C_{{\cal{V}}} = k_B \frac{1}{2\pi} \left ( \frac{k_B T}{\hbar v} \right )^2 \: 6 \: \zeta [3], 
  \label{calspec}
  \enea
  where $\zeta [n] $ is the Riemann function. When the velocity $v$  is inserted in this expression, we get an equation for $1/C_{{\cal{V}}}^2$, which can be related to Eq.(\ref{caco}) to give:
  \bea
 \left ( \frac{\kappa}{k_B \Gamma}\right )^4 \frac{1}{( h \Gamma )^2}�=  \frac{1}{C_{{\cal{V}}}^2T^2 }� = \frac{2\pi}{k_B} \: \frac{\hbar}{(k_BT )^3} \: \frac{\kappa }{6 \: \zeta [3] }. 
   \enea 
Inserting this result in  Eq.(\ref{ded}), with $v$ given by Eq.(\ref{coi1}) and   $\rho _0= \Gamma ^2/v^2$we obtain:
    \bea
  \tilde{D}_Q = \frac{ \kappa ^2}{ \Gamma ^2} \frac{ T}{ \hbar k_B } \tilde{a}^2 =  \frac{\hbar ^2\Gamma ^2}{\hbar} \frac{ 1}{k_B T} \left (\frac{ 1}{6 \: \zeta [3] }\right )^{1/3} \tilde{a}^2.
  \label{diq0}
  \enea

 Assuming again $\Gamma \sim T$,  Eq.(\ref{diq0})  shows that the diffusion constant is  in this case $\propto T$ as in the Einstein
-Smoluchowski formula. At least formally,  it can be put in the form of  a bound  on the diffusion rate, which has been conjectured  for strongly interacting  systems at zero chemical potential\cite{kovtun,hartnollNatPhys}:
\bea
 \tilde{D}_Q  \gtrsim \hbar  \frac{\Gamma ^2  \tilde{a}^2}{k_B T}.
 \label{bound}
 \enea 
 In this case  the velocity which arises here  is not $v_F$ but  $\tilde{v} \sim  \tilde{a} \Gamma $.  Eq.(\ref{bound}) is  non universal. 


 Now we proceed just by analogy with the previous case  and we assume that, just by  replacing  $T_0$ with $T_{coh}$, we can put here  
\bea 
 \tilde{a}^2_\ell \sim \tilde{D}_Q {\cal{T}}_{Q} = \tilde{a}^2 T_{coh} / T .
 \label{groppo}
 \enea 
 From Eq.(\ref{diq0})  it follows that:  
  \bea
 \tilde{a}^2 \frac{T_{coh}}{T}  \sim \frac{\hbar^2 \Gamma ^2}{\hbar} \frac{ 1}{k_B T} \left (\frac{ 1}{6 \: \zeta [3] } \right )^{1/3} \tilde{a} ^2{\cal{T}}_{Q}^{coh}, 
 \label{dtq}
 \enea
 what implies:
 \bea
  {\cal{T}}_{Q}^{coh} \sim  \left ( 6 \:\zeta [3]\right )^{1/3} \frac{ k_B T_{coh}}{\hbar \Gamma ^2}.
  \label{trans} 
  \enea
  Given $T_{coh} \propto t_0^2/N {\cal{J}}\sim {\cal{J}}/N $ , is  $ {\cal{T}}_{Q}^{coh}  \propto T^{-2}$ as in the Fermi liquid case. 
  
 \section{ superconductive coupling  at low temperature}
 In this Section we  present an Eliashberg approach to the superconducting instability of a quantum  electron liquid  that contains  the Q-excitations in its energy spectrum.  As explained in the Introduction, we consider a  model with  two components: a lattice  of local $0+1-d $ SYK dots  and an underlying FL which interacts with the dot lattice perturbatively. Higher dimensional complex SYK models with non-random intersite hopping  have been constructed  with fascinating  NFL properties\cite{haldarBanerjee,lantagne}. We use a perturbative approach\cite{chowdhurySenthil} in   Subsection A and  derive the selfenergy of the coherent phase of the quantum liquid, which turns out to be  a MFL with   short lived and ill defined quasiparticles. In Subsection B we assume an attractive pairing among the quasiparticles,  mediated by the  virtual Q-excitations, and we derive  the critical temperature $T_c$, which is  non BCS-like.
\subsection{Marginal Fermi liquid}
The quasiparticles of a low energy  $2-d$ FL have a quasiparticle residue $Z$ and a single particle energy $\epsilon_{\vec{k}}= \tilde{v_F} k $  in the continuum limit, with a renormalized physical velocity $ v_F^* = Z \tilde{v}_F $ and  a residual local interaction of strength $U_c$, which is dealt with perturbatively. 
The isotropic self-energy arising from the interaction, for $k$ on the Fermi surface, is: 
 \bea
   \Sigma (k_F, i\omega )\hspace{6cm}  \nonumber \\
   = U_c \sum _{\vec{q}} \int \frac{d\Omega}{2\pi} \:  G( \epsilon_{\vec{k}_F+\vec{q}}- \epsilon _{\vec{k}_F} , i\omega + i\: \Omega ) \: \Pi ( q,i\: \Omega )\nonumber\\
   \approx U_c \sum _{\vec{q}} \int \frac{d\Omega}{2\pi} \: \frac{ 1}{i Z^{-1} (\omega + \Omega ) - \tilde{v} _F q \cos\theta }\: \Pi ( q,i\: \Omega ). \label{sigo}
   \enea
 In Eq.(\ref{sigo}), $\theta $ is the angle between $\vec{q}$ and $\vec{k}= \vec{k}_F$ and, for $|q| << k_F$, we have approximated $ \epsilon_{\vec{k}_F-\vec{q}}- \epsilon _{\vec{k}_F} \approx  \tilde{v} _F q \cos\theta$.


  $ \Pi (q, i \: \Omega ) $ is the polarization function 
    \bea 
      \Pi (q, i \: \Omega ) = \sum _p  \sum _{\omega _n} G( \epsilon_{ p},i \omega_n )\: G( \epsilon_{p+q}, i\: \omega_n + i \: \Omega_m). \nonumber
\enea
 In the range of frequencies $ \Omega < \Omega^* =  W^2/ U_c$, where $W$ is the bandwidth, there are two contributions to the polarization, one (labeled by $i=1$) coming from the residual  FL  interaction and a second one ($i=2$) coming from  hybridization with the incoherent disordered  SYK clusters of $0+1 -d$ neutral fermions, interacting at energy ${\cal{J}}$, one at each lattice site (see Fig.\ref{model}).   While $ \Pi^1 (q, i \: \Omega )$ uses the Green's function which appears in Eq.(\ref{sigo}) with a simple pole, 
$ \Pi^2 ( i \: \Omega ) $ is evaluated from the single particle Green's of the SYK model, in the conformally symmetric limit, which is local in space (i.e.  $q-$ independent) and  reported in Eq.(\ref{galt}).
Approximately, is\cite{chowdhurySenthil}: 
  \bea
	  \Pi^1 (q, i \: \Omega )  \approx Z \nu_0 \: \left [1- \frac{\Omega \: sign (\Omega )}{\sqrt{  \Omega ^2 +\left (Z\tilde{v}_F \: q \right )^2 }} \right ], \\  \Pi^2 ( i \: \Omega ) \approx  - \frac{4}{ {\cal{J}}}\: \ln \left (\frac{ {\cal{J}}}{max[ \Omega,  \Omega^* ]} \right )\to - \frac{8}{ {\cal{J}}}\: \ln \left (\frac{ {\cal{J}}}{W} \right ). 
   \label{prim}
  \enea  
Here  $\nu_0 = k_F / ( 2\pi \hbar \tilde{v} _F )$ is the density of states at the Fermi surface and $Z\nu_0 \sim U_c^{-1}$.
  In performing the integral over momenta $p$, we have assumed that, at low temperatures $ T<< \Omega ^*$ , the difference in occupation numbers $n_F\left (Z \epsilon_{\vec{k}+\vec{q}}\right ) - n_F\left (Z \epsilon_{\vec{k}}\right) \approx - \delta \left ( \epsilon_{\vec{k}} \right ) \: \tilde{v} _F q \cos\theta$. 
 
Moving to real frequencies we get: 
 \bea
 \Sigma (k_F, \omega )= - \omega Z^{-1} -i \: \alpha \nu_0 |\omega |^2 \: \ln \frac{ Z \tilde{v}_F k_F }{|\omega | } \: sign (\omega )\nonumber\\
- i \: \frac{  \epsilon _F}{ 2 \Omega ^*}\: |\omega| \:\ln \left ( \frac{{\cal{J}}}{W}\right ) \: sign (\omega ).\:\:\:\: 
\label{rel}
\enea 
For $T> \Omega ^* = W^2 / U_c $ we should put $2\: \ln({\cal{J}}/ W )\to \ln({\cal{J}}/ T )$ in Eq.(\ref{rel}). 
 $\Sigma (k_F, \omega ) $ changes sign at $\omega =0$ when the quasi-particle becomes a quasi-hole. The first term is the real part, while the second term is the imaginary part, $\propto \omega ^2 \times \: \log |\omega |$, from the well known instability of the FL in $2-d$. The third term arises from the coupling to the high energy modes and is beyond the  Landau Fermi Liquid theory. Indeed, the quasiparticle relaxation rate is: 
\bea
 \frac{1}{\tau} \sim -Z \: \Im m  \Sigma (k_F, \omega )\nonumber\\
 = \left [|\omega| \frac{ \epsilon _F}{ \Omega ^*} \ln \left ( \frac{{\cal{J}}}{W }\right )\!+\! \frac{ \alpha }{Z} \frac{\nu_0 |\omega |^2}{\hbar }\ln \frac{ Z \tilde{v}_F k_F }{|\omega | } \right ]\: sign (\omega) \nonumber \\
 \label{tsca}
 \enea
($\alpha $ is a parameter of order one), which shows that to the lowest approximation, the perturbed FL is a Marginal Fermi Liquid. The interaction of the electronic quantum liquid (qL)  delocalized over the $2-d$ lattice with the SYK clusters, makes  the quasiparticles not well defined, but still with a well defined Fermi surface.  In the appendix \ref{app:secE}, we derive the lowest lying collective excitations in the present perturbative frame.  The hydrodynamic collective excitation, the would-be acoustic plasmon, is also rather well defined. At strong coupling, in the limit $U_c \to \caj$, its dispersion tends to the boundary of the p-h continuum and the imaginary part, which blurs the mode, vanishes. The acoustic plasmon is on the verge to emerge as a bound state at low energies, splitted off the p-h continuum.

 \subsection{Superconductive critical Temperature}
 We outline here the  derivation of  the superconducting critical temperature $T_c$, of a $2-d$ qL in interaction with the SYK  lattice, using the  Eliashberg approach\cite{eliashberg,marsiglio}.  Although we are unable to discuss the nature of the microscopic low temperature electron-electron interactions driven by the virtual Q-fluctuations, we assume that  Cooper pairing is induced in a qL of bandwidth $W$,  by virtual coupling with  the diffusive energy Q-modes in the lattice, which, in turn, are generated by  the $pGm$ of the SYK clusters, as discussed in the previous Sections. Three energy scales come into play in this context. The energy  scale $ t_0^2/ {\cal{J}}  \sim W^2/  {\cal{J}}  $, associated with  the temperature threshold  $ T_{coh}$,  below which coupling between the SYK clusters and the qL  is perturbative.   Two more energies associated to the coupling between the qL and the Q-excitations,   the  coupling strength $g$ and the energy cutoff of the interaction $U> U_c$, which also appears in the typical frequency for the attractive interaction  $\Omega^* = t_0^2/ U \sim W^2/ U $ (in this Subsection is   $\hbar =1$). This assumptions immediately implies an electronic energy scale,  as the reference scale for the superconducting transition.  Our standard approach to the superconducting transition within the Eliashberg theory\cite{chowdhury20B} gives rise to a non BCS-like phase transition. The non BCS critical temperature is a direct consequence of  the quantum liquid to be marginal and of  the  excitation modes to be diffusive.

 In a mean field superconducting  Hamiltonian, in the Nambu representation, the one electron Green's function and the electronic self-energy $\Sigma (p, i\: \omega _\nu )$ are $2 \times 2$ matrices defined by the Dyson equation
 \bea
[{\cal{G}}(p, i\: \omega _\nu)]^{-1} = [G_0 (p, i\: \omega _\nu )]^{-1} - \Sigma (p, i\: \omega _\nu), \label{filot}
\enea
 where $G_0 (p, i\: \omega _\nu)$ is the one-electron Green' s function for the non interacting system ($ \left [G_0 (p, i\: \omega _\nu ) \right ]^{-1} = i\: \omega _\nu - \xi _p \sigma_3$)
and  the approximation used for the self-energy is (see appendix \ref{app:secF}):
  \begin{widetext} 
  \bea
 \Sigma (p, i\: \omega _\nu)= - \frac{1}{\beta} \frac{1}{N_q} \sum_q \sum _{p' \nu'} \sigma _3 \: G(p', i\: \omega _{\nu'})\: \sigma _3\: \left | g ( p \: p';q) \right |^2 {\cal{D}} 
 \left (q, i\: \omega _\nu- i\: \omega _{\nu'} \right ) \hspace{1cm}\nonumber\\
 = - \frac{1}{\beta} \sum _{p' \nu'} \sigma _3 \: G(p', i\: \omega _{\nu'})\: \sigma _3\: \int d\Omega  \frac{ 1}{4\pi\:} \frac{\tilde{a}^2_\ell}{\tilde{D}_Q}\int d\Omega _q \left | g_{p,p'}(\Omega _q) \right |^2 {\cal{B}}(\Omega _q,\Omega) \: \left \{�\frac{1}{ i\: \omega _\nu- i\: \omega _{\nu'} -\Omega }-�\frac{1}{ i\: \omega _\nu- i\: \omega _{\nu'} +\Omega }\right \}. \hspace{1cm}\label{sis1}
  \enea
  \end{widetext} 
    Here $\vec{q} = \vec{p}-\vec{p}\:' $ is the transferred momentum and  $\Omega_q =  \tilde{D}_Q q^2$  is the energy of the  collective excitations. An isotropic coupling density  $g ( p \: p') $  is assumed  and, in place of the sum over $p'$ vectors, we integrate over  $\Omega_q $, with the energy density of the $q$ momenta  
 $ \frac{ 1}{4\pi}  \frac{\tilde{a}^2_\ell}{\tilde{D}_Q }$. The imaginary part of the retarded energy flux density response function is
  \bea
 {\cal{B}}(\Omega_q,  \Omega) = -\frac{1}{\pi} \Im m\left \{ {\cal{D}}^R  \left ( \Omega_q,\Omega  \right ) \right \} \nonumber\\
  = -\frac{1}{\pi }\: \left [ \frac{\Omega_q \:\Omega }{\Omega ^2+ \Omega _q^2} \right ]\: {\cal{T}}_Q. 
  \label{respov}
 \enea
   Note the difference, due to diffusivity,  with respect to the usual Eliashberg approach, in which $  {\cal{B}}(q,\Omega)\sim \frac{\Omega_q }{\Omega_q ^2+ 	\Omega ^2}$.   We take  $ \tilde{D}_Q {\cal{T}}^{coh}_{Q} =\tilde{a}_\ell^2 = \tilde{a}^2 {T_{coh}}/{T}$, as  in Eq.(\ref{groppo}) (we drop the label $^{coh}$ from ${\cal{T}}^{coh}_{Q}$ in the following).  
   
Using   Eq.(\ref{rel}),  Eq.(\ref{filot}),  we write  $\left [{\cal{G}} (p, i\: \omega _\nu )\right]^{-1} = i\: Z^{-1} \omega -\left (  \tilde{\xi}_p -i\:  \Im m  \Sigma (k_F, \omega )\right )\sigma_3 - \Xi \: \sigma_1$, in which the mean pairing field  $\Xi$  has to be self-consistently determined:
 \begin{widetext}
  \bea
 \left [ {\cal{G}} (p, \omega) \right]^{-1} =  Z^{-1} \: { \omega} {\bf 1}  -\left \{ \xi_p 
 - i\: Z^{-1} \left [|\omega| \: \frac{ \epsilon_F }{ \Omega ^*} \:\ln \left ( \frac{ {\cal{J}} }{W }\right )+ \frac{ \alpha }{Z} \: \nu_0 |\omega |^2 \: \ln \frac{ Z \tilde{v}_F k_F }{|\omega | } \right ]\: sign (\omega )\right \} \sigma _3 - \Xi ( \omega ) \sigma _1.
 \label{gigat}
 \enea
 
  The final result for Eq.(\ref{sis1}) is: 
  \bea
  \Sigma (k_F,  \omega ) = \nu_0 \int _{-\infty}^\infty d\omega '\: \Re e \left \{ \frac{  Z^{-1} \omega ' {\bf 1} -  \Xi ( \omega ') \: \sigma_1}{\left [ {\cal{P}} (\omega ')\right ]^{1/2}} \right \} \: \int _{0}^\infty d\Omega   \frac{\tilde{a}^2_\ell}{\tilde{D} }\int \frac{d\Omega _q}{4\pi} \left | g_{k_F,\omega '}(\Omega _q) \right |^2 {\cal{B}}(\Omega _q,\Omega) \nonumber\\
\times \left [ \frac{ f( -\omega ') }{ \omega -\omega ' -\Omega +i\: 0^+}+ \frac{ f( \omega ') }{ \omega -\omega ' +\Omega +i\: 0^+}+ \frac{ N( \Omega ) }{ \omega -\omega ' -\Omega +i\: 0^+}+ \frac{ N( \Omega ) }{ \omega -\omega ' +\Omega +i\: 0^+}\right ] \nonumber\\
{\cal{P}}(\omega ) = Z^{-2} \: { \omega} ^2  + Z^{-2} \:
\left [|\omega| \:\frac{ \epsilon_F }{ \Omega ^*} \:\ln \left ( \frac{ {\cal{J}} }{W }\right )+ \frac{ \alpha }{Z} \: \nu_0 |\omega |^2 \: \ln \frac{ Z \tilde{v}_F k_F }{|\omega | } \right ]^2 - \Xi ^2( \omega ) \nonumber\\
  N( \Omega )= \frac{1}{e^{\beta \Omega }-1},\:\:\:  f( \omega )= \frac{1}{e^{\beta \omega }+1},
  \label{filoti}
  \enea
  \end{widetext}
where $ N( \Omega )$ and $f( \omega )$ are the Bose and Fermi occupation probabilities. The term in curly brackets arises from $ \Im m \left \{ \nu_0 \int _{-\infty}^{+\infty} d \xi_{p'} \: \sigma _3 
 {\cal{G}} (p', \omega) \: \sigma_3 \right \}$ which turns into a real part by working out  the inverse of Eq.(\ref{gigat}). 
 A limited region contributes to the integral $ \int d \xi_{p'} $, but we can extend the integration limits  to infinity  with no big error.

\begin{widetext}


 Following McMillan \cite{mcmillan}, we want to find an approximate solution to the gap equation ($ \Delta (\omega ) = Z (\omega ) \Xi (\omega ) $). At the critical temperature, 
  $\Delta \sim 0 $ and can be dropped in the denominator, but the gap equation has to be satisfied. 
From Eq.s(\ref{gigat}), the term multiplied by $\sigma _1 $ gives 
\bea
\Delta (\omega)= Z  \nu_0 \int _{0}^\infty  \frac{ d\omega '}{ \left [{\omega '}^2 + \left (\frac{1}{\tau(\omega ')}\right )^2\right ]^{1/2} } \: \Re e \left \{ \Delta (\omega' )\right \} \: \int _{0}^\infty d\Omega\: \frac{\tilde{a}^2_\ell}{\tilde{D}_Q }\int   \frac{d\Omega _q}{4\pi} \left | g_{k_F,\omega '}(\Omega _q) \right |^2 {\cal{B}}(\Omega _q,\Omega) 
\nonumber\\
\times \left \{ 
\left [ N(\Omega ) + f(-\omega' ) \right ]
 \left [ \frac{1}{ \omega +\omega ' +\Omega }+ \frac{1}{ -\omega +\omega ' +\Omega }\right ] - \left [ N(\Omega ) + f(\omega') \right ]
\left [ \frac{1}{ \omega -\omega ' +\Omega }+ \frac{1}{ -\omega -\omega ' +\Omega }\right ] \right\}.
\enea
\end{widetext}
 $\tau(\omega ')$ is the lifetime of the quasiparticles from Eq.(\ref{tsca}). 
 
 In the rest of the calculation we neglect the thermal excitations and drop $N(\Omega)$. Two energy ranges contribute to $\Delta (\omega)$:
 \bea
 \Delta (\omega ) = \left \{ \begin{array}{cc} \Delta _0 & \:\:\:	\:\: 0 < \omega < \Omega^* \\ \Delta _\infty & \:\:\:	\:\: \Omega^* < \omega \end{array} \right . .
 \enea
 The first, $ \Delta_a (\omega)$, arises from integration over $0< \omega' < \Omega^*$ and the second, $\Delta_b (\omega)$, from  the integration over $\Omega^* < \omega' < U$ ($ U$ is the cutoff energy) . Hence  $\Delta (\omega) =\Delta_a+\Delta_b$.

 While $ \Delta_0$,  can be assumed as the usual order parameter in the lattice, it is unclear what  $ \Delta _\infty $ is, when $\omega > \Omega^*$ and incoherence is established at these energies.  In the mean field approach, $ \Delta _\infty $ can be thought of some kind of intradot field induced by the ordering of the low energy system. Of course we concentrate on the ordering transition for $\omega < \Omega^*$, but both $\Delta$'s should be non vanishing. 
 
\begin{figure}
   \includegraphics[height=82mm]{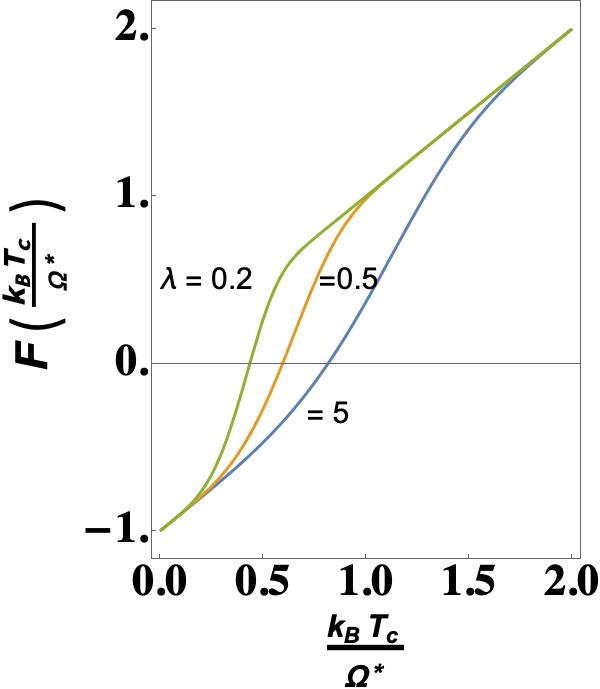}
\caption{ Plot of the function $F[u]= u- \Theta^{\left ({u^4}/{\lambda}\right )}$ of Eq.(\ref{lambda}), vs  $u= \frac{k_BT_c}{\Omega ^*}$, for various values of the exponent $u^4/\lambda$, where $\lambda \propto W \: |g|^2 {\cal{T}}_Q ^2 /U_c $ $(= 5, 0.5, 0.2)$ is defined in the text. The plot is for $ \Theta=0.1$. The zeros of $F[u] $ provide the critical temperature $T_c$. } 
  \label{zeze}
\end{figure}

  Observing that the integration variable $\Omega _q$ has the meaning of the diffusive energy (see Eq.(\ref{respov})), it is clear that it cannot be integrated at energies above $\Omega$. We also use the parameter equality $Z \nu _0 = U_c^{-1}$ and we take $ \left | g_{k_F,\omega '}(\Omega _q) \right |^2 = g^2 $ constant ($[ g]^{-1} \sim  \: time \: \: (\hbar =1\: $  here)):
 \begin{widetext}
\bea
\Delta (\omega )\approx \frac{\tilde{a}^2_\ell}{\tilde{D}_Q } \int _{0}^\infty  \frac{ d\omega '}{ \left [{\omega '}^2 + \left (\frac{1}{\tau(\omega ')}\right )^2\right ]^{1/2} }\: \Re e \left \{ \Delta (\omega' )\right \}  \:\frac{1}{U_c} \int _{0}^{U} d\Omega \: \int_0^\Omega  \frac{d\Omega _q}{4\pi} \left | g_{k_F,\omega '}(\Omega _q) \right |^2   \nonumber\\
\times  \left ( \frac{1}{\pi}\: \left [ \frac{ \Omega _q\:\Omega }{\Omega ^2+ \Omega _q^2} \right ]\: {\cal{T}}_Q \right ) \: 2 \left \{  f(-\omega' ) \frac{1}{ \Omega +\omega ' }- f(\omega') 
\frac{1}{\Omega -\omega ' }\right\} .
 \enea
 \end{widetext}
In the case of $ \Delta_a (\omega ')$, the range of $\omega '$ values  cannot be larger than $\Omega^* $, as well. However,  Fermi functions select $\omega ' \sim 0$ and  we neglect $ \omega '$ in the denominators of the curly bracket, obtaining \cite{tinkham}: 
 \bea
 \Delta_a (0 ) = \frac{\tilde{a}_\ell^2}{\tilde{D}_Q } \int _{0}^{\Omega^*} d\omega '\: \frac{\Delta _o }{ \omega ' \: \frac{ \epsilon _F}{ \Omega^*} \:\ln \left ( \frac{ {\cal{J}} }{W }\right )} \: 
 \nonumber\\
 \times  \: \frac{|g|^2 \ln 2}{4\: U_c} \int _{0}^{U} \frac{d\Omega}{2\pi} \:
  \: {\cal{T}}_Q  \left \{  f(-\omega' ) - f(\omega') \right\} \nonumber\\
  \approx \frac{ \tilde{a}_\ell^2 }{\tilde{D}_Q} {\cal{T}}_Q \: \frac{ |g|^2 W\ln 2 }{8\pi U_c}  \frac{ \Delta _o }{\ln \left ( \frac{ {\cal{J}} }{W }\right )} \: \ln { \beta _c\Omega^*}.
 \enea
 Now the contribution that is coming from $ \Delta_b (\omega ')$ . We neglect $\Omega $ in the denominator in the curly bracket and we keep the FL contribution to the lifetime for large $\omega '$:
  \bea
 \Delta_b (0 ) 
=\frac{\tilde{a}_\ell^2}{ \tilde{D}_Q } \: \frac{ |g|^2 \ln 2 }{4\pi} \: {\cal{T}}_Q \hspace{4cm} \nonumber\\
 \: \times  \: \int ^{U}_{\Omega^*} d\omega '\: \frac{\Delta _{\infty} }{ \frac{ \alpha }{Z} \: \nu_0 |\omega' |^3 \: \left |\ln \frac{|\omega' | }{ Z \tilde{v}_F k_F }\right |} \: 
 \: \frac{1}{U_c}
 \int _{0}^{U}d\Omega \: \Omega\: , \hspace{1cm}\nonumber\\
  \approx  \frac{\tilde{a}_\ell^2}{ \tilde{D}_Q }  \:\frac{ |g|^2\: \ln2}{8\pi}  \: {\cal{T}}_Q \: \frac{ \Delta _{\infty} }{\alpha  } \left (Z \frac{  U^2}{ W^2}\right)^2\: \frac{1}{ \ln \frac{\Omega^*}{Z \tilde{v}_F k_F }}, \hspace{2cm}
 \nonumber
\enea
where  $\ln \frac{\Omega^*}{Z \tilde{v}_F k_F } \approx  \ln \frac{U_c}{U} < 0 $, as  $  \nu_0W \sim 1 $. Summing the two contributions together $ \Delta _0 = \Delta _a(0) + \Delta _b(0) $ we have:
\bea
 \Delta _0 \left [ \frac{8\pi}{ |g|^2 \ln 2 } \: \frac{\tilde{D}_Q}{\tilde{a}_\ell^2} \: \frac{1}{{\cal{T}}_Q } \: -\frac{  W}{U_c \:\ln \left ( \frac{ {\cal{J}} }{W }\right )} \: \ln {\beta_c\Omega^*}  \right ] \nonumber\\
 =\: \frac{ \Delta _{\infty} }{\alpha  } \left (Z \frac{U^2}{W^2}\right)^2\: \frac{1}{ \ln \frac{\Omega^*}{Z \tilde{v}_F k_F }}. 
 \label{cond}
 \enea
 Using the definition of ${{\cal{T}}_Q }\equiv {{\cal{T}}_Q }^{coh}$ given by Eq.(\ref{trans}), the pairing parameter takes the form: 
\bea
 \frac{ \tilde{D}_Q} {|g|^2 \tilde{a}_\ell^2} \frac{1}{{\cal{T}}_{Q}}  \sim \frac{1}{|g|^2 {\cal{T}}_{Q} ^2 } =\left [\frac{ \Gamma  }{|g| } \frac{\hbar \Gamma }{ k_B T_{coh}} \frac{1}{ \left ({6\: \zeta [3] }\right )^{1/3}}\right ]^2.
 \label{kappo}
 \enea
As $ {\cal{T}}_{Q}, T_{coh} \sim {\cal{O}}\! \left ( \frac{\beta {\cal{J}}}{  N \pi \alpha _S \: }\right )$,  it follows that $|g| \sim {\cal{O}}\! \left ( \frac{  N \pi \alpha _S \: }{\beta {\cal{J}}}\right )$, so that $|g| {T}_{coh}  \sim {\cal{O}} (1)$.
 Assuming both $\Delta _0 $ and ${\Delta _{\infty} }$ to be non zero,  Eq.(\ref{cond}) gives: 
  \bea
  \frac{ k_B T_c}{\Omega ^*}= \left ( \frac{ {\cal{J}}  }{W }\right )^{ \frac{ 1}{\alpha \ln \frac{\Omega^*}{Z \tilde{v}_F k_F } }  \frac{U_c}{W}\frac{\Delta _{\infty} } { \Delta _0}\: \left (\frac{1}{\nu_0\Omega^*} \right )^2\: -\frac{8\pi  U_c}{W}\frac{1}{\left ( |g| {\cal{T}}_Q\right )^2 \ln 2 } }. \nonumber\\
   \label{lambda}
\enea
Eq.(\ref{lambda}) provides the value of $T_c$ on a scale of $ \Omega^*$, which is a power of  $ {\cal{J}} /W $, which is difficult to determine,  because it requires the  full  quantitative charcterization of the model. However, qualitatively, the non-BCS behavior is fully apparent. Indeed,  $ {\cal{T}}_Q $ itself is a function of  the temperature, because  the energy width of the mode relaxation, $\Gamma $, appearing in Eq.(\ref{kappo}), is expected to be $\sim T$. In this case, Eq.(\ref{lambda}) defines $T_c$ only implicitely. Dropping the first negative exponent and writing the second exponent as $ u^4/{\lambda }$ where $u= \frac{k_BT_c}{\Omega ^*}$, the zeros of the function $ F[u] = u- \Theta^{\left ({u^4}/{\lambda}\right )} $ give the $T_c$ value.  In the prefactor $\Theta \sim W/ {\cal{J}} $,  all the unknown features of the  pairing interaction  is lumped. $\Theta$ strongly depends on the cutoff energy  $U_c/W$ and on  ${\cal{J}}/W$, as well as on  the lifetime of the quasiparticles in $2-d$ at higher energy $ \sim W$ (see Eq.(\ref{gigat})).  $F[u]$ is plotted in Fig.\ref{zeze} vs $u$, at $\Theta = 0.1$, for $\lambda = 5,0.5,0.2$. Increasing the pairing strength $\propto |g|^2$, $\lambda$ increases, and so does $T_c$. 

\section {Conclusions}
Hopefully, the intriguing  high temperature  "strange metal" phase of  materials undergoing a   HTc superconducting transition is at a turning point, since  attention was drawn to the "universal" linear dependence on $T$ of the resistivity and to features like the possible violation of the  Wiedemann and Franz law\cite{hill,lavasani}. The  Wiedemann and Franz universal ratio unambiguously rests on the coexistence of heat and charge transport typical of weakly interacting electronic  Fermi Liquids.  It is accepted that interactions in these systems are strong and play a nonperturbative role. This gives credit to a  Non Fermi Liquid (NFL) perspective for the high temperature normal "bad metal" phase. Consensus  in the physics community increases on the use of  the doped Mott insulator paradigm  as an interpretation ground  for  the copper oxide HTc materials\cite{carlson}. On the other hand, strong crystal  anisotropy  and doping tend to privilege  the role of copper-oxide planes and the role of collective fluctuations. 
Even when clean single crystals are available\cite{ sreedhar}, the doping and the chemistry of the charge reservoir layers separating  the CuO2 plane(s) from one another could induce inhomogeneities. 

 It was really a twist the discovery that the Su-Ye-Kitaev model, in the limit of  strong interactions and strong disorder,  can be solved exactly in 0+1 dimension displaying  a  NFL  incoherent  toy system, with non trivial properties as a zero temperature finite entropy and a chaotic behavior at long times.  Moreover, hydrodynamical extensions to higher space dimensions provides the linear dependence of the resistivity which  has attracted a flurry of interest from the condensed matter community. By contrast, a  conventional Eliashberg approach, typical of  FL systems has not been seriously questioned\cite{chowdhury20B}. 
 
 On one side we  enquire on the influence of a hopping term  between neighbouring  SYK clusters organized in a  $2-d$ lattice. Hopping is assumed to be marginal in a   $1/ N$ expansion  and  strong coupling  ${\cal{J}}$ limit, with $ \beta {\cal{J}}/N $ kept finite.  On the other side we study the perturbative effect that the SYK lattice  exerts on a  FL with delocalized  electrons  (in the continuum hydrodynamical  limit),  displaying  a well defined Fermi surface and a large Fermi energy.  The aim is to characterize the collective excitations of the SYK system in view of identifying the latter as responsible for the superconducting instability  via an unidentified mechanism.

 There are various ways to extend  the SYK model to a lattice and we use one of them\cite{song,chowdhurySenthil,guQi}. All of them rest on a disorder average and we assume that self-averaging allows for a translationally invariant approach with wavevector $k \tilde{a} << 1$, where $ \tilde{a} $ is the lattice spacing. We focus on the role of the collective fermionic excitations of a SYK dot $\delta g_m$ of  Matsubara frequency $\omega_m$. Among these, there are also incipient Goldstone Modes,  which originate from the spontaneous breaking of the conformal symmetry. However, they acquire mass when  the first  UV correction   $\sim {\cal{O}} \left ( \frac{ N}{\beta {\cal{J}}}\right )$ is included. They are denoted as pseudo  Goldstone modes ($pGm$'s) in the text\cite{kitaev,maldacena}. The  UV correction forces  locality  in space and time.  The real  fermionic propagator $ \delta g_m$ of the IR limit  acquires a complex local  phase  in the extended SYK action,  due to  minimal  compact  $U(1)$ coupling. The energy fluctuations driven by these dressed excitations  across the lattice  can be monitored by investigating  the correlations of a local space-time  UV "order parameter" of the incoherent phase, an extension of  the bilocal two-point propagator $G_c$ of the conformal symmetric limit.  In a NFL system they can be interpreted as energy density excitations, better than chemical potential fluctuations. In this work our focus was on the nature of these dressed  bosonic fluctuations which we nickname as "Q-excitations" and on the response $D^\beta$ of the lattice system to perturbations which excite them. In the recent  past,  the scaling of U(1) RVB  models with a  gap to  both  charge and spin excitations  has been studied\cite{ye}. 
 
  We take advantage of the fact that two time (or temperature) scales come into play,  the "fast" intradot  fermionic  $\delta g_m$ modes and the "slow" interdot  bosonic energy density fluctuations   originating from the Q-excitations. This allows us to characterize the dynamics in a range of energies $\sim T_0$, where $T_0$ is a threshold  temperature for efficient thermalization.  In this temperature range  the Q-excitations are proved to be diffusive when the dynamics induced by the UV correction  is appropriately accounted for. Diffusivity arises from the combination of disorder in the SYK dots and hopping in the superlattice. We find the  mode-mode correlations  in imaginary time  $   \propto G_cG_c {\cal{F}} ^{-1}$ 
where  ${\cal{F}} $ is the bilocal 4-point propagator, which diverges  in the conformal limit, but is made finite when the UV dominant correction is included. 
 The presence of ${\cal{F}}^{-1}$ in the diffusion parameter is the signature of the presence of the  $pGm$ and is the main result of this work. 
The  corresponding  retarded response function in real time can be  derived from the correlations, by analytic continuation  to real frequencies $i \: \omega _m \to \omega +i 0^+$. A similar result was derived directly  in real time\cite{song}, 
but without including the role of the $pGm$ and is reproduced  in the appendix \ref{app:secC}.  In the real time approach,
   the factor ${\cal{F}} ^{-1} $ does not appear as part of the diffusive pole.  The scaling rinormalizes the thermalization temperature $T_0$ and  the diffusion constant $\tilde{D}_Q$, by  introducing a lattice length $\tilde{a}_\ell \sim \tilde{a}\: T_0/T$ and a diffusion time   $ {\cal{T}}_{Q}$, such that   $\tilde{D}_Q {\cal{T}}_{Q} \sim \tilde{a}_\ell$. A  simple quantum  approach to the dynamics of the energy fluctuations in presence of damping $\Gamma$ allows for their explicit determination. $\Gamma \sim T$ is the energy broadening of the Q-fluctuation excitation due to relaxation in the lattice. If we resort to the Einstein relations which connects the diffusion coefficient $\tilde{D}_Q $ to  the transport coefficients\cite{hartnollNatPhys},  we derive the temperature dependence of these quantities and obtain  Eq.(\ref{bound}) which  refers to $\tilde{v} \sim \tilde{a} \Gamma $ as a physical (non universal) diffusion velocity.  Eq.(\ref{bound}) has to be contrasted with   a bound for incoherent systems that has been conjectured\cite{kovtun}. 
  
 
 In the study of the correlations, it emerges clearly (see Fig.\ref{overM}) that  our approach to the partition function and to the generating functional  is only valid for $ T\sim T_0$, an energy range which  we conclude to be  well separated with respect to  the one  $\sim T_{coh}$, the  temperature which marks the prevail of the low energy Fermi Liquid. For $T < T_{0}$, entanglement of the dynamics of the $pGm$'s in the SYK dots with  the  dynamics of the energy fluctuations across the lattice  require more sophisticated methods  than the factorization used here, in the calculation of the thermodynamic functionals.  
 Still, some qualitative hint  is presented   in Subsection V.B.
 
  In Section VI.B we assume that the Q-excitations have a role in the superconductive instability at low temperature.  A dispersive self-energy for an electronic quantum liquid perturbed  by a  SYK dot with a  local interaction ${\cal{J}}$ turns the FL  into a marginal FL\cite{shekhter}, with inverse lifetime of the quasiparticles close to  the Fermi surface $\propto \omega $. The quasiparticle lifetime   influences the mean field superconductive order parameter  $\Delta _{\infty}$ at energies  above $\Omega ^* \sim W^2/U_c$, where $U_c$  is cutoff energy for the pairing interaction.  
 The topic, whether 
 the Q-excitations  could   really play the role of virtual excitations inducing pairing, provided an appropriate attractive coupling is active \cite{lee, kopnicky, chowdhury20A}, is beyond the present state of the art.  It is an old idea that an  incipient Goldstone mode of an ordered phase can accomplish this task.  This possibility was examined in the past and it was concluded that the  fluctuations  involved would lead to a depression of $T_c$\cite{schrieffer}. We think that this pattern may   not work here for various reasons.  Here, indeed, the vertex corrections vanish to lowest approximation order. However,    the fluctuations driving the transition do not arise from an incipient order but  are  non local in time,  in a fully disordered system.   What we call "order parameter" here  is energy  relaxational modes which are effectively non local in space and  non number conserving  in nature, as phonons would be. 
We have also omitted the influence of long-range Coulomb interactions, which certainly modifies the spectrum of boson density fluctuations\cite{carlson}.

 Of course, if  the Q-modes play a role, the temperature scale of the superconducting $T_c$ is of electronic origin, $\sim \Omega _c^*$, defined in Eq.(\ref{lambda}).   $T_c$, as derived  using the Eliashberg\cite{eliashberg} and McMillan\cite{mcmillan} approaches,  is not BCS-like and appears as the zeros of a function $F\left [\frac{k_BT_c}{\Omega _c^*}\right ]$, which is plotted in Fig.\ref{zeze}. It also depends  on the "low" energy scale $ \sim {{\cal{T}}_{Q}^{coh}} ^{-1}$,  on the lifetime at higher energies of the Cooper-pairing electron charges and on the diffusion length of the Q-excitations.  Indeed, the correlation length of the pairs $\xi$ depends on the effective mean square length $ \tilde{a}_\ell ^2 \sim D_Q {\cal{T}}_Q$ which identifies  the $2-d$ range  of the pairing attractive potential.  In our model, its  temperature dependence is $ \tilde{a}_\ell ^2 \sim  T_{coh}/T$. This suggests   a possible experimental check for the surmise  that the superconductive instability is driven by the  Q-modes in the CuO2 planes. Two possibilities arise: $a)$ multiple order parameters could provide different  intervortex interactions for different magnetic field strengths in lowering the temperature. However this possibility requires a two-component  Ginzburg-Landau  formalism, even when only one divergent length scale is associated with the transition at $T_c$\cite{babaev05,babaev12}. 
 $b)$  a second superconducting phase transition  to  $Type \: I$ superconductivity takes place, a  rather unlikely possibility.  Discussion related to case $a)$  arose  in connection with  superconductivity in  the two band $MgB_2$\cite{brandt, kogan,babaevComment}.

 \section*{Acknowledgements}
 The authors acknowledge useful discussions and correspondence  with Elisa Ercolessi, Antony Leggett, Andrej Varlamov, Francesco Tafuri and Rosario Fazio. They also thank D. Chowdhury for communicating ref. [\onlinecite{chowdhury20A}],[\onlinecite{chowdhury20B}]. A.T. benefitted of the lectures by A.Altland at the 15th Capri Spring School on Transport in Nanostructures, 2019 and acknowledges financial support of Universit\`{a} di Napoli, project "PlasmJac", E62F17000290005  and project "time crystal", E69C20000400005

\vspace{0.3cm}
\clearpage 
\onecolumngrid
\appendix
\section*{Appendices}

 \section{Expansion of the action up to second order}
\label{app:secA}
We expand Eq.(7) of the Main Text (MT),
 \bea 
\frac{I_{a}}{N} = \sum _{x}\left[ -\ln Det ( G_o^{-1} - \Sigma _x) + \met \int d\theta _1 d\theta_2 \left ( \Sigma_x (\theta _1 ,\theta _2 )\: G_x^*( \theta _1 ,\theta _2 ) - \frac{(\beta {\cal{J}})^2}{4} \left | G_x( \theta_1 ,\theta _2 ) \right |^4 \right ) \right . \nonumber\\
\left .+\frac{ (\beta t_0)^2}{N} \sum _{x' nn} \int d\theta _1 d\theta _2 \: G_x (\theta _1 ,\theta _2 )\: G_{x'}^*( \theta _1 ,\theta _2 ) \right ]
\label{acaC}
\enea
 to second order in $\delta \Sigma_x, \: \delta G_x,\: \partial _{\tau} \varphi _x $:
 \bea
G_{x}(\theta_1,\theta_2) = \left (G_{c }(\theta_1-\theta_2)+ \delta G(x,\theta_1-\theta_2, \theta_+\right)) \: e^{i \: \varphi _x(\theta_+) },\nonumber\\
 \Sigma_{x}(\theta_1,\theta_2) = \left ( \Sigma_{c }(\theta_1-\theta_2)+ \delta \Sigma (x,\theta_1-\theta_2, \theta_+)\right ) \: e^{i \varphi _x(\theta_+) }.
 \nonumber
\enea
where $\theta _+ = (\theta_1+\theta_2)/2$. Gauge invariance is exploited, transforming $\Sigma_{x}(\tau_1,\tau_2) $ in such a way that the time derivative $\partial _\tau \varphi _x(\tau) $ appears in the Det, so that the variation of the Det term reads:
\bea
 \met \ln \: Det\left [ G_o^{-1}-\Sigma_c \right ] - \frac{1}{4} {\bf Tr} \left ( \left [ G_o^{-1}-\Sigma_c \right ] ^{-1} \left | i \: \partial _{\tau} \varphi _x + \delta \Sigma \right |\right )^2\hspace{7cm}\nonumber\\
 \to \met \ln\: det \left [ G_o^{-1}-\Sigma_c \right ] - \frac{1}{4}{\bf Tr} \left [ 2\:\Re e \left \{ G_c \: i \: \partial _{\tau} \varphi _x G_c \: \delta \Sigma\right \} - \left ( G_c \partial _{\tau} \varphi _x \right )^2 + \left ( G_c \delta \Sigma \right )^2 \right ]
\label{pro3}
\enea
Integrals in time are intended in the second term. Close to the conformal symmetry point (using the saddle point equality $ \Sigma_c = {\cal{J}}^2 G_c^3$) the second term in the action of Eq.(\ref{acaC}) gives:
\bea
( \Sigma_c + \delta \Sigma ) ( G_c + \delta G) - \frac{ {\cal{J}}^2}{4} ( G_c + \delta G)^4 \nonumber\\
 \sim \frac{ 3 {\cal{J}}^2}{4} G_c^4 + \delta \Sigma \delta G - \frac{3 {\cal{J}}^2}{2} G_c^2 \delta G^2.
\label{ecco2}
\enea
Only second order terms will be retained. 
 
 We introduce a dimensionless approach: $ \theta = 2\pi \tau /\beta $ with the substitutions proposed by Kitaev \cite{kitaev} and define $K_c ( \theta _1 ,\theta _2, \theta _3, \theta _4) $:
 \bea
 g(\tau_1,\tau_2) = R_c(\tau_1,\tau_2)\: G(\tau_1,\tau_2), \nonumber\\
 f(\tau_1,\tau_2)= R_c^{-1}(\tau_1,\tau_2)\: \Sigma(\tau_1,\tau_2) \nonumber
 \enea
 \bea
 R_c (\theta _1 ,\theta _2)
 = \beta J \sqrt{(q-1)} \left | \tilde{G}_c (\theta _1 ,\theta _2) \right |^{\frac{q-2}{2}},\nonumber\\
 K_c = R_c (\theta _1 ,\theta _2) \tilde{G}_c( \theta _1, \theta _3) \tilde{G}_c( \theta _4, \theta _2) R_c(\theta _3, \theta _4) 
 \label{trasf}
 \enea 
 For $q=4$, the saddle point bilocal field is: 
 \bea
 G_c( \tau) = b \left [ \frac{ \pi}{\beta \sin \frac{\pi \: \tau }{\beta }} \right ]^{�1/2} \!\!\!\!sign( \tau), \:\: \:\: b^2 = \frac{1}{2\: J \pi^{1/2}}.
 \enea
 Excluding for the time being the hopping term, the action $I_a$ of Eq.(\ref{acaC}), expanded up to second order, reads:
 \bea
 \frac{ \tilde I_2}{N} = -\frac{1}{2} \langle \delta f | \delta g\rangle -\frac{1}{4} \left [ \langle \delta f | K_c | \delta f \rangle +2\: \Re e \left \{ \langle \delta f |K_c| i\:R_c^{-1}\partial _{\tau} \varphi _x \rangle \right \} + \langle R_c^{-1}\partial _{\tau} \varphi _x| K_c|R_c^{-1}\partial _{\tau} \varphi _x \rangle \right ] -\frac{1}{4} \langle \delta g | \delta g \rangle 
 \label{secC}
 \enea
  We integrate out $\delta f' = \delta f + i\: R_c^{-1} \: \partial _{\tau} \varphi _x $:
 \bea
 \sim e^{-\frac{N}{4} \sum _x \left [\langle \delta f ' | K_c | \delta f'\rangle +2 \langle \delta f'| \delta g \rangle - 2 \langle i\: R_c^{-1} \: \partial _{\tau} \varphi _x | \delta g \rangle \right ]} e^{ -\frac{1}{4} \langle \delta g | \delta g \rangle 
}
 \sim \prod _x\: e^{\frac{N}{2} \langle i\: R_c^{-1} \:\partial _{\tau} \varphi _x | \delta g \rangle }\: 
 e^{\frac{N}{4} \left [\langle \delta g | K_c ^{-1}-1| \delta g \rangle \right ]}.
 \enea
 The second variation of the hopping term (third term in Eq.(\ref{acaC}) ), to lowest order  can be expanded as follows:
 \bea
  \delta G_{c,x} (\tau_{12}, \tau_+) \delta G_{c,x'}^*(\tau_{34},\tau_+') =
 \left [  G_{x}(\tau_1,\tau_2)\: G_{x'}^*(\tau_3,\tau_4) -G_{c }(\tau_1-\tau_2)\: G_{c }(\tau_3-\tau_4) \right ] \nonumber\\
   \approx \frac{1}{2}\: G_{c }(\tau_{12}) \:\left ( e^{-i   {\tilde a}\cdot  \nabla_x \left [\varphi _x(\tau_+)-\varphi _x(\tau_+')\right ]} -1 \right ) \: G_{c }(\tau_{34}) 
 + c.c. \hspace{6cm}  \nonumber
 \enea
\bea
\to - \frac{1}{2}\: G_{c }(\tau_{12}) \left ( {\tilde a}\cdot  \nabla_x \left [\varphi _x(\tau_+)-\varphi _x(\tau_+')\right ]  \right )\!\!^2\: G_{c }(\tau_{34}) ,
     \label{pro2}
     \enea
     where only quadratic terms  of the exponential have been included, to account for the additional complex conjugate contribution.    We approximate $ \left [\varphi _x(\tau_+)-\varphi _x(\tau_+')\right ]  = \left  (\tau_+-\tau_+' \right )\partial _\tau \varphi _x(\tau_+)$ and define $ R_c^{-1}\Lambda_c  R_c^{-1}= 
     \frac{1}{2} ( {\tilde a}\cdot   \overset{\leftarrow}{ \nabla}_x  )G_{c }(\tau_{12})(\tau_+-\tau_+')^2 G_{c }(\tau_{34}) ( {\tilde a}\cdot   \overset{\rightarrow}{ \nabla}_{x'} )$, so that, with relative  time integrals traced out, 
  \bea
 \int \!d\tau \!\! \int \! d\tau' \:\partial _\tau \varphi _x(\tau ) R_c^{-1} \:\Lambda_c (\tau - \tau ') \: R_c^{-1} \partial _\tau \varphi _x(\tau ' ).
  \label{reti}
 \enea 
  Fourier transform of  the time dependences gives  $\ \Lambda_c\left (\omega _m,\omega _{m'}; \Omega _\ell \right )$.
 Fourier transforming the hopping term w.r.to $x$ we obtain the kinetic term added to the action of Eq.(\ref{secC}) so that the full functional integral is:
 \bea
 \propto \prod _p \int \left ( \Pi _{m} \delta g^*_m \right ) \left (\Pi _{m'} \delta g_{m'}\right )\: e^{\frac{N}{2} \langle i\: R_c^{-1} \:\partial _{\tau} \varphi _x | \delta g \rangle }\: 
 e^{\frac{N}{4} \left [\langle \delta g | K_c ^{-1}-1| \delta g \rangle \right ]}\:
 e^{-\frac{N}{2} { t_0}^2 k^2\tilde{a}^2 \left [\langle - i\: R_c^{-1} \partial _{\tau} \varphi _p | \Lambda_c | - i\: R_c^{-1} \partial _{\tau} \varphi _p \rangle\right ]}
 \label{gentu}
 \enea
where the forks $\langle ...\rangle$ denote integration over $(\tau(12)_{+}-\tau(34)_{+})$. 

     In the conformal limit $i\: \omega_m$ can be dropped in $G_0^{-1}$in Eq.(\ref{acaC}) and the functional integral of Eq.(\ref{gentu}) is just a gaussian form,
so that, integrating out the $\delta g$'s we have:
\bea
  \sim \prod _p e^{-\frac{N}{2} \left \langle - i\: R_c^{-1} \: \partial _{\tau} \varphi _p \left | \left [K_c ^{-1}-1 \right ]^{-1} \right |- i\: R_c^{-1} \: \partial _{\tau} \varphi _p \right \rangle }\: 
 e^{\frac{N}{4} \: \frac{ { t_0}^2}{N}\: p^2 \left [\langle - i\: R_c^{-1} \partial _{\tau} \varphi _p | \Lambda_c | - i\: R_c^{-1} \partial _{\tau} \varphi _p \rangle\right ]}\nonumber\\
 \sim \prod _p e^{-\frac{N}{2} \left \langle - i\: R_c^{-1} \: \partial _{\tau} \varphi _p \left |\left (1 - \frac{ { t_0}^2}{N}p^2 \Lambda _c [1- K_c ]\right ) K_c \left [1- K_c \right ]^{-1} \right |- i\: R_c^{-1} \: \partial _{\tau} \varphi _p \right \rangle }\: \nonumber\\
  \sim \prod _p e^{-\frac{N}{2} \left \langle - i\: \partial _{\tau} \varphi _p \left | \left (1 - \: \ \frac{t_0^2}{N} {p^2} R_c^{-1} \Lambda_cR_c^{-1} {\cal{F}}^{-1} \right ) {\cal{F}}\right | - i\: \partial _{\tau} \varphi _p \right \rangle }.
\label{gen2}
\enea

 Here we have defined ${\cal{F}} = R_c^{-1} \: K_c [K_c-1]^{-1} \: R_c^{-1} $, and in the Kernel of Eq.(\ref{gen2}) we write: 
\bea
 R_c^{-1} K_c [1-K_c]^{-1} R_c^{-1} - \frac{t_0^2}{N}R_c^{-1} \Lambda_c [1-K_c] K_c^{-1}K_c[1-K_c]^{-1} R_c^{-1}
 = \left (1 - \: \ \frac{t_0^2}{N} {p^2} R_c^{-1} \Lambda_cR_c^{-1} {\cal{F}}^{-1} \right ) {\cal{F}}.\hspace{1cm}
 \label{grif}
 \enea
 The numerical evaluation of this kernel is discussed in \ref{sec:numericalkernel}.

\section{Numerical evaluation of the Kernel of Eq.(\ref{grif})}
\label{app:secB}
\label{sec:numericalkernel}
The correlator ${\cal{F}} = R_c^{-1} \: K_c [K_c-1]^{-1} \: R_c^{-1} $ has been defined after Eq.(\ref{gen2}). It diverges in the conformal limit because $ [K_c-1]$ has zero eigenvalues due to the spontaneous symmetry breaking. We will keep just the series of these eigenvalues (usually denoted by $h=2$), which make the largest contribution to the correlator functions. The spectral representation in imaginary time of the regularized form of the kernel, obtained by shifting the zero eigenvalues, $k(h=2,n)\approx 1- \frac{\alpha _K }{ \beta{\cal{J}}} |n|+ ... $($\alpha _K\approx 3 $), thus including the UV correction at ${\cal{O}} \left ( N/\beta {\cal{J}}\right )$ is \cite{maldacena}
 \bea 
 K_c[1-K_c]^{-1} = \met R_c( \tau_{12})\: {\cal{F}}\left ( \tau_{12} , \tau_{34}\right ) \: R_c( \tau_{34}) \nonumber\\
 = \sum _{h,n} \Psi_{h,n} (\tau_{12}) \frac{ k(h,n)}{1-k(h,n)} \Psi_{h,n}^* (\tau_{34}). \nonumber
\enea
 
The basis functions for $h=2$ are
\bea
\Psi_{2\: n} ( x,y) =\gamma _n \frac{ e^{-i\: n y}}{ 2 \sin\frac{x}{2}} \: f_n(x) ,\nonumber\\
 f_n(x) = \frac{ \sin\frac{nx}{2}}{ \tan\frac{x}{2}} - n \: \cos\frac{nx}{2}, \:\: \: \gamma _n^2 = \frac{3}{\pi^2 |n| ( n^2-1) },
 \label{funz1} 
\enea
with $ x_{12} = \tau _{1}- \tau_2, \:\:\:\: y_{12}=\frac{ \tau_1+\tau_2}{2} $. 
We Fourier transform the variables $\tau_{12}$ and $\tau_{34}$. In full generality:
\bea
 {\cal{O}}_{m,m'} = \sum _{n \geq 2} e^{i\: n (y_{12}-y_{34})} \Phi_{n} (\omega_m)\left \langle 2,n \left | {\cal{O}} \right | 2,n \right \rangle \Phi_{n}^* (\omega _{m'}), 
\nonumber
 \enea
where $\omega _m,\omega _{m'}$ are the fermionic Matsubara frequencies. We ridefine variables in such a way that $\omega _m \to m$ with $m$ integer. 
 The basis functions are:
\bea
 \Phi_n(m) = \int_0^{2\pi} \frac{d\tau}{2\pi} \: \frac{ 1}{ \sin\frac{\tau}{2}} \: f_n(\tau ) \: e^{i\: m\: \tau}. \label{funz2} 
 \enea
It can be shown that the Fourier transforms $\Phi_n(m)$ have a factor $ -\met \left (1 + e^{2 \: i\: m \pi}\right )$ or $ -\met \left (1 - e^{2 \: i\: m \pi}\right )$, depending on $n$ being even or odd, respectively. It follows that odd $m$ imply even $n$ as expected because the $\tau _+ $ time dependence has to be with $n$ even, i.e. bosonic-like. $ \Phi_n(m) $ have a maximum at increasing values of $m$ when $n$ increases and eventually go to zero. 

 The largest contribution ${\cal{O}}( \beta {\cal{J}}/N )$ of $ \left [1- K_c \right ]^{-1} $ appearing in Eq.(\ref{gen2}), gives $R_c^{-1}K_c \left [1- K_c \right ]^{-1} R_c^{-1} \sim {\cal{O}}(1)$ and, as $ G_cG_c \sim 1/\beta {\cal{J}}$, to have the same order of the kinetic term $t_0 \sim {\cal{O}}( [\beta {\cal{J}}/N]^{1/2} )$. 
 
 In the definition of the matrix function $ R_c^{-1} \Lambda_cR_c^{-1} $ from Eq.(\ref{lalo}), the matrix $ \widehat{G_cG_c } \hat{ {\cal{ F}}}^{-1}$ appears, which is the inverse of $\frac{{\cal{F}}}{G_cG_c}$.
 
 The dominant expression for $\frac{{\cal{F}}}{G_cG_c} ( \tau_1...\tau_4)$ in imaginary times, on the subspace orthogonal to the $pGm$ fluctuations is \cite{maldacena}: 
 \bea 
\frac{{\cal{F}}}{G_cG_c} ( \tau_1...\tau_4) = \frac{6 \alpha _0}{\pi^2 \alpha _K} \beta {\cal{J}}\!\!  \sum _{n \geq 2} \! \frac{e^{i\: n (y_{12}-y_{34})}}{n^2 (n^2-1)}\: f_n(\tau_{12}) \: f_n(\tau_{34}).
\label{funz2}
\enea
Its Fourier transform requires the transformed basis functions: 
\bea \phi_n(m) = \int_0^{2\pi} \frac{d\tau}{2\pi} \: f_n(\tau ) \: e^{i\: m\: \tau}. \nonumber
\enea
 All of them have a factor $ \sin {m\pi }$ which vanishes for $m$ odd integer. However, this zero can be compensated by a zero in the denominator. Consider the case $n=2$ for example:
 \bea
 \phi ( n=2, m) = \frac{\sin {m\pi }}{\pi \: m ( m-1)(m+1)} \: e^{i m\pi },
 \nonumber
 \enea
 We give a finite expression to this vector element using the limit:
 \bea 
 \lim_{x =\pm 1} \frac{\sin x \pi} {(x-1)(x+1)} = \frac{1}{2 }( \mp \pi ), 
 \enea
However, all $m > 1$ give zero for $n=2$, because there is no other factor of the kind $(m-3), (m-5), ...$(only odd $m$ are considered) in the denominator. We get  $\phi _n(m) = \: sign(m)  \times$:
\bea
 n= 2 \:\:\:\freccia & \left \{ \begin{array} {cc} m=1 & \frac{1}{2 \: m^2} \: e^{i m\pi } , \\ 
 m=3,5,7, ... & 0 \end{array} \right . \nonumber\\
 n=4 \:\:\: \freccia & \left \{ \begin{array} {cc} m=1 & - 2\: \frac{ ( 5 m^2 -2)}{ (m- 2)(m+2)} \: \frac{1}{2\: m^2}\: e^{i m\pi } \\
m=3,5,7, ...  & 0 \end{array} \right . \\
 n=6 \:\:\: \freccia & \left \{ \begin{array} {cc} m=1 & - \: \frac{ ( 36-119 m^2 + 35 m^4 )}{ (m- 3)(m- 2)(m+2)(m+3)}\: \frac{1}{2\: m^2}\: e^{i m\pi } \\
 m=3 & - \: \frac{( 36-119 m^2 + 35 m^4 )}{ (m- 2)(m- 1)(m+1)(m+2)} \: \frac{1}{2\: m^2}\: e^{i m\pi } \\
m=5,7,9, ...  & 0 \end{array} \right . \nonumber\\
  n=8 \:\:\: \freccia &  \left \{ \begin{array} {cc} m=1 & - \: \frac{2 \:  ( -288+1068 m^2 -462 m^4+42 m^6 )}{ (m- 4)(m- 3)(m- 2)(m+2)(m+3)(m+4)}\: \frac{1}{2\: m^2}\: e^{i m\pi } \\
 m=3 &  - \: \frac{2\:  ( -288+1068 m^2 -462 m^4+ 42 m^6 )}{ (m- 4)(m- 2)(m- 1)(m+1)(m+2)(m+4)}\: \frac{1}{2\: m^2}\: e^{i m\pi } \\
 m=5,7,9,... & 0 \end{array} \right . \nonumber\\
  n=10 \:\:\: \freccia &
  \left \{ \begin{array} {cc} m=1 & - \: \frac{ 3 \: ( -4800+18964 m^2 -9735 m^4+1386 m^6 +55 m^8)}{ (m- 5) (m- 4)(m- 3)(m- 2)(m+2)(m+3)(m+4)(m+5)} \:\frac{1}{2\: m^2}\: e^{i m\pi } \\
 m=3 &  - \: \frac{ 3\: ( -4800+18964 m^2 -9735 m^4+1386 m^6 +55 m^8)}{(m- 5) (m- 4)(m- 2)(m- 1)(m+1)(m+2)(m+4)(m+5)} \: \frac{1}{2\: m^2}\: e^{i m\pi } \\
  m=5 &  - \: \frac{ 3\: ( -4800+18964 m^2 -9735 m^4+1386 m^6 +55 m^8)}{ (m- 4)(m- 3)(m- 2)(m- 1)(m+1)(m+2)(m+3)(m+4)} \:\frac{1}{2\: m^2}\: e^{i m\pi } \\
 m=7,9, 11,... & 0 \end{array} \right . \nonumber\\
  n=12 \:\:\: \freccia & \left \{ \begin{array} {cc} m=1 & =- \: \frac{2 (-259200 + 1066104 m^2 - 603746 m^4 + 105963 m^6 - 6864 m^8 + 143 m^{10})}{ (m- 5)(m- 4)(m- 3)(m- 2)(m+2)(m+3)(m+4)(m+5)}\: \frac{1}{2\: m^2}\: e^{i m\pi } \\
 m=3 &  - \: \frac{ 2 (-259200 + 1066104 m^2 - 603746 m^4 + 105963 m^6 - 6864 m^8 + 143 m^{10})}{ (m- 5)(m- 4)(m- 2)(m- 1)(m+1)(m+2)(m+4)(m+5)}\: \frac{1}{2\: m^2}\: e^{i m\pi } \\
  m=5 &  - \: \frac{ 2 (-259200 + 1066104 m^2 - 603746 m^4 + 105963 m^6 - 6864 m^8 + 143 m^{10})}{ (m- 4)(m- 3)(m- 2)(m- 1)(m+1)(m+2)(m+3)(m+4)}\: \frac{1}{2\: m^2}\: e^{i m\pi } \\
 m= 7,9, 11,... & 0 \end{array} \right . \nonumber\\
 \label{lista}
  \enea
 Also the polynomials in the numerator could have been factorized but the roots are non-integer. We normalize each $n-$vector of the basis but we do not orthogonalize these basis vectors. We define matrices $W^n_{m, m'} $, by multiplying $ column \times row $ each $n-$vector.To the elements with $m,m'=1$, all vectors $n=2,4,6,8,10,12$ contribute. To the elements with $m,m'=3$, vectors with $n=4,6,8,10,12$ contribute. To the elements with $m,m'=5$ only vector with $n= 10,12$ contributes. The final result for $n\leq 12 $ is a $3\times 3 $matrix. Each $3\times 3 $ matrix $\hat{W}^n $ has eigenvalues $1,0,0$. Computations leading to Fig.2,3,4 of the MT  have been performed with $3\times 3$ matrices up to $n=12$. The starting point is Eq.(\ref{zozo1}) of the text: 
 \bea
 \left . \frac{{\cal{F}}}{G_cG_c} \left ( y_{12} - y_{34} \right ) \right |_{m,m'} \label{28} = \frac{6 \alpha _0}{\pi^2 \alpha _K} \beta {\cal{J}} \sum _{n \geq 2, even}^{10} \frac{\cos { n (y_{12}-y_{34})}}{n^2 (n^2-1)}\: W^n_{m, m'}. \nonumber
 \enea
  where $y$ are center-of-mass times: $ y = y_{12}-y_{34} $.
  
  In Fig.4 of the MT we plot an interpolated smoothed curve of the (approximate) lowest $\overline{M}$ value which fulfills unitarity of the partition function of Eq.(\ref{tracca}) of MT vs. $T_0/T $, for $ k \tilde{a} =1$. Precision is up to $ > 10^{-5}$. The trend is only meaningful for $T_0/T \sim 1 $, because larger values require $n >12 $ in the spectral representation of Eq.(\ref{funz2}) and matrices $ m\times m$ of rank $ \tilde{r}>3$, i.e. higher than the ones used here. 
 
 
 \section{Real time approach to the correlation functions}
\label{app:secC}
  The response function in real time is derived from extension of the imaginary time along the Keldysh contour.
The present approach\cite{song} is only semiclassical. In setting up the functional we disregard the $q-q$ term and neglect variation with respect to the saddle point solutions for the $ G$ and $\Sigma$ of the SYK dots. This means that, in going back to the action in Eq.(\ref{secC}), terms with $\delta g$ and $\delta f$ are disregarded, so that only an extra term is added here, taken from the imaginary time action, that is $-\frac{N}{4} \langle R_c^{-1}\partial _{\tau} \varphi _x| K_c|R_c^{-1}\partial _{\tau} \varphi _x \rangle = -\frac{N}{4} \langle \partial _{\tau} \varphi _x| G_c G_c| \partial _{\tau} \varphi _x \rangle $, which we denote as the kinetic energy term.

 As for the non local phase fluctuations, Fourier transforming with respect to time the non local fluctuation part of the action (i.e. the "hopping" term of Eq.(\ref{reti})) is of a similar form:
 \bea
{\scriptstyle \int dp \: \tilde{D}\:p^2 \: \sum _{ss'} \int d\omega \:\varphi_s(\omega )\: s s'\: G_{ss''}(\omega )\: G_{s'' s'}(-\omega ) \: \varphi_{s'}(-\omega)}.
\label{kel}
\enea 
With respect to the Fourier transform of the kinetic energy term, the present one has a factor $\omega ^2$ lacking, so that we merge the two together, by defining a function 
\bea
 \left [ \frac{ \tilde{D}\: p^2\: h(\omega) }{\omega ^2} -1 \right ] \: G_cG_c \equiv \zeta \: G_cG_c .
 \label{deak}
 \enea
 which defines the function $\zeta(\omega )$. $h(\omega)$ excludes the $\omega =0$ term. Its retarded form is defined as
 \bea
 \tilde{D}\: \: h_{R} (\omega ) \: G^{R}_\omega G^K_{-\omega} = \tilde{D}\:\int _{-\infty}^{\infty} dt\: G^{R}(t) G^K(-t) \: \left [ e^{i\: \omega t }
 -1\right ] \nonumber\\
 \sim i\: \tilde{D}\: \: \omega \: \int _{-\infty}^{\infty} dt\: G^{R}(t) G^K(-t) 
 \approx  i\: \tilde{D}\: \: \omega \: G^{R}_0 G^K_0. 
 \nonumber
 \enea
When writing $h(\omega )$, we will not specify the label $R/A$ for the retarded or advanced form, in the following, as long as no ambiguity arises. 
 Transforming from the branches $s,s' = +,-$ to the combined $\alpha, \beta \equiv cl, q$\cite{kamenev,belzig}, we get:
 \bea
 \varphi^{cl/q}( t_j) = \frac{1}{\sqrt{2}} \left ( \varphi_{+}( t_j)\pm \varphi_{-}( t_j) \right ), \nonumber\\
 \met \left ( \begin{array}{cc} 1 & 1 \\1&-1 \end{array} \right ) \: \left ( \begin{array}{cc} G_{++} & G_{+-} \\G_{-+} &G_{--} \end{array} \right ) \: \left ( \begin{array}{cc} 1 & 1 \\1&-1 \end{array} \right ) = \left ( \begin{array}{cc} 0 & G^A \\G^R & G^K \end{array} \right ).
 \enea
  The $cl-cl $ component is zero in the matrices on the right. It reflects the fact that for a pure classical field configuration ($ \varphi^q =0$), the action is zero. Indeed, in this case $ \varphi_+ = \varphi_-$ and the action on the forward part of the contour is canceled by that on the backward part (safe for the boundary terms, that may be omitted in the continuum limit), because the circuit is closed\cite{kamenev}.
  
 The integrand of Eq.(\ref{kel}) becomes \cite{song}:
 \bea 
 \begin{array}{cc} \left ( \: \varphi^c_\omega \right . &\left . \varphi ^q_\omega \right )\\ & \end{array} \: \left ( \begin{array}{cc} G^A_\omega \: G^R_{-\omega} & G^A_\omega G^K_{-\omega} \\G^K_{\omega}G^R_{-\omega} & G^R_{\omega}G^A_{-\omega} +G^K_{\omega} G^K_{-\omega} \end{array} \right ) \left ( \begin{array}{c} \varphi^c_{-\omega} \\ \varphi ^q_{-\omega} \end{array} \right ) 
 \nonumber
  \enea
  The resulting matrix can be rewritten as matrix of the self-energies $ {\Sigma}_{D}$ due to the $\tilde{D}\:$ coupling, which shows the same causality structure:
  \bea
   \left ( \begin{array}{cc} G^A_\omega \: G^R_{-\omega} & G^A_\omega G^K_{-\omega} \\G^K_{\omega}G^R_{-\omega} & G^R_{\omega}G^A_{-\omega} +G^K_{\omega} G^K_{-\omega} \end{array} \right ) = \left ( \begin{array}{cc} G^A_\omega \: G^R_{-\omega} & {\Sigma}_{{D}} ^A \\ {\Sigma}_{{D}} ^R & {\Sigma}_{{D}}^K \end{array} \right ).
   \nonumber
  \enea

  
   We neglect the ${qq}$ term and write the functional: 
   \bea 
   \int {\cal{D}}\varphi^c_\omega {\cal{D}}\varphi^q_\omega \:e^{\begin{array}{cc} -\frac{i}{2} \int d\omega \: \left (\omega \: \varphi^c_\omega \right . &\left . \omega\: \varphi ^q_\omega \right )\\ & \end{array} \: \left ( \begin{array}{cc} G^A_\omega \: G^R_{-\omega} & \zeta^A\: G^A_\omega G^K_{-\omega} \\ \zeta^R\: G^K_{\omega}G^R_{-\omega} & 0 \end{array} \right ) \left ( \begin{array}{c} \omega \varphi^c_{-\omega} \\ \omega \varphi ^q_{-\omega} \end{array} \right ) }, 
   \label{actphi}
  \enea
where we have also added the kinetic energy term of the semiclassical approach. 

 The field ${\dot{\cal{N}}}(x,t)$ is the source of $ \partial _t \varphi (x,t) $. We want the response written along the Keldysh contour:
  \bea
   D_{{\dot{\cal{N}}}{\dot{\cal{N}}}}(x,t) = i\: \theta (t) \langle \left [{\dot{\cal{N}}}(x,t), {\dot{\cal{N}}}(0,0) \right ] \rangle \nonumber\\
   =
   \frac{i}{2} \langle \left \{ {\dot{\cal{N}}}^{c}(x,t) {\dot{\cal{N}}}^{q}(0,0) +{\dot{\cal{N}}}^{q}(x,t) {\dot{\cal{N}}}^{c}(0, 0) \right \}\rangle.
 \label{reta}
 \enea

To get the generating functional of the $\varphi-\varphi $ fluctuations we invert the kernel of Eq.(\ref{actphi}), obtaining the matrix: 
\bea
\left ( \begin{array}{cc} 0 & ( \zeta^R G^K_\omega G^R_{-\omega})^{-1} \\(\zeta^A G^A_{\omega} G^K_{-\omega}) ^{-1}& - ( G^K_{\omega}G^K_{-\omega} )^{-1} \end{array} \right )
\nonumber
 \enea
 The $[G^{-1}]^K$ component for the free field is only a regularization factor, originating from the (time) boundary terms. It is, in general, non-local in $x$ and $x'$, however, being a pure boundary term, it is frequently omitted\cite{kamenev}.
In our case this should apply because $ [G^{-1}]^K(t,t') = [G^{-1}]^R \circ F -F\circ [G^{-1}]^A = [G^R ]^{-1}\circ F -F\circ [G^A]^{-1} = [G^K]^{-1} $ .
Integrating out the $\varphi$ fields and ignoring again the $q-q$ term, we get:
 \bea 
  \propto \exp \left \{ \begin{array}{cc} -\met \int d\omega \: \left ({\dot{\cal{N}}}^c(\omega) \right . &\left . {\dot{\cal{N}}}^q(\omega) \right )\\ & \end{array} \left ( \begin{array}{cc} 0 & ( \zeta^R G^K_\omega G^R_{-\omega})^{-1} \\ ( \zeta^A G^A_{\omega} G^K_{-\omega}) ^{-1}& 0 \end{array} \right ) \: \left ( \begin{array}{c} {\dot{\cal{N}}}^c(-\omega) \\ {\dot{\cal{N}}}^q(-\omega) \end{array} \right ) \right \}. 
  \enea
Functional derivation with respect to the sources provides the cross contributions (we keep just the lowest order in $\omega$). Using the definition of Eq.(\ref{deak}):
 \bea
N \omega ^2 \langle  \varphi^{cl }( \omega ) \: \varphi^{q }( -\omega ) \rangle \approx \: (G^K_0 G^R_{0})^{-1} \: \frac{\omega ^2 }{ i\: \tilde{D}\: p^2 \omega -\omega^2 } 
\nonumber\\
N \omega ^2\langle  \varphi^{q }( \omega ) \: \varphi^{cl}( -\omega ) \rangle \approx \left (G^A_{0} G^K_{0}\right ) ^{-1}\: \frac{\omega^2} { - i\: \tilde{D}\: p^2\omega -\omega^2 }.\nonumber
\enea 
 Now, the retarded energy flux density response of Eq.(\ref{reta}) can be estimated, considering that $\delta \dot{ \varphi}^{c,q}$ and $ {\dot{\cal{N}}}^{c/q} $ are conjugate variables 
 $ {\dot{\cal{N}}}^{c/q} (t) = \frac{ \delta S_\varphi}{\delta \dot{ \varphi}^{c,q}},$ so that, 
 keeping just the $\omega ^2-$term in Eq.(\ref{actphi}), 
 \bea
 {\dot{\cal{N}}}^{cl} (\omega )= - N \: G^{A}_0 G^K_{0} \omega \varphi _{q} (-\omega), \nonumber\\
 {\dot{\cal{N}}}^{q} (\omega )= - N \: G^K_{0} G^{R}_0 \omega \varphi _{cl} (-\omega) \nonumber
 \enea
we get 
 \bea
 \langle {\dot{\cal{N}}}^{cl} (\omega )\: {\dot{\cal{N}}}^{q} (-\omega )\rangle = - N^2 \: G^{A}_0 G^K_{0} \: G^K_{0} \: G^{R}_{0}\: \omega ^2 \times \langle \varphi ^{q} (-\omega) \varphi ^{cl} (\omega) \rangle \: \nonumber \\ 
 = - N^2 \: \frac{\omega ^2 }{i \: t_0^2 p^2 \omega -\omega^2 } \: G^{A}_{0} G^K_{0},  \: \nonumber \\ 
 \langle {\dot{\cal{N}}}^{q} (\omega )\: {\dot{\cal{N}}}^{cl} (-\omega )\rangle  = -N^2 \: G^{K}_{0 } G^R_{0} \: \omega ^2 \times \:\langle \varphi ^{cl}(-\omega) \varphi ^{q} (\omega) \rangle \: G^A_{0} G^{K}_{0}  \: \nonumber \\
 = - N^2 \: \frac{\omega ^2 }{-i\: t_0^2 p^2 \omega -\omega^2 } \: G^{K}_{0} G^R_{0}.
 \nonumber
 \enea
 
 The symmetrized correlation is: 
 \bea
 \met \: \left ( \langle {\dot{\cal{N}}}^{cl} (\omega )\: {\dot{\cal{N}}}^{q} (-\omega )\rangle + \langle {\dot{\cal{N}}}^{q} (\omega )\: {\dot{\cal{N}}}^{cl} (-\omega )\rangle 
 \right ) = N^2 \: \Re e \left \{ \frac{\omega ^2 }{i \: t_0^2 p^2 \omega +\omega^2 } \: G^K_{0} G^{R}_{0}\right \} \nonumber\\
\approx N^2 \: \frac{ t_0^2 p^2 \omega \: }{( t_0^2 p^2 )^2 + \omega ^2} \: \Im m [ G^K_{0} G^{R}_{0}]. 
\label{simcorr}
 \enea 
 This result can be rewritten as 
 \bea
 sign(\omega )\: \Im m \left \{\frac{\omega }{i \: t_0^2p^2 - \omega } \: \Im m [ G^K_{0} G^{R}_{0}] \right \} . 
\nonumber
\enea
 Subtracting the $p=0$ term, we recognize $\Im m \{ D^R (p, \omega )\}$, the imaginary part of the density response function\cite{wen} 
 \bea
\Im m \left \{ D^R_{{\dot{\cal{N}}}{\dot{\cal{N}}}}(p,\omega ) \right \}  = - sign(\omega )\: \Im m \left \{ \left ( \frac{ D\: p^2 }{ i\: \omega - D\: p^2 } \right ) \: \Im m [ G^K_{0} G^{R}_{0}] \right \}. 
\label{reD}
 \enea
 This result should be compared with Eq.(51)  of the MT. Apart for the matrix structure of the function in Eq.(47)  of the MT, the important point is that $ {\cal{F}} ^{-1} $ is absent here in the definition of the diffusion parameter. 
 
We add here the important consequence on the electrical conductivity. In the conformal limit, the electrical conductivity is \cite{song}:
\bea
Re\{ \sigma \} = -\lim _{\omega \to 0} \: \frac{1}{\omega } \Im m \left \{D^{R\beta} ( \omega )\right \} \propto - \frac{\beta}{2\pi^2} \: \frac{ t_0^2}{N } \int d\omega \: {\rm sech}^2 \frac{\beta \omega}{2} \: \left ( \Im m \left \{ G^{R\beta}(\omega)\right \}\right )^2 \nonumber\\
= - \frac{1}{2\pi^{3/2}} \: \frac{ t_0^2}{N\epsilon } \: \int dx \: {\rm sech}^2 (\pi x) \:\left [ \Re e \left \{ \frac{ \Gamma \left ( \frac{1}{4} - i\: x\right )}{ \Gamma \left ( \frac{3}{4} - i\: x \right )}\right \}\right ]^2 
 \propto \left ( \frac{ \beta {\cal{J}}} { \pi \alpha _S N }� \right ). \hspace{5.5cm}
 \label{codty}
\enea
Resistivity is $\propto T$ in this approach. 
\section{Quantization of gapless diffusive excitation mode}
\label{app:secD}
$ J_ {\cal{Q}} = - \kappa \nabla T $ is a classical diffusion equation of a non conserving system. We now construct a Hamiltonian of the excitation modes which is conserving but we ask that, introducing a relaxation time $\tau _0 = \hbar /\Gamma$ for these modes, the equation of motion reproduces $ J_ {\cal{Q}} = - \kappa \nabla T $. We will quantize this Hamiltonian and derive the response function from the fluctuations of these modes. The canonical conjugate variables and the corresponding Lagrangian (in 2-d) are:
 \bea
 \dot \theta = \left ( \frac{\kappa C}{\hbar T}\right )^{1/2} \!\! \frac{ J_ {\cal{Q}}\tau_0}{k_B},\:\:
 \nabla \theta = \left ( \frac{\hbar}{\kappa C\: T}\right )^{1/2} \!\! \frac{\kappa}{T} \nabla T \label{can}
 \enea
 Here $C$ is the thermal capacitance, $\kappa $ is the thermal conductance and $J_Q$ is is the thermal energy current . 
 The corresponding Lagrangian is 
 \bea
 {\cal{L}}= \met \int d^2 x\: \left [\frac{k_B}{T}\: \left (\frac{ J_ {\cal{Q}}\tau_0}{k_B}\right )^2+ \frac{\hbar }{\kappa C} \: \left (\frac{ \kappa}{T} \nabla T \right ) ^2\right ]
 \nonumber\\ \equiv \met \int d^2 x\: \left [ A\: \dot \theta ^2+ B \: ( \nabla \theta )^2 \right ]\label{lagx}
 \enea
 with $ A = \frac{\hbar k_B}{\kappa C}$ and $ B=T $. These choices provide 
terms in the square brackets which have dimension $ {\cal{E}}/\ell ^2 $ ($ {\cal{E}} \equiv \: energy$).

 With the approximation $ \tau _0 \dot {J}_ {\cal{Q}} \approx {J}_ {\cal{Q}}$, the equation of motion,
 \bea
 \frac{d}{dt} \left (\frac{\partial L}{\partial \dot \theta }\right ) -\frac{\partial L}{\partial \nabla \theta }= A \ddot \theta + B \nabla \theta =0, \enea
 boils down to the diffusion equation: 
 \bea
 \frac{\hbar }{\kappa C} \: J_ {\cal{Q}} = - \frac{\hbar }{\kappa C} \: \kappa \nabla T . \nonumber
 \enea
 Although $\hbar$ is already in the Lagrangian, we proceed with quantization of the theory \cite{wen}. Fourier transforming, the canonical momentum for $\theta _k$ is
 \bea
 \pi_k = \frac{ \partial {\cal{L}}}{\partial \dot \theta _k} = A \dot \theta _{-k},\:\:\: \left [ \theta _k, \pi _{k'} \right ] = i\: \delta _{ k k'}.
 \enea
The Hamiltonian
\bea 
 {\cal{H}} = \met \sum _k \left [\frac{1}{A} 
 \pi_{-k} \pi_k + B k^2 \: \theta_{-k} \theta_k\right ].\nonumber
 \enea
 is second quantized according to $ \hat{a} _k = u_k \theta _k + i\: v_k \: \pi _{-k} $, with 
 \bea 
 u_k = \frac{1}{ \sqrt{2} } \left ( A B \right ) ^{ 1/4}, \:\: v_k = \frac{1}{ \sqrt{2} } \left ( A B \right ) ^{ -1/4}, \nonumber\\
 \epsilon _k = \sqrt{ \frac{ B\: k^2}{ A}} = \left [\frac{ \kappa }{k_B} \frac{ C T}{\hbar }\right ]^{1/2} |k| \equiv v\: |k| , \label{coi0}\\
 \pi _k = -i\: T^{1/2} \frac{1}{ (2 \epsilon_k)^{1/2}} |k| \: \left ( \hat a_{-k} - {\hat a}^\dagger _k \right ), \nonumber\\ 
 \theta _k = T^{-1/2} \frac{ (2 \epsilon_k)^{1/2}}{ |k|} \: \left ( \hat a_{k} + {\hat a}^\dagger _{-k} \right )\label{coi}\\
 {\cal{H}} = \sum _k \epsilon_k \: {\hat a}^\dagger _{k}\hat a_{k} + cnst 
 \enea
 In Eq.(\ref{coi0}) we have defined the velocity $v$ of these modes. 
 The approach is similar to the one for phonons. $\pi (x) $ plays the role of the space displacement $d(x,t )$, while $\nabla \theta $ plays the role of the phonon impulse $\Pi (x,t) $. The thermal conductance used in the text is given by 
 \bea
 � \Re e \{\kappa (\omega)\} = - \frac{1}{\omega } \Im m\!\! \left \{ \met \left \langle \left \{ J^{ {\cal{Q}}}_{-k}(-\omega) ,J^{ {\cal{Q}}}_{k}(\omega ) \right \} \right \rangle \right \}_{(\omega , k=0)} \!\!\! \!\!\! \!\!\! \!\!\!,\: \: 
 \label{corc}
 \enea
 where 
  \bea
 \met \left \langle \left \{ J^{ {\cal{Q}}}_{-k} ,J^{ {\cal{Q}}}_{k} \right \} \right \rangle_\omega = \frac{1}{\tau _0^2} \frac{ \kappa CT}{2 \hbar} 
 \left \langle \left \{ \pi_{-k} ,\pi _{k} \right \} \right \rangle_\omega. 
 \label{simo}
 \enea
 The symmetrized correlation on the right hand side, $ D^{\beta}(\omega ) =\left \langle \left \{ \pi_{-k} ,\pi _{k} \right \} \right \rangle_{\omega, k=0}$, apart of the prefactor $T$, can be evaluated at zero temperature in a standard way \cite{wen}. Eq.(\ref{corc}) gives: 
  \bea
  \Re e \{\kappa (\omega)\} = \pi^2\: k_B \frac{\omega \tau _0}{v^2 \tau_0^2}. 
  \label{vir}
  \enea
If $ \Re e \{\kappa (\omega)\} \sim \kappa \: \tau _0$ we get, from Eq.(\ref{coi0}), 
\bea
v^2 \tau _0^2 = \frac{ \kappa \tau _0 }{k_B} \: \frac{ CT}{\hbar / \tau _0} \freccia \kappa = \frac{ \pi k_B}{ \tau _0} \left ( \frac{ \hbar \omega } { C T }�\right )^{1/2} .
\label{cas}
\enea

However, introducing the damping of the mode in $D^{\beta}$, by adding an energy broadening $\Gamma$, we get
 \bea
 D^{\beta}(\omega ) =
 \frac{ \pi T }{v^4\Gamma} \int _0^{+\infty} \frac{2 \: e^{-\epsilon 0^+}\: \epsilon}{\epsilon ^2- \left (\omega + i \: \Gamma\right )^2} \: \epsilon ^2 d\epsilon 
\label{dbet}
\enea
and
\bea
\Im m \left \{ D^{\beta}(\omega )\right \}= - \frac{2 \pi T\: \omega ^2}{ v^4 \Gamma } \: \left \{ \left (1- \frac{\Gamma ^2}{\omega ^2 } \right ) \arctan \frac{\Gamma}{\omega }� + \frac{ \Gamma}{\omega} \: \ln \left ( 1+ \frac{\Gamma^2}{\omega^2 } \right ) \right \},\nonumber
\enea
which, in the limit $ \Gamma /\omega << 1$ gives:
\bea
 \Im m \left \{ D^{\beta}(\omega )\right \}= - \frac{2 \pi T }{v^4} \: \omega .
 \label{imdamp}
 \enea
 Posing again $ \Re e \{\kappa (\omega)\} \sim \kappa \: \tau _0$, in place of Eq.(\ref{cas}) we have: 
\bea
v^2 \tau _0^2 = \frac{ \kappa \tau _0 }{k_B} \: \frac{ CT}{\hbar / \tau _0} \freccia \kappa = k_B \Gamma 
 \left ( \frac{ h \: \Gamma } { C T }�\right )^{1/2}. 
 \label{kdis}
\enea


 Using the $\kappa = C \: v_F \ell = C \: v_F^2 \tau_0$, the gapless bosonic excitations of energy $\hbar v\: k$ generate a specific heat at fixed $2-d$ volume:
 \bea
\label{heat}
 \frac{ C_{{\cal{V}}}}{\tilde{a}^2} =\frac{d}{d T } \frac{1}{2\pi} \left ( \frac{ k_B T }{\hbar v } \right )^3 \hbar v \int _0^{+\infty} \frac{z^2 \: dz}{e^z-1 } \nonumber\\
 = k_B \frac{1}{2\pi} \left ( \frac{k_B T}{\hbar v} \right )^2 6 \:\zeta [3] 
 \enea
 and the thermal conductivity 
 \bea
 \kappa =\frac{ k_B }{\pi} \left ( \frac{k_B T}{\hbar v} \right )^2 3 \:\zeta [3] \: v \: \ell.
 \enea
 Here $\zeta[n]$ is the Riemann function\cite{abramovitz}. This is the Stefan-Boltzmann relation in two dimensions\cite{pientka}. 
 
 In the case of the SYK model, based on the saddle point contribution to energy\cite{maldacena}: 
 \bea
  \ln {\cal{Z}} = -\beta E_0 + S_0 + \frac{c}{2\beta} + ... ,\:\: \mbox{with }\:\: \frac{c}{2}= \frac{ 2\pi^2 \alpha _S N}{{\cal{J}}},
  \nonumber
  \enea 
 the first energy correction in temperature is $E = c/(2\beta^2)$ ($c\approx0.396 N/{\cal{J}}$), so that,
 by taking $CT \sim c/ (2 \beta^2) $ in Eq.(\ref{kdis}), we get:
\bea
 \kappa = k_B \: \Gamma \: \left ( \frac{ h \: \Gamma } { C T }� \right )^{1/2} \to k_B \: \Gamma ^{3/2} \left ( \hbar \beta \right )^{1/2} \: \left (2\pi \frac{ 2\: \beta } { c }� \right )^{1/2} \nonumber\\
 = k_B \: \Gamma ^{3/2} \left ( \hbar \beta \right )^{1/2} \: \left ( \frac{ \beta {\cal{J}}} { \pi \alpha _S N }� \right )^{1/2} 
 \enea
 (dimensions are $[C/t]$ as always). 
 
 The thermal conduction response in the conformal limit \cite{song} requires the energy current response function ${G_c}^{R\beta}_Q(\omega)$
 \bea
\frac{\Re e\{ \kappa \}}{NT} = -\lim _{\omega \to 0} \: \frac{1}{\omega NT} \Im m \left \{D_ {\cal{Q}}^{R\beta} ( \omega )\right \} \nonumber\\
= \frac{1}{NT \pi^2} \: \int d\omega \: \tanh \frac{\beta \omega}{2} \: \frac{ \partial }{\partial \omega } \left ( \omega \: \Im m \left \{ {G_c}^{R\beta}_Q(\omega)\right \} \right )^2\nonumber \propto \left (\frac{ \beta {\cal{J}}}{N}\right ). \hspace{6cm}
 \enea 
  In the Fermi liquid case, $ \tau_0 \sim T^{-2} $ and $ C \sim T$, so that $\kappa \sim T^{-1} $. 
 \section{The acoustic plasma mode in the marginal Fermi Liquid}
 \label{app:secE}
 
 To characterize the MFL phase, it  is important to check the nature of the collective excitations, in particular the particle-hole continuum,  under the action of the  increasing coupling to the high energy localized modes. We will show that, within our approximations, the real part of the $\omega(q) $ dispersion of the density excitations is linear, but with a small reduction of the physical velocity $ d\omega /dq$ at small $q$, and, most of all, a peculiar imaginary part. We also find that, at large couplings, the interaction pulls a linearly dispersed, well defined acoustic plasmon mode out of the particle-hole continuum.

  When the residual interaction is turned on, the vertex function $\Gamma (p,p-q;q,i\: \Omega )$ satisfies the Bethe-Salpeter equation\cite{fetter}, 
  \begin{align}
   \: \Gamma (p,p+q;q, \omega ) &=   \ n_{-\vec{q}} + i\:g\: U_c \frac{1}{ {\cal{V}} }\:\sum _{p'}D_{p',q}( \omega ) \: \Gamma (p', p+q,q; \omega ). 
 \label{pi-1}\\
  D_{p,q}( i\: \Omega ) &= \sum _{\omega _n} G( \epsilon_{ p-q},i\: \omega_n )\: G( \epsilon_p, i\: \omega_n + i\: \Omega_m) \nonumber\\
  G( \epsilon _{\vec{p}} , i\: \Omega ) &= \frac{ 1}{i Z^{-1} (\omega + \Omega ) - \epsilon _{\vec{p} }}.\hspace{2.5cm}
\nonumber
   \end{align}
  The functions $D_{p',q}( \omega )$ are related to the polarization functions of Eq.(65,66)  of the MT, when frequency is continued to real values and $p' \sim p_F$. We define 
  \bea
   \tilde{ \Pi}^{1,2}(q, \omega ) =-i\: \nu_0D^{1,2}_{q}( \omega ) \: \Gamma_{1,2} (p_F,p_F-q;q,\omega )
   \label{into}
   \enea
    where, in place of the $\sum _{p'} $ appearing in Eq.(\ref{pi-1}) we multiply by $ \nu _0$ after having put $|\vec{p'}|=p_F$. The resulting  functions $D^{1,2}_{q}$ of Eq.(\ref{into}) are redefined as ($ \frac{\omega }{ Z\tilde{v}_F \: q } < 1$)
    \bea
    \nu_0D^{1}_{q} (\omega ) =Z \nu_0 \: \left [1-i\: \frac{\omega}{Z\tilde{v}_F \: q \sqrt{ 1- \frac{\omega ^2}{ \left (Z\tilde{v}_F \: q \right )^2 }}} \right ]
   , \nonumber\\ 
    \nu_0{\cal{D}}^2_{q}( \omega ) = \left . \nu_0 \int \frac{d\theta}{2\pi}\:  (1-\cos\theta )\: D^2_{\vec{k}, \vec{q} } (\omega ) \right |_{k=p_F}.
   \label{fo20}
 \enea
 In fact,  following [\onlinecite{chowdhurySenthil}], we consider two ranges of energy values: a low energy one ($\omega < \Omega _c^*$), ($i=1$), and an high energy one ($\omega < \Omega _c^*$), ($i=2$), with 
 $\nu_0 D^i_{k_F,q }(q,\omega ) = \Pi ^i( q ,\omega )$.
 
 Limiting ourselves to the FL energy range, $i=1$, for the moment, Eq.(\ref{into}), becomes 
 \bea
 i\: \tilde{ \Pi}^1(q,\omega )= - i\: \nu_0D^{1}_{p_F,q}( \omega ) \Gamma (p_F,p_F-q;q,\omega ), 
 \label{vafa2} 
 \enea
  \bea
{\rm where} \:\:\:\:\:\: \nu_0 \: {\cal{A}}= \frac{1}{(2\pi)^2} \:2\pi k_F \frac{{\cal{A}}}{\hbar v_F^*} = \frac{1}{\pi }\frac{k _F\:{\cal{A}}}{\hbar v_F^*}.\nonumber
 \enea
$ \nu_0 $ is the 2-d density of states at $\epsilon_F$ per unit volume $ {\cal{A}}$ .

 Here we are assuming that, in this energy range, $ \: \Gamma (p,p-q;q,i\: \Omega )$ does not depend on the angle $ \theta_{p',q} =\widehat {p' q}$ except for an average of $ \sin^2 \theta/2 \sim ( q/(2k_F ))^2 \sim 1/2 $ . We have also put $ |p|,|p' |=k_F$, so that the only $ \Gamma $ dependence is $ \Gamma (q,i\: \Omega )$. 
 This choice, together with that of the onsite interaction $U_c$, provides the reference result we are looking for.


 
 We expect a collective mode of compressibility type embedded in the 
 particle-hole excitation continuum. In the low temperature Fermi liquid limit, the p-h continuum has a boundary of the kind $ min\left \{ \epsilon _{k+q} -\epsilon _{k} \right \}= Z \tilde{v}_F q$. We find a collective mode $\omega = Z \: z\tilde{v}_F q$ with $z$ complex and $\Re e \{z\} <1$ and negative imaginary part which is related to the lifetime of the mode. 
 
 Here
 \bea 
 \Pi ^1 ( q, \omega ) = Z \nu_0 \: \left [1-i\: \frac{\omega}{Z\tilde{v}_F \: q \sqrt{ 1- \frac{\omega ^2}{ \left (Z\tilde{v}_F \: q \right )^2 }}} \right ], \nonumber
 \enea

 So that 
 \bea
   \left \{ {\left [ \nu_0D^{1}_{p_F,q}( \omega ) \right ] }^{-1} - i\: g\: U_c\right \}\left [ i\: \tilde{ \Pi}(q,\omega ) \right ] = -i\: n_{-\vec{q}} 
   \nonumber\\
 \enea
 
 This provides the equation for the FL collective excitation mode ( $ Z \nu_0 = U_c^{-1}$):
 \bea 
 \left [ 1-i\: \frac{ z}{\sqrt{1-z^2}} \right ] ^{-1} -i\: g =0 , \:\:\:\: z=\frac{ \omega }{Z\tilde{v}_F \: q } <1
 \nonumber\\
\freccia 1-i\: g = g\frac{ z}{\sqrt{1-z^2}}. 
 \label{seq}
\enea
The homogeneous equation can be cast in the form: $ 1/\chi^{(0)} -U =0$ if
\bea
i\: \nu_0 D^1_{k_F,q }(q,\omega ) = i\: \Pi^{1(0)} (q,\omega ) = - i\: \chi ^{(0)}. \label{free}
\enea
 The contribution to the polarization function from high-energy excitations $\Pi^2(q, i\: \Omega )$ has a completely local q-independent form and is given by 
 \bea
 \Pi^2 ( i \: \Omega ) \approx - \frac{8}{ {\cal{J}}}\: \ln \left (\frac{ {\cal{J}}}{W} \right ). \:\:\:\: \:\:\:\: \:\:\:\: (\Omega^*_c >> \Omega>0)
   \nonumber
 \enea

 In the case of $ \nu_0D^{1}_{p',q}$, it was enough to put $ p'\to p_F$, but, in the case of $  \tilde{ \Pi}^{2}_{q}( \omega )$, it is important to keep a dependence on the scattering angle $ \theta_{\vec{k},\vec{q}} =\widehat {\vec{k},\vec{q}}$ explicitely, and to integrate over it. In fact, we cannot neglect the dependence of the vertex $\Gamma$ appearing in the Bethe-Salpeter Eq.(\ref{pi-1}) on the scattering angle $ \theta_{\vec{k},\vec{q}} =\widehat {\vec{k},\vec{q}}$. As it is usual when calculating the dc conductivity in metals\cite{ashcroft}, we assume that this angular dependence is $ \sim (1-\cos\theta )$ in Eq.(\ref{fo20}). The factor $\left ( 1-\cos \theta _{\vec{k},\vec{q}}� \right ) $  expresses the growing predominance of forward scattering with declining temperature, which  contributes less than wide angle scattering, to the effective "collision rates". A $q-$dependence has been added in $ D^{2}_{\vec{k}, \vec{q} }( \omega )$ of Eq.(\ref{fo20}) by including a $k-$dependent correction  to the conformal local SYK Green function $G_c$:

 Here we use $G_c (i\: \omega _n) $, extended to include a $k$ dependence 
 \bea
G^2(\vec{k},i\omega_n ) \approx -\frac{1}{ \Sigma _c^{0}(i\:\omega_n )} + \frac{ \epsilon_{\vec{k}}}{\left |  \Sigma _c^{0}(i\:\omega_n )\right |^2},
\label{nfl0}
  \enea 
  Here  $\Sigma _c^{0}(i\:\omega_n ) = \left [G _c^{0}(i\:\omega_n ) \right ]^{-1} $  is the self-energy corresponding to the zero order approximation. 
Eq.(\ref{nfl0}) can be viewed as an expansion of the high energy total Green function to lowest order in $\epsilon _k$\cite{chowdhurySenthil}, or can be obtained, by assuming random hopping between sites.
 so that, according to Eq.(\ref{into}), the propagator is: 
 \bea 
 \nu_0 D^2_{ ,q} ( i \: \Omega _m ) = \left ( \frac{1}{{\cal{V}} }\:\sum _{k}\right ) \sum_{\omega_n}\frac{1}{ \Sigma _c^{0}(i\:\omega )} \frac{1}{ \Sigma _c^{0}(i\:\omega+i\Omega )} \nonumber\\
 \left [ 1+ \left ( \frac{\epsilon_{k-q} }{ {\Sigma _c^{0}}^*(i\:\omega )} + \frac{\epsilon_{k} }{ {\Sigma _c^{0}}^*(i\:\omega +i\Omega )} \right )+ \frac{\epsilon_{k-q} }{{ \Sigma _c^{0}}^*(i\:\omega )}\frac{\epsilon_{k} }{ {\Sigma _c^{0}}^*(i\:\omega +i\Omega )} \right ] \nonumber
 \enea
 \bea
  \label{pion1}
\enea
The $\omega_n $ sum can be transformed into an integral. A lengthy but straightforward calculation provides a function of $k$,$q$, $\vec{k}\cdot \vec{q} = k\: q \cos\theta $ and $i\: \Omega$: 
\bea
 f(k,q,\cos\theta ; i\Omega )= -\frac{8}{U_c} \: \ln \frac{U_c}{W} +\frac{ \epsilon_k \: \epsilon_{k-q}}{U_c^2} \: \frac{2}{\Omega } \: {\rm arctanh} \frac{\Omega }{\Omega _c^* } 
 + \frac{2\: i}{U_c }\: \left \{ - \frac{\left (\epsilon_{k-q}-\epsilon_{k} \right )}{W}+ \frac{\left (\epsilon_{k-q}+\epsilon_{k} \right )}{W} \sqrt{\frac{\Omega _c^*}{\Omega }} \: {\rm arctanh} \frac{\Omega }{\Omega _c^* } \right \} \nonumber
 \enea 
 
 We neglect the second and the fourth term ( for a p-h pair in a p-h symmetric system is $\epsilon_{k-q}+\epsilon_{k} = 0 $), obtaining 
 \bea 
 f(k,q,\cos\theta ; \Omega )\approx -\frac{8}{U_c} \: \ln \frac{U_c}{W} + \frac{2\: i}{U_c }\: \left \{ - \frac{\left (\epsilon_{k-q}-\epsilon_{k} \right )}{W} \right \}.
\nonumber
\enea
 

 We finally get:
 \bea
 \nu _0 {\textit{D}}^2_{,q} ( \omega ) = 
 -\nu_0 W
 \frac{8}{U_c} \: \ln \frac{U_c}{W} \: \left [1 +i\: \frac{1}{8\: \ln \frac{U_c}{W} }\: \frac{ \tilde{v}_F q}{W} \right ].
 \enea 
 We now rewrite Eq.(\ref{pi-1})  for the vertex as follows ( $i = 1,2$):
 \bea
   \: \Gamma (p_F,p_F-q;q,\omega ) = n_{-\vec q} + U_c\: \left [ g_{i1}\: \nu_0D^1_{,q}( \omega ) + g_{i2} \: \nu_0{\cal{D}}^2_{,q}( \omega ) \right ] \: \Gamma (p_F,p_F-q;q,\omega ).\nonumber
   \enea
  By posing
  \bea
  \tilde{ \Pi}^{1,2}(q, \omega ) =-i\: \nu_0D^{1,2}_{,q}( \omega ) \: \Gamma (p_F,p_F-q;q,\omega )
   \label{vert1}
 \enea
we get a system, which, using $Z\nu_oU_c=1$ and $ z = \omega /(Z \tilde{v}_F \: q)$, takes the form:

\bea 
\left ( \begin{array}{cc}\left [ 1-i\: \frac{z}{\sqrt{1-z^2}} \right ]^{-1} -i\: g_{11} & i\: g_{12} \\
 i\: g_{12} &- \frac{Z}{8 \: \frac{W}{U_c} \: \ln \frac{U_c}{W} } \left [1 +i\: \frac{1}{8\: \ln \frac{U_c}{W} }\: \frac{ \tilde{v}_F q}{W} \right ]^{-1} -i\: g_{22} \end{array} \right ) \left ( \begin{array}{c}  i \: \tilde{ \Pi}^{1}(q, \omega )\\ i\: \tilde{ \Pi}^{2}(q, \omega )\end{array} \right ) = -i\: n_{-q} \left ( \begin{array}{c} 1 \\ 1\end{array} \right ) 
\label{seq2}
\enea
The results are given in Fig.1,2 for real and imaginary part of $\omega/( Z \tilde{v}_F q ) $ vs $g= U/U_c$.

 When $q$ increases there is a monotonic flattening of $ \Im m \{\omega _q\} $ with $ Z\tilde{v}_F q/W$ with a saturation at large $q$, as can be seen by plotting $ d\omega _q/ d q$ vs. $q$. 
As $ \Re e \{\omega \} $ is strictly linear with $q$ in a large range of values of $q$, the plot of Fig.(\ref{deriv}) shows the behavior of this derivative. 
 

The polarization function of the coupled system, given by Eq.(\ref{into}), satisfies the equation:
 \bea 
\left ( \begin{array}{cc}\left [ \nu_0D^{1}_{q}( \omega ) \right ]^{-1} -i\: g_{11} \:U_c& i\: g_{12} \: U_c\\
 i\: g_{12} \: U_c & \left [ \nu_0{D}^{2}_{q}( \omega ) \right ]^{-1} -i\: g_{22} \:  U_c \end{array} \right ) \left ( \begin{array}{c}   i \: \tilde{ \Pi}^{1}(q, \omega )\\  i\: \tilde{ \Pi}^{2}(q, \omega )\end{array} \right ) = -i\: n_{-q} \left ( \begin{array}{c}  1 \\  1\end{array} \right ). 
\label{sys0}
\enea
Given a transferred momentum $q$, the energy of the corresponding collective excitation makes the determinant of the matrix on the left hand side of Eq.(\ref{sys0}) vanish. 
Here, $g_{11} U_c$ is assumed to be the residual interaction within the low energy FL due to the SYK cluster, $g_{22} U_c \sim \caj$ parametrizes the interaction within the SYK cluster, while $g_{12} U_c $ provides the coupling between the two.

 To zero order in perturbation, $ \chi^{(0)} =  \nu_0D^{1}_{q} (\omega ) $, in the limit $ g \to 0 $ the solution is $z\to 1$, giving a strictly linear dispersion, $ \omega \propto v_F^* q$.

When the couplings are non vanishing, the mode dispersion keeps being substantially linear, but the physical velocity is renormalized and an imaginary part arises. The real and  imaginary part of the function $ z= \omega/( {v}_F^* q ) $ are plotted in Fig.s(\ref{frerexo},\ref{fimrexo}) as a function of $g_{11}$ and $g_{22} $, for $ {v}_F^* q /W= 0.1$.  Plots are reported for $ U_c/W = 10$ and $Z=0.1$. The limitation $z=\frac{ \omega }{{v}_F^* \: q } <1$ implies that we track the collective excitation mode inside the p-h continuum only. In Fig.(\ref{frerexo}) we plot the real part  $\Re e\{ z\} $ vs $ g_{11}$, choosing $ g_{22} =2$, when the ratio of the couplings between the two systems $ b = g_{12} / g_{11} $is $b= 0,0.4, 0.6, 0.8$. 



 The effective velocity of the excitation mode, $\Re e\{ z\} $, decreases with respect to the unperturbed value $ v_F^*$ and saturates to about $90\%$ of the unperturbed value when $ g_{12}$ increases. As shown in the inset, the saturation is even faster, when $g_{11} $ is kept fixed (in our case at the value $ g_{11} = 1 $) and $ g_{22} > g_{11} $ is increased. 

The imaginary part of the energy of the mode $ \Im m \{\frac{ \omega }{{v}_F^* \: q }\}$ is zero at $ g_{11} = 0$ and increases mildly, in absolute value, with increasing of $ g_{11} $, as reported in Fig.({\ref{fimrexo}) vs $ g_{11} $ at  $ g_{22} = 1.8 $. It is remarkable that, when $ g_{11} > g_{22} $, $ \Im m \{z\} $ vanishes. Simultaneously the slope of the mode increases up to value one, for $b \to 1$. This appears in Fig.({\ref{fimrexo}) and, more explicitely, in Fig.({\ref{vsb}) which is a plot vs $b$ with $ g_{11} =g_{22} $ for various values $g_{ii} = 0.8, 1.2,1,8$ and $ {v}_F^* q /W= 0.8$. When $g_{12}$ increases in Eq.(\ref{sys0}), a real term  $- g_{12}^2$ grows in the determinant, which reduces $\Im m\{ z\}$. In these conditions, the dispersion tends to the boundary of the particle-hole continuum ($\Re e\{z\}\to 1$), while the corresponding imaginary part in Fig.\ref{fimrexo} vanishes. This feature appears clearly in Fig.s \ref{vsb} {\bf a),b)} and is the signature of the splitting of a bound state out of the particle-hole continuum with linear dispersion and velocity $ > {v}_F^*$. We interpret this as an acoustic plasmon which, however, requires strong coupling of the MFL to the SYK cluster, to be tackled even further, far beyond the present perturbative approach. 
 When $q$ increases, there is a monotonic increase of $ \Im m\{\omega _q\} $ with $ {v}_F^* q/W$ and a saturation at large $q$, as can be seen by plotting $ d\omega _q/ d q$ vs. $q$ (see Fig. \ref{deriv} ). 
 \begin{figure}
 \centering
\def\big{\includegraphics[height=4.8cm]{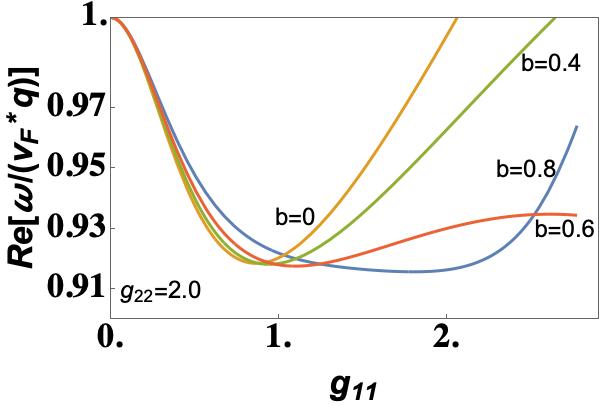}}
\def\little{\includegraphics[height=3.4cm]{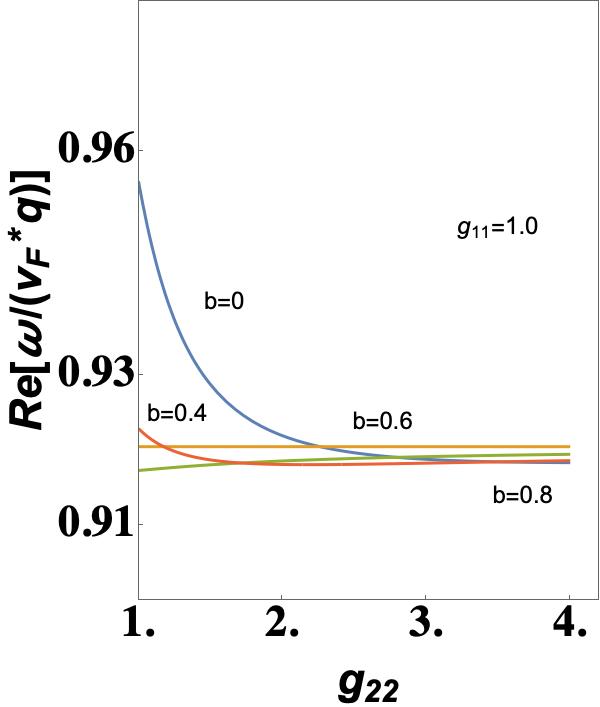}}
\def\stackalignment{l}
\topinset{\little}{\big}{-30pt}{+65pt}
\caption{Renormalization of the velocity of the linear dispersion $v_F^*$ in presence of coupling to high energy correlations. Here we plot the real part of $\omega/( v_F^* q ) $ versus $g_{11}$ at $g_{22} =1.8$ and $ v_F^* q /W= 0.1$. Here $g_{12} = b \: g_{11}$ is the coupling strength to the high energy SYK fluctuations ($b=0, 0.4, 0.6, 0.8$, respectively, with $ U_c/W = 10$ and $Z=0.1$).  } 
\label{frerexo}
\end{figure}

 \begin{figure}
   \includegraphics[height=48mm]{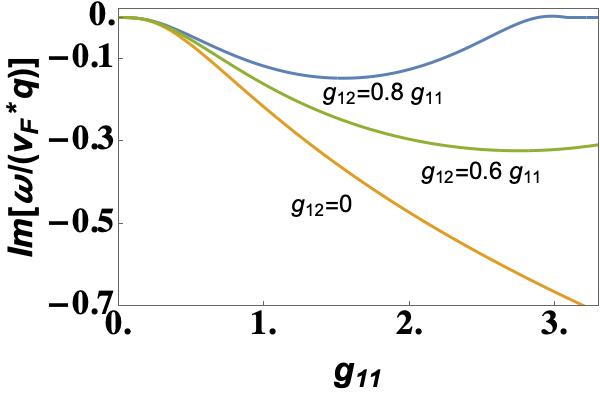}
\caption{ Imaginary part of $\omega/( v_F^* q ) $ versus $g_{11}$ corresponding to the real part of Fig.\ref{frerexo} ($g_{12} = b g_{11}$, with $b=0, 0.6, 0.8$, respectively), at $g_{22} =1.8$ and $ v_F^* q /W= 0.1$. Note the flattening at zero for $g_{11} > 3.0$, when $b \to 1$. } 
  \label{fimrexo} 
\end{figure}

\begin{figure}
\begin{minipage}{.3\textwidth}
 \includegraphics[width=0.9\linewidth]{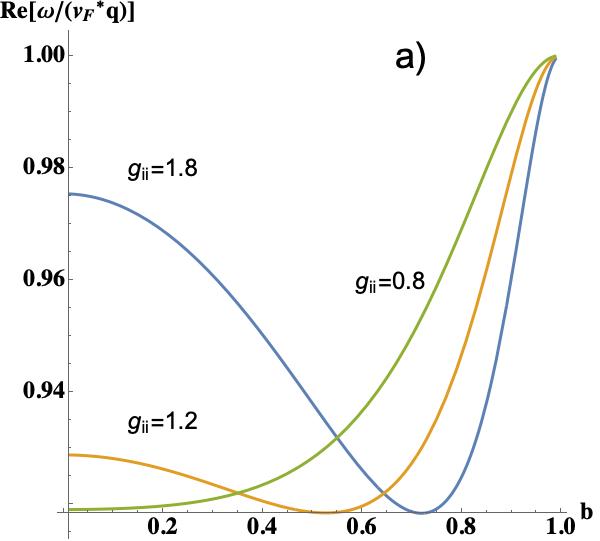}
\end{minipage}
\begin{minipage}{.3\textwidth}
 \includegraphics[width=0.9\linewidth]{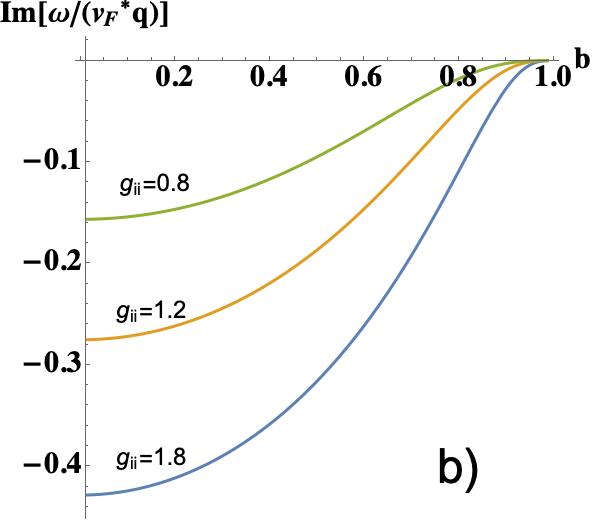}
\end{minipage}
\hspace{0.5cm}
 \caption{ Real [{\bf a)}] ad Imaginary [{\bf b)}] part of $\omega/({v}_F^* q ) $ versus $b= g_{12}/g_{11}$ for various values of $g_{ii} \equiv g_{11} = g_{22}$, with $g_{ii}=0.8,1.2,1.8$, respectively, at $ {v}_F^* q /W= 0.8$. } 
  \label{vsb}
\end{figure}
\begin{figure}
  \includegraphics[height=70mm]{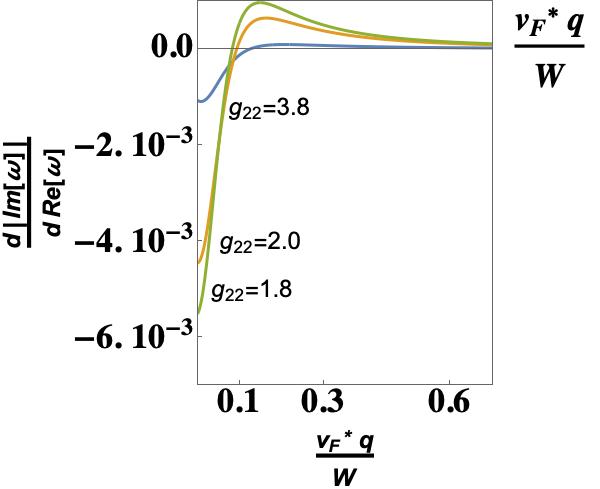}
\caption{ Plot of $ \frac{\partial \Im m \{\omega \} }{\partial \Re e \{\omega \} } $ vs ${v}_F^* q / W $. As $ \Re e \{\omega \} $ is strictly linear with $q$ in a large range of values of $q$, the derivative with respect to $ \Re e \{\omega \} $ can be interpreted as derivative with respect to $q$. 
Here $g_{11} =g_{22} = 1.8$ and $g_{12} =0.8 \times g_{11}$ ($Z=0.1, \: U_c/ W =10.$, as always)} 
 \label{deriv} 
\end{figure}

To sum up the results of this Appendix, we can conclude that the low energy FL on the lattice appears quite robust with respect to interaction with incoherent local disordered SYK clusters, when only the lowest perturbative order is included and no disorder, in the continuum, $k\to 0$, limit. The Fermi surface is still well defined, but the liquid becomes a MFL. The hydrodynamic collective excitation, the would-be acoustic plasmon, is also rather well defined. At strong coupling, in the limit $U_c \to \caj$, its dispersion tends to the boundary of the p-h continuum and the imaginary part, which blurs the mode, vanishes. We expect that a well defined acoustic plasmon is on the verge to emerge as a bound state at low energies, splitted off the p-h continuum. 



 \section{ The superconducting critical temperature at low temperature}
 \label{app:secF}
 In this Appendix we use an Eliasberg\cite{eliashberg} approach to the superconducting critical temperature, $T_c$, assuming that pairing is driven by the diffusive excitation modes introduced in Section D. As explained in the main text, $\Omega _c^*$ is the energy scale of $T_c$ and the dependence on the coupling strength turns out to be non BCS-like. We report here the derivation. 

 In the mean field Hamiltonian, in the Nambu\cite{nambu} representation, 
 \bea
 H (k) \equiv \sum _{ij} \: \left [ h_{ij} \: \hat{c}^\dagger_i\: \hat{c}_j + \Delta _{ij} c^\dagger _i c^\dagger _j + h.c. \right ] \nonumber\\
 = \sum _k \begin{array}{c} \left ( c_{\vec{k}} \: ^\dagger \:\: c_{- \vec{k}}  \right ) \\ \hspace{1cm} \end{array} \: \left ( \begin{array}{cc} \xi_{k} & \Delta \\ \Delta & -\xi_k \end{array}\right ) \left ( \begin{array}{cc} c_{\vec{k}} \\ c_{-\vec{ k}} \: ^\dagger \end{array} \right ) 
 \label{helicalH}
 \enea
 the one electron Green's function ${\cal{G}} (p, i\: \omega _\nu )$ and the electronic self-energy $\Sigma (p, i\: \omega _\nu )$ are $2 \times 2$ matrix defined by the Dyson equation
 \bea
\left [{\cal{G}} (p, i\: \omega _\nu )\right]^{-1} = [G_0 (p, i\: \omega _\nu)]^{-1} - \Sigma (p, i\: \omega _\nu), \label{mop}\\ 
= i\: Z^{-1} \omega _\nu- \tilde{\xi}_p \sigma_3 - \phi (p, i\: \omega _\nu) \: \sigma_1, 
 \enea
 where $G_0 (p, i\: \omega _\nu)$ is the one-electron Green' s function for the non interacting system, $\sigma _i$ are Pauli matrices and $\xi_k= \epsilon_k-\mu$. The $\epsilon_k$'s are single particle electron energies and $\mu $ is the chemical potential.
We do not include Coulomb electron-electron interaction, so that the self-energy $\Sigma \propto \sigma _1 $ is just offdiagonal. The approximation used for the self-energy is \cite{sara}:
 \bea
 \Sigma (p, i\: \omega _\nu)= - \frac{1}{\beta} \sum _{p' \nu'} \sigma _3 \: G(p', i\: \omega _{\nu'})\: \sigma _3\: \left | g ( p, \: p') \right |^2 {\cal{D}} 
 \left ( p-p', i\: \omega _\nu- i\: \omega _{\nu'} \right ),
 \enea
 where $g ( p, \: p') $ is the coupling with the bosonic modes and $ {\cal{D}} \left ( p-p', i\: \Omega _n\right )$ is 
 the response function in imaginary frequency to the bosonic modes.
 The latter can be represented in terms of its imaginary part, $ {\cal{B}}(q, i\: \Omega _n) = -\frac{1}{\pi} \Im m\left \{ {\cal{D}} \left ( q, i\: \Omega _n\right ) \right \} $ as 
\bea
 {\cal{D}} \left ( q, i\: \omega _\nu- i\: \omega _{\nu'} \right ) = \int _0^{\infty} d\Omega \: {\cal{B}}(q,\Omega) \left \{�\frac{1}{ i\: \omega _\nu- i\: \omega _{\nu'} -\Omega }-�\frac{1}{ i\: \omega _\nu- i\: \omega _{\nu'} +\Omega }\right \} 
 \nonumber
 \enea
where ( $ \Omega _q=\tilde{D}_Q q^2 $ ) and, in our case 
 \bea
 {\cal{B}}(q, i\: \Omega ) = -\frac{1}{\pi}\: \left [ \frac{ \Omega _q\:\Omega }{\Omega ^2+ \Omega _q^2} \right ]\: {\cal{T}}_Q 
 \label{diffo}
 \enea
 To make matches with the usual Eliashberg theory, the isotropic gap model which we consider here provides the dimensionless coupling function : 
 \bea
 \alpha^2F(\Omega )= \!\frac{1}{N_q}\!\! \sum_q \! \delta ( \Omega -\Omega _q) \nu(0)\!\! \!\int d\omega ' \left | g ( k_F, \omega ';q) \right |^2\!\!\!,
 \enea
 as an integral over the transferred momentum $q$. We have to integrate over all $q 's$:
\bea
\frac{1}{N_q} \: \sum _q \freccia \frac{\tilde{a}^2}{(2\pi)^2} 2\pi \: \int q \frac{ d\: q}{ d \Omega_q } \: d\Omega_q = \frac{\tilde{a}^2}{2\pi} \frac{1}{2 \: \tilde{D}_Q } \int d\Omega _q, \nonumber
\enea
where $\tilde{a}^2$ is the average diffusion area. We obtain
\bea
 \Sigma (p, i\: \omega _\nu)= - \frac{1}{\beta} \sum _{p' \nu'} \sigma _3 \: {\cal{G}}(p', i\: \omega _{\nu'})\: \sigma _3\: \int d\Omega \frac{ 1}{4\pi\:} \frac{\tilde{a}^2}{\tilde{D}_Q }\int d\Omega _q \left | g_{p,p'}(\Omega _q) \right |^2 {\cal{B}}(\Omega _q,\Omega) \: \left \{�\frac{1}{ i\: \omega _\nu- i\: \omega _{\nu'} -\Omega }-�\frac{1}{ i\: \omega _\nu- i\: \omega _{\nu'} +\Omega }\right \} \nonumber
 \enea
 with 
 \bea
 \left [ {\cal{G}} (p, \omega) \right]^{-1} = Z^{-1} \: { \omega} {\bf 1} -\left \{ \xi_p 
 - i\: Z^{-1} \left [|\omega| \: \frac{\epsilon_F }{\Omega ^* } \:\ln \left ( \frac{{\cal{J}} }{W }\right )+ \frac{ \alpha }{Z} \: \nu_0 |\omega |^2 \: \ln \frac{ Z \tilde{v}_F k_F }{|\omega | } \right ]\: sign (\omega )\right \} \sigma _3 - \Xi ( \omega ) \sigma _1
 \label{giga}
 \enea
  The mean field $ \Delta (\omega ) = Z (\omega ) \Xi (\omega ) $ has to be determined in the following. 
  
In Eq.(\ref{giga}) the inverse lifetime of the quasiparticles of Eq.(\ref{tsca}) of MT appears. Here $\alpha $ is a numerical factor of order one. 
 
  Note the difference with the usual approach: diffusivity implies Eq.(\ref{diffo}) here, while usually is $ {\cal{B}}(q,\Omega)\sim \frac{\Omega_q }{\Omega_q ^2+ 	\Omega ^2}$ in the Eliashberg approach. 
  
  On the other hand, from Eq.(\ref{mop}): 
 The final form of the selfenergy is:
 \bea
 \Sigma (p, \omega ) = \nu_0 \int _{-\infty}^\infty d\omega '\: \Re e \left \{ \frac{ Z^{-1} \omega ' {\bf 1} - \Xi ( \omega ') \: \sigma_1}{\left [ {\cal{P}} (\omega ')\right ]^{1/2}} \right \} \: \int _{0}^\infty d\Omega \int \frac{d\Omega _q}{4\pi} \left | g_{p,p'}(\Omega _q) \right |^2 {\cal{B}}(\Omega _q,\Omega) \nonumber\\
\times \left [ \frac{ f( -\omega ') }{ \omega -\omega ' -\Omega +i\: 0^+}+ \frac{ f( \omega ') }{ \omega -\omega ' +\Omega +i\: 0^+}+ \frac{ N( \Omega ) }{ \omega -\omega ' -\Omega +i\: 0^+}+ \frac{ N( \Omega ) }{ \omega -\omega ' +\Omega +i\: 0^+}\right ], \nonumber\\
{\cal{P}}(\omega ) = Z^{-2} \: { \omega} ^2 + Z^{-2} \:
\left [|\omega| \:  \frac{\epsilon_F }{\Omega ^* } \:\ln \left ( \frac{ {\cal{J}} }{ W}\right )+ \frac{ \alpha }{Z} \: \nu_0 |\omega |^2 \: \ln \frac{ Z \tilde{v}_F k_F }{|\omega | } \right ]^2 - \Xi ^2( \omega ), 
 \label{filo}
 \enea
where $ N( \Omega ) = \left [e^{\beta \Omega }-1\right ]^{-1}$ and $f( \omega ) = \left [e^{\beta \omega }+1\right ]^{-1}$ are the Bose and Fermi occupation probabilities. The term in curly brackets arises from $ \Im m \left \{ \nu_0 \int _{-\infty}^{+\infty} d \xi_{p'} \: \sigma _3 
 {\cal{G}} (p', \omega) \: \sigma_3 \right \}$, which turns into a real part from the inverse of $ \left [ {\cal{G}} (p', \omega) \right]^{-1} $ given in Eq.(\ref{giga}). The critical temperature is the one at which $\Delta \sim 0 $ and can be dropped in the denominator, but the gap equation has to be satisfied. 


 In all the further calculations we neglect the thermal excitations and drop $N(\Omega)$. Observing that the integration variable $\Omega _q$ has the meaning of the diffusive energy (see Eq.(\ref{diffo})) it is clear that it cannot be integrated at energies above $\Omega$. We also use the parameter equality $Z \nu _0 = U_c^{-1}$ and we take $ \left | g_{k_F,\omega '}(\Omega _q) \right |^2 = g^2 $ constant  ($[ g]^{-1} \sim  \: time \: \: (\hbar =1\: $  here)). We get:
\bea
\Delta (\omega )\approx \frac{\tilde{a}^2}{\tilde{D}_Q } \int _{0}^\infty \frac{ d\omega '}{ |\omega '| \: \frac{ U_c \epsilon _F}{ W^2} \:\ln \left ( \frac{U_c}{W }\right )} \: \Re e \left \{ \Delta (\omega' )\right \} \:
 \:\frac{1}{U_c} \int _{0}^{U_c} d\Omega \nonumber\\
\times \int_0^\Omega \frac{d\Omega _q}{4\pi} \left | g_{k_F,\omega '}(\Omega _q) \right |^2\left ( \frac{1}{\pi}\: \left [ \frac{ \Omega _q\:\Omega }{\Omega ^2+ \Omega _q^2} \right ]\: {\cal{T}}_Q \right ) \: 2 \left \{ f(-\omega' ) \frac{1}{ \Omega +\omega ' }- f(\omega') 
\frac{1}{\Omega -\omega ' }\right\} . \label{ferro}
 \enea

 We concentrate on $\omega =0$ and we deal with two contributions to $\Delta (\omega ')$ separately, $\Delta (\omega' ) =\Delta_a+\Delta_b$, where the first arises from integration over $0< \omega' < \Omega^* \sim W^2/U_c$ and the second from integration over $\Omega^* < \omega' < U_c$. 

In the first case, observing that the range of $\omega '$ cannot be larger than $\Omega^* $, but the Fermi function selects $\omega ' \sim 0$, we neglect $ \omega '$ in the denominators of the curly bracket obtaining \cite{tinkham}: 
 
 \bea
  \Delta_a (0 ) = \frac{\tilde{a}_\ell^2}{\tilde{D}_Q } \int _{0}^{\Omega^*} d\omega '\: \frac{\Delta _o }{ \omega ' \: \frac{ \epsilon _F}{ \Omega^*} \:\ln \left ( \frac{ {\cal{J}} }{W }\right )} \: 
  \: \frac{|g|^2 \ln 2}{4\: U_c} \int _{0}^{U_c} \frac{d\Omega}{2\pi} \:
  \: {\cal{T}}_Q  \left \{  f(-\omega' ) - f(\omega') \right\}  \nonumber\\
  \approx \frac{ \tilde{a}_\ell^2 }{\tilde{D}_Q} {\cal{T}}_Q \: \frac{ |g|^2 \ln 2 }{8\pi}  \frac{ \Delta _o}{ \frac{  \epsilon _F}{ \Omega^*} \:\ln \left ( \frac{ {\cal{J}} }{W }\right )} \:  \int _{0}^{\Omega^*} \frac{d\omega ' }{ \omega ' } \: \tanh \frac{\beta \omega '}{2}
 \approx \frac{ \tilde{a}_\ell^2 }{\tilde{D}_Q} {\cal{T}}_Q \: \frac{ |g|^2 \ln 2 }{8\pi}  \frac{ \Delta _o}{ \frac{  \epsilon _F}{ \Omega^*} \:\ln \left ( \frac{ {\cal{J}} }{W }\right )} \: \ln { \beta _c\Omega^*}.
\nonumber
 \enea
 
 Now the $\Delta _b$ contribution. We neglect $\Omega $ in the denominator in the curly bracket of Eq.(\ref{ferro}) and we include the ${\omega '}^2$ term of the inverse lifetime, only.
 \bea
 \Delta_b (0 ) = \frac{\tilde{a}^2}{\tilde{D}_Q} \int ^{U_c}_{\Omega_c^*} d\omega '\: \frac{\Delta _{\infty} }{ \frac{ \alpha }{Z} \: \nu_0 |\omega' |^2\: \ln \frac{ Z \tilde{v}_F k_F }{|\omega' | } } \: \: \frac{1}{U_c} \int _{0}^{U_c} d\Omega \int_0^\Omega \frac{d\Omega _q}{4\pi} \left | g_{k_F,\omega '}(\Omega _q) \right |^2 \nonumber\\
\times \left ( \frac{1}{\pi}\: \left [ \frac{ \Omega _q\:\Omega }{\Omega ^2+ \Omega _q^2} \right ]\: {\cal{T}}_Q \right ) 
\: 2 \left \{ f(-\omega' ) \frac{1}{ \Omega +\omega ' }- f(\omega') 
\frac{1}{\Omega -\omega ' }\right\} \nonumber\\
=\frac{\tilde{a}^2}{ \tilde{D}_Q } \: \frac{ |g|^2 \ln 2 }{4\pi^2} \: {\cal{T}}_Q \: \int ^{U_c}_{\Omega_c^*} d\omega '\: \frac{\Delta _{\infty} }{ \frac{ \alpha }{Z} \: \nu_0 |\omega' |^2 \: \ln \frac{ Z \tilde{v}_F k_F }{|\omega' | } } \: \: \frac{1}{U_c}
 \int _{0}^{U_c}2 \Omega\: d\Omega \: \frac{1}{ \Omega +\omega ' } \nonumber
\enea
where we have disregarded the terms $\propto e^{- \beta \omega '}$ in the last integral.  The rest of the derivation can be found in the main text.

\twocolumngrid
\bibliography{biblioscotto2}

\begin{thebibliography}{68}
\bibitem[{ord()}]{ordini}
\bibinfo{note}{Checking ${\cal{O}} ( \beta{\cal{J}}) $: Green's function $G_c $
  has a $b$ with $b^2 = 1/( 2 \pi^{1/2} J) $ so that $ G_c \sim b
  /\sqrt{\beta}\sim 1 /\sqrt{ \beta {\cal{J}} }$. $K = R_c G_cG_c R_c \sim
  \sqrt{\beta {\cal{J}} }\: \frac{1}{\beta {\cal{J}} }\: \sqrt{\beta {\cal{J}}
  } \sim 1$. ${\cal{F}}= R_c^{-1} K_c [ K_c-1]^{-1} R_c^{-1} \sim \frac{1}{
  {\cal{J}}} \times {\cal{J}} \sim 1 $ because $ [ K_c-1]^{-1} \sim \beta
  {\cal{J}}$.}
\end{thebibliography}

\end{document}